\documentclass[
    reprint,
    onecolumn,
    superscriptaddress,
    amsmath, amssymb,
    aps,
    12pt
]{revtex4-2}

\usepackage[usenames, dvipsnames]{xcolor}
\usepackage{graphicx, overpic, rotating}
\usepackage{dcolumn} % Align table columns on decimal point
\usepackage{xpatch}
    \makeatletter \xpretocmd \start@align{\linenomathWithnumbers}{}{\fail}
\usepackage{mathtools, url}

\renewcommand{\vec}[1]{\boldsymbol{#1}}

\newcommand{\DD}{\mathrm{D}}
\newcommand{\p}{\partial}

\newcommand{\bn}{\vec{\nabla}}
\newcommand{\widebar}[1]{\mskip.5\thinmuskip\overline{\mskip-.5\thinmuskip {#1} \mskip-.5\thinmuskip}\mskip.5\thinmuskip}
\newcommand{\ee}{\mathrm{e}}
\newcommand{\ii}{\mathrm{i}}

\newcommand{\Od}[1]{\mathcal{O}(#1)}
\newcommand{\EQ}{\begin{equation}}
\newcommand{\EN}{\end{equation}}

\newcommand\Rey{\mathit{Re}}  % Reynolds number
\newcommand{\Real}{\mathrm{Re}}
\newcommand{\Imag}{\mathrm{Im}}
\renewcommand\Re[1]{\mathrm{Re}\left[ #1 \right]}

\renewcommand{\le}{\leqslant}
\newcommand{\ket}[1]{| #1 \rangle}

\newcommand{\dphi}{\Delta\phi}

\begin{document}
% \preprint{APS/123-QED}

\title{Quantum computing of fluid dynamics using the hydrodynamic Schr\"odinger equation}

\author{Zhaoyuan Meng}
    \affiliation{State Key Laboratory for Turbulence and Complex Systems, College of Engineering, Peking University, Beijing 100871, PR China}
\author{Yue Yang}
    \email{yyg@pku.edu.cn}
    \affiliation{State Key Laboratory for Turbulence and Complex Systems, College of Engineering, Peking University, Beijing 100871, PR China}
    \affiliation{HEDPS-CAPT, Peking University, Beijing 100871, PR China}

\date{\today}

\begin{abstract}
Simulating fluid dynamics on a quantum computer is intrinsically difficult due to the nonlinear and non-Hamiltonian nature of the Navier--Stokes equation (NSE).
We propose a framework for quantum computing of fluid dynamics based on the hydrodynamic Schr\"odinger equation (HSE), which can be promising in simulating three-dimensional turbulent flows in various engineering applications.
The HSE is derived by generalizing the Madelung transform to compressible/incompressible flows with finite vorticity and dissipation.
Since the HSE is expressed as a unitary operator on a two-component wave function, it is more suitable than the NSE for quantum computing.
The flow governed by the HSE can resemble a turbulent flow consisting of tangled vortex tubes with the five-thirds scaling of energy spectrum.
We develop a prediction-correction quantum algorithm to solve the HSE.  %$\Od{(n2^n+\mathrm{poly}(2^n))/\mathrm{poly}(n)}$.
This algorithm is implemented for simple flows on the quantum simulator Qiskit with exponential speedup.
%
%Besides, we investigate two complex incompressible Schr\"odinger flows to illustrate the similarities and differences between the HSE and the NSE.
%With the rapid development of hardware and algorithms for quantum computing, the HSE framework appears to be promising in computational fluid dynamics applications.
\end{abstract}

%\keywords{Suggested keywords}%Use showkeys class option if keyword
                              %display desired
\maketitle

%\tableofcontents

\section{Introduction}\label{sec:intro}
Quantum computing has emerged to be the next disruptive technology since Feynman pointed out the enormous potential of quantum simulation~\cite{Feynman1982_Simulating}.
Compared to conventional digital computing, quantum computing can dramatically reduce the execution time, memory usage, and energy consumption~\cite{Nielsen2010_Quantum}.
There are various hardware techniques for quantum logic gates~\cite{Sleator1995_Realizable, Makhlin2001_Quantum, Kok2007_Linear, Saffman2010_Quantum, Leibfried2011_Microwave, Zhang2021_Supercompact}, quantum algorithms for specific tasks~\cite{Shor1994_Algorithms, Ekert1996_Quantum, Shor1997_polynomial, Grover1996_A}, and applications~\cite{Harrow2009_Quantum, Reiher2017_Elucidating, Schuld2019_Quantum, McArdle2020_Quantum} implemented on a noisy intermediate-scale quantum computer~\cite{Bharti2022_Noisy}.

%e.g., many-body physics and computational chemistry~\cite{Aspuru2005_Simulated, Reiher2017_Elucidating, McArdle2020_Quantum}, nuclear physics~\cite{Hauke2013_Quantum, Martinez2016_Real, Kokail2019_Self}, machine learning~\cite{Benedetti2017_Quantum, Mitarai2018_Quantum, Schuld2019_Quantum}, combinatorial optimization~\cite{Wang2018_Quantum, Morales2018_Variational, Arrazola2018_Using}, singular value decomposition~\cite{Bravo2020_Quantum}, linear system problem~\cite{Harrow2009_Quantum, Wen2019_Experimental} and differential equations~\cite{Childs2020_Quantum, Liu2021_Efficient, Liu2021_Variational, Childs2021_High}.
%

% 2. Introduce quantum computational fluid dynamics
Quantum computing is not only for simulating quantum systems~\cite{Somaroo1999_Quantum, Georgescu2014_Quantum, Ju2014_NV, McArdle2020_Quantum, Zhong2020_Quantum, Yuan2020_A, Han2021_Experimental, Monroe2021_Programmable}, but also possible to simulate classical systems~\cite{Bharadwaj2020_Quantum, Dodin2021_On, Giannakis2022_Embedding, Jin2022_Quantum}.
Fluid dynamics, described by the Navier--Stokes equation (NSE), is notoriously difficult to be fully simulated on a classical computer at a large Reynolds number ($\Rey$), because the high-$\Rey$ turbulent flow involves length and time scales over a wide range of orders of magnitude.
%As a rough estimate~\cite{Pope2000_Turbulent}, the direct numerical simulation (DNS) of homogeneous isotropic turbulence (HIT) at $\Rey=10^7$ [justify why choose this Re] on a supercomputer of 1 PFlops [check the latest pflops of the top supercomputer] requires 491 PB of memory and 17375 years of CPU time, which is unacceptable in engineering applications.
The computational cost with $\Od{\Rey^3}$ operations for the direct numerical simulation (DNS) of turbulence~\cite{Pope2000_Turbulent} is unaffordable in engineering applications~\cite{Moin1998_Direct, Ishihara2009_Study}.
Therefore, the combination of computational fluid dynamics (CFD) and quantum computing can be promising for the next-generation simulation method~\cite{Givi2020_Quantum}.

To date, quantum computing has been demonstrated to be effective to handle some linear problems~\cite{Harrow2009_Quantum, Clader2013_Preconditioned, Cao2013_Quantum, Montanaro2016_Quantum, Costa2019_Quantum}, but remains intrinsically difficult in solving nonlinear differential equations~\cite{Lloyd2020_Quantum, Lubasch2020_Variational, Liu2021_Efficient} due to the linear nature of quantum mechanics.
Thus, it appears to be challenging to efficiently solve the highly nonlinear NSE on a quantum computer.
The current studies on the quantum computation of fluid dynamics can be divided into three categories.

First, quantum computing was performed for a specific simplified problem, e.g., 1D steady inviscid Laval nozzle~\cite{Gaitan2020_Finding}, 1D steady channel flow~\cite{Ray2019_Towards}, 1D Burgers equation~\cite{Oz2022_Solving}, and 2D thermal convection~\cite{Pfeffer2022_Hybrid}, to avoid dealing with the intractable full 3D NSE. %, 2D unsteady scalar convection-diffusion problem~\cite{Budinski2021_Quantum}, 2D cavity flow~\cite{Ljubomir2022_Quantum} and predicting reactant conversion in non-premixed homogeneous turbulence~\cite{Xu2019_Quantum}.
These works demonstrated the feasibility of quantum computing in CFD, but cannot be simply extended to complex 3D flows.

Second, quantum algorithms were applied to solving linear systems~\cite{Harrow2009_Quantum, Wen2019_Experimental}, e.g., the quantum linear solver~\cite{Chen2022_Quantum,Lapworth2022_A,Demirdjian2022_Variational} and Poisson solver~\cite{Steijl2018_Parallel}, to replace a part of a classical CFD algorithm.
These hybrid quantum-classical algorithms involve frequent data exchanges between classical and quantum hardware.
Since the conversion can take even much longer time than the computational time for solving the equations~\cite{Aaronson2015_Read}, only steady problems were considered in these works to avoid data exchange.

Third, fluid dynamics was described by the approaches that are more suitable than the NSE for quantum computing, e.g., the Madelung transform~\cite{Zylberman2022_Quantum}, the generalized Koopman--von Neumann (KvN) representation~\cite{Joseph2020_Koopman}, the lattice Boltzmann method~\cite{Yepez2001_Quantum, Keating2007_Entropic, Todorova2020_Quantum}, and the tensor network-based method inspired from quantum many-body physics~\cite{Gourianov2022_A, Fukagata2022_Towards}.
On the other hand, each of these methods has certain limitations, e.g., the Madelung transform can only describe inviscid potential flows, and the KvN representation encounters the non-closure problem of the probability density function.

% 3. Briefly introduce our work
The present study adopts the third approach to describe fluid dynamics using the hydrodynamic Schr\"odinger equation (HSE).
The HSE is derived by generalizing the Madelung transform to compressible/incompressible flows with finite vorticity and dissipation.
It can be expressed as a unitary operator on a two-component wave function, so it is more natural than the NSE for quantum computing.
We develop a quantum algorithm for solving the HSE with a notable speedup, and implement the algorithm for simple flows on IBM's quantum simulator~\cite{Qiskit}.

% 4. Outline
The outline of the present paper is as follows.
Section~\ref{sec:Theo_SF} introduces the HSE.
Section~\ref{sec:ISF_DNS} compares the flows governed by the HSE and NSE.
Section~\ref{sec:QA_ISF} develops and validates the quantum algorithm.
Some conclusions are drawn in Section~\ref{sec:conclusion}.

\section{Theoretical framework of the Schr\"odinger flow}\label{sec:Theo_SF}
\subsection{Madelung transform} %Probabilistic current and the
In quantum mechanics, the probabilistic current for a wave function $\psi(\vec{x},t)$ is defined as \cite{Griffiths2005_Introduction}
\begin{equation}\label{eq:j0}
    \vec{J}(\vec{x},t)
    \equiv \frac{1}{2m}\left(\widebar{\psi}\widehat{\vec{p}}\psi - \psi\widehat{\vec{p}}\widebar{\psi}\right)
\end{equation}
with the momentum operator $\widehat{\vec{p}}$ and particle mass $m$, where $\widebar{f}$ denotes the complex conjugate of $f$, and $\widehat{f}$ denotes an operator.
In the coordinate representation, we have $\widehat{\vec{p}}=-\ii\hbar\bn$ with the imaginary unit $\ii$ and Planck constant $\hbar$.
%The probabilistic current $\vec{J}(\vec {x},t)$ tells you the rate at which probability is ``flowing" past the point $\vec{x}$ at time $t$.
Considering a particle moving in a potential field $V\in\mathbb{R}$, its motion satisfies the Schr\"odinger equation~\cite{Schrodinger1926_An}
\begin{equation}\label{eq:SchEq0}
    \ii\hbar\frac{\p}{\p t}\psi(\vec{x},t)
    %= \left(\frac{|\widehat{\vec{p}}|^2}{2m} + V \right)\psi(\vec{x},t)
    = \left(-\frac{\hbar^2}{2m}\nabla^2 + V \right)\psi(\vec{x},t).
\end{equation}
Without loss of generality, we set $m=1$.
From Eqs.~\eqref{eq:j0} and \eqref{eq:SchEq0}, the conservation of the probability density $\rho\equiv \widebar{\psi}\psi$ reads
\begin{equation}\label{eq:contEq0}
    \frac{\p \rho}{\p t} + \bn\cdot\vec{J} = 0.
\end{equation}
The form of Eq.~\eqref{eq:contEq0} is identical to the continuity equation in fluid mechanics, with a ``velocity" $\vec{u}\equiv\vec{J}/\rho$.

The Madelung transform \cite{Madelung1927_Quantentheorie} shows an analogy between quantum mechanics and fluid mechanics.
Table~\ref{tab:SymbolMeaning} explains the physical meanings of the same symbol in different contexts.
Using the Madelung transform, the momentum equation
\begin{equation}\label{eq:mom_Mdl}
    \frac{\p\vec{u}}{\p t} + \vec{u}\cdot\bn\vec{u} = -\bn V + \frac{\hbar^2}{2}\bn\frac{\nabla^2\sqrt{\rho}}{\sqrt{\rho}}
\end{equation}
of a fluid flow is obtained from Eq.~\eqref{eq:SchEq0}, where the fluid velocity is
\begin{equation}
    \vec{u} =  \frac{\ii\hbar}{2}\frac{\psi\bn\widebar{\psi} - \widebar{\psi}\bn\psi}{\psi\widebar{\psi}}
    %= \frac{\ii\hbar}{2}\left(\frac{\bn\widebar{\psi}}{\widebar{\psi}} - \frac{\bn\psi}{\psi} \right)
    = \frac{\ii\hbar}{2}\bn\ln\frac{\widebar{\psi}}{\psi}
    = \bn\phi
\end{equation}
with $\psi=\sqrt{\rho}\ee^{\ii\phi/\hbar}$.
Equation~\eqref{eq:mom_Mdl} corresponds to the Euler equation for a potential flow with vanishing vorticity.
It has very limited applications for general viscous flows with finite vorticity~\cite{Schonberg1954_On, Sorokin2001_Madelung, Love2004_Quaternionic}.

\begin{table}
\caption{Meanings of the same symbol in different contexts.}
\begin{ruledtabular}
    \renewcommand\arraystretch{1}
    \begin{tabular}{lcccc}
      & $\rho$ & $\vec{u}$ & $\vec{J}=\rho\vec{u}$ & $\hbar$  \\
    \hline
    quantum mechanics & probability density & - & probabilistic current & Planck constant \\
    fluid mechanics & mass density & velocity & momentum & arbitrary constant \\
    \end{tabular}
\end{ruledtabular}
\label{tab:SymbolMeaning}
\end{table}

\subsection{Schr\"odinger flow}
To introduce the finite vorticity into the hydrodynamic representation of the Schr\"odinger equation,
we use a two-component wave function~\cite{Mueller2002_Two, Kasamatsu2003_Vortex, Werner2012_General} represented by a quaternion as
\begin{equation}
    \vec{\psi}(\vec{x},t) = a(\vec{x},t) + \vec{i}b(\vec{x},t) + \vec{j}c(\vec{x},t) + \vec{k}d(\vec{x},t)
\end{equation}
with the basis vectors $\{\vec{i},\vec{j},\vec{k}\}$ of the imaginary part of the quaternion and real-valued functions $a$, $b$, $c$, and $d$.
This quaternion facilitates deriving governing equations of the fluid flow below, and it is essentially the same as the two-component spinor~\cite{Bergmann1957_Two, Sachs1982_Spinor, Adler1995_Quaternionic, Dreiner2010_Two}.

The probabilistic current in Eq.~\eqref{eq:j0} is generalized to
\begin{equation}\label{eq:j1}
    \vec{J} \equiv \frac{\hbar}{2}\left((\bn\widebar{\vec{\psi}})\vec{i\psi} - \widebar{\vec{\psi}}\vec{i}\bn\vec{\psi} \right).
\end{equation}
Similarly, the fluid mass density and velocity become $\rho \equiv \widebar{\vec{\psi}}\vec{\psi}$ and
\begin{equation}\label{eq:IHSE_Vel}
    \vec{u} \equiv \frac{\vec{J}}{\rho} = \frac{\hbar}{2}\frac{\bn\widebar{\vec{\psi}}\vec{i\psi} - \widebar{\vec{\psi}}\vec{i}\bn\vec{\psi}}{\widebar{\vec{\psi}}\vec{\psi}},
\end{equation}
respectively. Then, we obtain
\begin{equation}
    \frac{\p\rho}{\p t} + \bn\cdot(\rho\vec{u})
    = \left(\frac{\p\widebar{\vec{\psi}}}{\p t} + \frac{\hbar}{2}\nabla^2\widebar{\vec{\psi}}\vec{i}\right)\vec{\psi}
    + \widebar{\vec{\psi}}\left(\frac{\p\vec{\psi}}{\p t} - \frac{\hbar}{2}\vec{i}\nabla^2\vec{\psi}\right)
\end{equation}
after some algebra.
With identities $\widebar{\vec{i}\vec{\psi}}=-\widebar{\vec{\psi}}\vec{i}$ and $\widebar{\vec{i}\nabla^2\vec{\psi}}=-\nabla^2\widebar{\vec{\psi}}\vec{i}$, we derive a sufficient condition
\begin{equation}\label{eq:SchEq1}
    \vec{i}\hbar\frac{\p\vec{\psi}}{\p t} = \left( -\frac{\hbar^2}{2}\nabla^2 + V \right)\vec{\psi}
\end{equation}
with a real-valued potential $V$ for the continuity equation
\begin{equation}\label{eq:contEq}
    \frac{\p\rho}{\p t} + \bn\cdot(\rho\vec{u})=0.
\end{equation}
%where $V\in\mathbb{R}$ is a real-valued function.
%
Note that Eq.~\eqref{eq:SchEq1} is the Schr\"odinger--Pauli equation (SPE) in a quaternion form, which describes the motion of a spin-$1/2$ particle without an external electromagnetic field in the non-relativistic limit~\cite{Bjorken1964_Relativistic, Davydov1965_Quantum, Messiah1968_Quantum}.
%Each component of the two-component spinor $\vec{\psi}$ is decoupling, i.e. each satisfies the Schr\"odinger equation because there is not a magnetic field present.

After some algebra (detailed in Appendix~\ref{app:derive_SF}), we derive the momentum equation
\begin{equation}\label{eq:Du/Dt_SF}
    \frac{\p\vec{u}}{\p t} + \vec{u}\cdot\bn\vec{u}
    = -\frac{1}{\rho}\bn p - \bn V_F
    - \frac{{\hbar}^2}{4\rho}\bn\vec{s}\cdot\left[\bn\cdot\left(\frac{1}{\rho}\bn\vec{s}\right)\right]
\end{equation}
for $\vec{\psi}$, along with an equation of state
\begin{equation}\label{eq:p_SF}
    p = -\frac{{\hbar}^2}{4}\vec{s}\cdot\left[\bn\cdot\left(\frac{1}{\rho}\bn\vec{s}\right)\right].
\end{equation}
Here,
\begin{equation}\label{eq:potential}
    V=V_F(\vec{x})+\frac{\hbar^2}{8\rho^2}|\bn\vec{s}|^2
\end{equation}
is a nonlinear potential, where $V_F$ corresponds to conservative body forces and
\begin{equation}\label{eq:SpinVector}
    \vec{s} \equiv \widebar{\vec{\psi}}\vec{i\psi}
\end{equation}
denotes a spin vector.
The last term in the right-hand side (RHS) of Eq.~\eqref{eq:Du/Dt_SF} does not appear in the momentum equation of practical fluid flows.
It can be considered as an external body force involving a dissipation effect, and it degenerates to the ``Landau--Lifshitz force'' (LLF)~\cite{Chern2016_Schrodinger, Chern2017_Fluid} for constant $\rho$.
%
%However, the corresponding flow differs from the real fluid flows because the external body force in Eq.~\eqref{eq:Du/Dt_SF} is a nonlinear term that depends on the physical quantities of the flow field such as velocity.

In sum, we convert the compressible flow with finite vorticity in Eqs.~\eqref{eq:contEq}, \eqref{eq:Du/Dt_SF}, and \eqref{eq:p_SF}, into a hydrodynamic Schr\"odinger equation (HSE)
\begin{equation}\label{eq:HSE}
    \vec{i}\hbar\frac{\p\vec{\psi}}{\p t}
    = \left(-\frac{\hbar^2}{2}\nabla^2 + V_F(\vec{x}) + \frac{\hbar^2}{8\rho^2}|\bn\vec{s}|^2 \right)\vec{\psi}.
\end{equation}
The HSE can be considered as a SPE with a specific potential in Eq.~\eqref{eq:potential}.
The fluid flow governed by the HSE is then called the Schr\"odinger flow (SF).
Comparing with the Gross--Pitaevskii equation~\cite{Gross1961_Structure, Pitaevskii1961_Vortex}
\begin{equation}
    \ii\hbar\frac{\p\psi}{\p t}
    = \left(-\frac{\hbar^2}{2}\nabla^2 + V(\vec{x}) + g|\psi|^2 \right)\psi,
\end{equation}
with an external potential $V(\vec{x})$ and a coupling constant $g$, which is a well-known model equation describing the dynamics of the Bose--Einstein condensate, the HSE~\eqref{eq:HSE} has a more complex nonlinear potential and incorporates the spin effect of a particle.

\begin{figure}
    \centering
    \includegraphics[width=\linewidth]{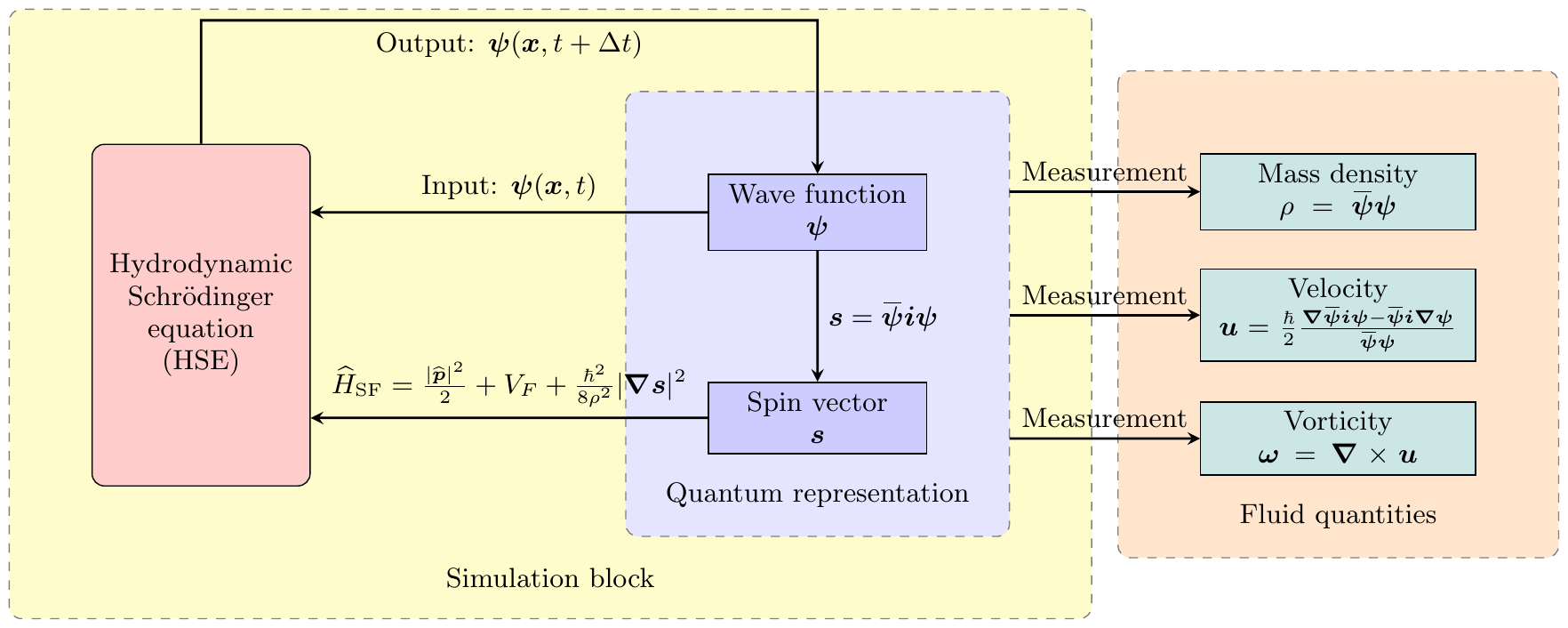}
    \caption{Schematic for quantum computing of the SF.}
    \label{fig:SystemConv}
\end{figure}

Since the real-valued Hamiltonian
\begin{equation}\label{eq:Hamt_SF}
    \widehat{H}_{\mathrm{SF}}
    = \frac{|\widehat{\vec{p}}|^2}{2} + V_F + \frac{\hbar^2}{8\rho^2}|\bn\vec{s}|^2
\end{equation}
of the SF is Hermitian, the evolutionary operator
\begin{equation}
    \exp\left(-\frac{\ii}{\hbar}\widehat{H}_{\mathrm{SF}}\Delta t \right)
    \equiv \sum_{n=0}^\infty \frac{1}{n!}\left(-\frac{\ii}{\hbar}\Delta t \right)^n\widehat{H}_{\mathrm{SF}}^n
\end{equation}
is unitary, with a time increment $\Delta t$.
We are able to use it to obtain $\vec{\psi}(\vec{x},t)$ at a given time from an initial wave function in quantum computing.
The procedure of the simulation and measurement for the SF is sketched in Fig.~\ref{fig:SystemConv}.
This simulation of the SF only involves the wave function and its derivatives without fluid quantities, so it is equivalent to a Hamiltonian simulation for the motion of a particle.

\subsection{Incompressible Schr\"odinger flow}
We consider a special SF for a constant-density incompressible flow with
\begin{equation}\label{eq:ConstRho}
    \rho = \rho_0.
\end{equation}
The wave function on the sphere $\mathbb{S}^3$ with radius $\sqrt{\rho_0}$ and the spin vector in Eq.~\eqref{eq:SpinVector} on $\mathbb{S}^2$ with radius $\rho_0$ for this flow are linked by the Hopf fibration \cite{Hopf1931_Uber}.
%, where $\mathbb{S}^n$ denotes a $n$-dimensional unit sphere. %the spherical surface of
Taking the material derivative $\DD/\DD t\equiv \p/\p t+\vec{u}\cdot\bn$ of $\rho$ yields
\begin{equation}\label{eq:DrhoDt_0}
    \frac{\DD\rho}{\DD t} = \frac{\DD\widebar{\vec{\psi}}}{\DD t}\vec{\psi} + \widebar{\vec{\psi}}\frac{\DD\vec{\psi}}{\DD t}
    %= 2\Re{\widebar{\vec{\psi}}\frac{\DD\vec{\psi}}{\DD t}}
    = 2\Re{\frac{\DD\vec{\psi}}{\DD t}\widebar{\vec{\psi}}}
    = 0.
\end{equation}
Setting $(\DD\vec{\psi}/\DD t)\widebar{\vec{\psi}} = \vec{f}^\psi$ as an pure quaternion in Eq.~\eqref{eq:DrhoDt_0} yields
\begin{equation}\label{eq:Dpsi/Dt_0}
    \frac{\DD \vec{\psi}}{\DD t} = \frac{1}{\rho_0}\vec{f}^\psi \vec{\psi}.
\end{equation}
After some algebra, the momentum equation
% the spin vector
% \begin{equation}\label{eq:Ds/Dt_0}
%     \frac{\DD\vec{s}}{\DD t} = 2\vec{s}\times\vec{f}^s
% \end{equation}
\begin{equation}\label{eq:Du/Dt_ISF0}
    \frac{\p\vec{u}}{\p t} + \vec{u}\cdot\bn\vec{u} = -\bn\left(\frac{|\vec{u}|^2}{2} - \frac{\hbar}{\rho_0^3}\vec{s}\cdot\vec{f}^s \right)
        - \frac{\hbar}{\rho_0^3}\bn\vec{s}\cdot\vec{f}^s
\end{equation}
in an incompressible flow is obtained, where $\vec{f}^s\equiv \widebar{\vec{\psi}}\vec{f}^\psi\vec{\psi}$ is also a pure quaternion.
Note that the flow governed by Eqs.~\eqref{eq:ConstRho} and \eqref{eq:Du/Dt_ISF0} has some physically interesting properties, such as the helicity conservation~\cite{Moreau1961_Constantes, Moffatt1969_The, Meng2023_Evolution} and the Lagrangian-like evolution of vortex surfaces~\cite{Yang2010_On, Hao2019_Tracking}.

Letting
\begin{equation}\label{eq:fpsi0}
    \vec{f}^\psi = -\frac{\vec{i}}{\hbar}\left(\frac{p}{\rho_0} - \frac{|\vec{u}|^2}{2} + V_F \right)
\end{equation}
and substituting it into Eqs.~\eqref{eq:Dpsi/Dt_0} and \eqref{eq:Du/Dt_ISF0}, we obtain the Euler equation
\begin{equation}\label{eq:InCompEuler}
    \frac{\p\vec{u}}{\p t} + \vec{u}\cdot\bn\vec{u} = -\bn\left(\frac{p}{\rho_0} + V_F \right)
\end{equation}
and the corresponding nonlinear Schr\"odinger equation
\begin{equation}\label{eq:InComp_EulerSchEq}
    \vec{i}\hbar\frac{\p\vec{\psi}}{\p t}
    = \left(-\frac{{\hbar}^2}{2}\nabla^2 + \frac{p}{\rho_0} + V_F + \frac{5{\hbar}^2}{8\rho_0^2}|\bn\vec{s}|^2 - \frac{{\hbar}^2}{4\rho_0^2}\vec{\psi}(\nabla^2\vec{s})\widebar{\vec{\psi}}\vec{i} \right)\vec{\psi}.
\end{equation}
However, the non-Hermitian Hamiltonian
\begin{equation}\label{eq:H_Euler}
    \widehat{H}_{\mathrm{Euler}} = \frac{|\widehat{\vec{p}}|^2}{2} + \frac{p}{\rho_0} + V_F + \frac{5{\hbar}^2}{8\rho_0^2}|\bn\vec{s}|^2 - \frac{{\hbar}^2}{4\rho_0^2}\vec{\psi}(\nabla^2\vec{s})\widebar{\vec{\psi}}\vec{i}
\end{equation}
in Eq.~\eqref{eq:InComp_EulerSchEq} can inhibit an efficient quantum computation.
%, and it can be slightly modified to be Hermitian.
%Therefore, in order to make potentially ultra-high performance computing on quantum computers possible, we consider a small modification of the Hamiltonian~\eqref{eq:H_Euler} to become Hermitian. This is done by modifying

To make the Hamiltonian Hermitian, Eq.~\eqref{eq:fpsi0} is modified to
\begin{equation}
    \vec{f}^\psi
    = -\frac{\vec{i}}{\hbar}\left(\frac{p}{\rho_0} - \frac{|\vec{u}|^2}{2} + V_F \right)
    - \frac{\hbar}{4\rho_0^3}\vec{\psi s}(\vec{s}\times\nabla^2\vec{s})\widebar{\vec{\psi}}.
\end{equation}
Then we have the modified momentum equation
\begin{equation}\label{eq:Du/Dt_ISF1}
    \frac{\p\vec{u}}{\p t} + \vec{u}\cdot\bn\vec{u}
    = -\bn\left(\frac{p}{\rho_0} + V_F \right)
    - \frac{\hbar^2}{4\rho_0^2}\bn\vec{s}\cdot\nabla^2\vec{s},
\end{equation}
and the incompressible hydrodynamic Schr\"odinger equation (IHSE)
\begin{equation}\label{eq:ISF_SchEq}
    \vec{i}\hbar\frac{\p\vec{\psi}}{\p t}
    = \left(-\frac{{\hbar}^2}{2}\nabla^2 + \frac{p}{\rho_0} + V_F(\vec{x}) - \frac{{\hbar}^2}{8\rho_0^2}|\bn\vec{s}|^2\right)\vec{\psi}.
\end{equation}
% and the evolution equation for the spin vector
% \begin{equation}\label{eq:Ds/Dt_ISF}
%     \frac{\DD\vec{s}}{\DD t}
%     = \frac{\hbar}{2}\vec{s}\times\nabla^2\vec{s}.
% \end{equation}
%
The physical meaning of the last term in the RHS of Eq.~\eqref{eq:Du/Dt_ISF1}, i.e., the LLF~\cite{Chern2016_Schrodinger, Chern2017_Fluid}, is further discussed in Appendix~\ref{app:LLF}.
The flow governed by Eq.~\eqref{eq:Du/Dt_ISF1} with $\rho_0=1$ has been called the incompressible Schr\"odinger flow (ISF) \cite{Chern2016_Schrodinger, Chern2017_Fluid} and studied by numerical simulations~\cite{Tao2021_Construction}.
The Hamiltonian of the ISF
\begin{equation}\label{eq:Halm_ISF}
    \widehat{H}_{\mathrm{ISF}}
    = \frac{|\widehat{\vec{p}}|^2}{2} + \frac{p}{\rho_0} + V_F - \frac{\hbar^2}{8\rho_0^2}|\bn\vec{s}|^2
\end{equation}
is Hermitian, so the ISF is suitable for quantum computing.
Moreover, the form of the IHSE can be obtained by setting $\rho=\rho_0$ in the HSE.
%Eq.~\eqref{eq:Du/Dt_SF}.
%Substituting Eq.~\eqref{eq:p_SF} into Eq.~\eqref{eq:Hamt_SF} with $\rho=\rho_0$ yields
%\begin{equation}\label{eq:Halm_SF_rho1}
%    \widehat{H}_{\mathrm{SF}}|_{\rho=\rho_0}
%    = \frac{|\widehat{\vec{p}}|^2}{2} + \frac{p}{\rho_0} + V_F + \frac{\hbar^2}{4\rho_0^2}\vec{s}\cdot\nabla^2\vec{s} + \frac{\hbar^2}{8\rho_0^2}|\bn\vec{s}|^2.
%\end{equation}
%%
%For a smooth function with $|\vec{s}|=\rho_0$, the identity $(\vec{s}\times\nabla^2\vec{s})\times\vec{s} = \rho_0^2\nabla^2\vec{s}+|\bn\vec{s}|^2\vec{s}$ implies $\vec{s}\cdot\nabla^2\vec{s}+|\bn\vec{s}|^2=0$. Substituting this into Eq.~\eqref{eq:Halm_SF_rho1} yields
%\begin{equation}\label{eq:Halm_SF_rho2}
%    \widehat{H}_{\mathrm{SF}}|_{\rho=\rho_0}
%    = \frac{|\widehat{\vec{p}}|^2}{2} + \frac{p}{\rho_0} + V_F - \frac{\hbar^2}{8\rho_0^2}|\bn\vec{s}|^2,
%\end{equation}
%which is identical with Eq.~\eqref{eq:Halm_ISF}.
%It should be emphasized that Eqs.~\eqref{eq:Halm_SF_rho1} and \eqref{eq:Halm_SF_rho2} are only used to show that the ISF is a particular SF.
%The additional constraint ($\rho=\rho_0$) requires an alternative route to rigorously derive the ISF in Eqs.~\eqref{eq:ConstRho}--\eqref{eq:Halm_ISF}.
%

Similar to the incompressible Navier--Stokes equation (INSE), the pressure $p$ in the IHSE~\eqref{eq:ISF_SchEq} is coupled with Eq.~\eqref{eq:Du/Dt_ISF1} to ensure the divergence-free velocity.
Thus, the mathematical natures of the HSE~\eqref{eq:HSE} with non-constant $\rho$ and IHSE~\eqref{eq:ISF_SchEq} are very different, which is similar to the difference between compressible and incompressible NSEs, so they have to be solved by different methods.
In the present study, we focus on the physical property and quantum algorithm for the ISF.
%Since \mzy{the SF may have the shock wave which is difficult to accurately simulate numerically due to the discontinuity caused by the possible shock wave}, we only study the physical property and quantum algorithm for the ISF \mzy{in the present paper}.
%\mzy{In the future work, the physical property of the SF can be further studied and the quantum algorithm for the SF can subsequently be designed.}

\section{Turbulent ISFs}\label{sec:ISF_DNS}
We investigate two ISFs, the Taylor--Green (TG) flow and decaying homogeneous isotropic turbulence (HIT), to illustrate the similarities and differences between the ISF and the real viscous flow.
Additionally, the evolution of vortex knots was investigated in the ISF~\cite{Tao2021_Construction}.

Since quantum hardware and algorithms for simulating such complex flows are still under development, the DNS of the ISF was carried out to solve Eq.~\eqref{eq:ISF_SchEq} with $\rho_0=1$ on a classical computer.
The standard pseudo-spectral method~\cite{Xiong2019_Identifying, Hao2019_Tracking, Shen2022_Topological} was adopted in a periodic cube of side $\mathcal{L}=2\pi$ on $512^3$ uniform grid points.
%Aliasing errors are removed using the two-thirds truncation method with the maximum wavenumber $k_{\max}\approx N/3=170$.
The numerical implementation was described in detail in Ref.~\cite{Tao2021_Construction}.

\subsection{TG flow}\label{sec:TG_Flow}
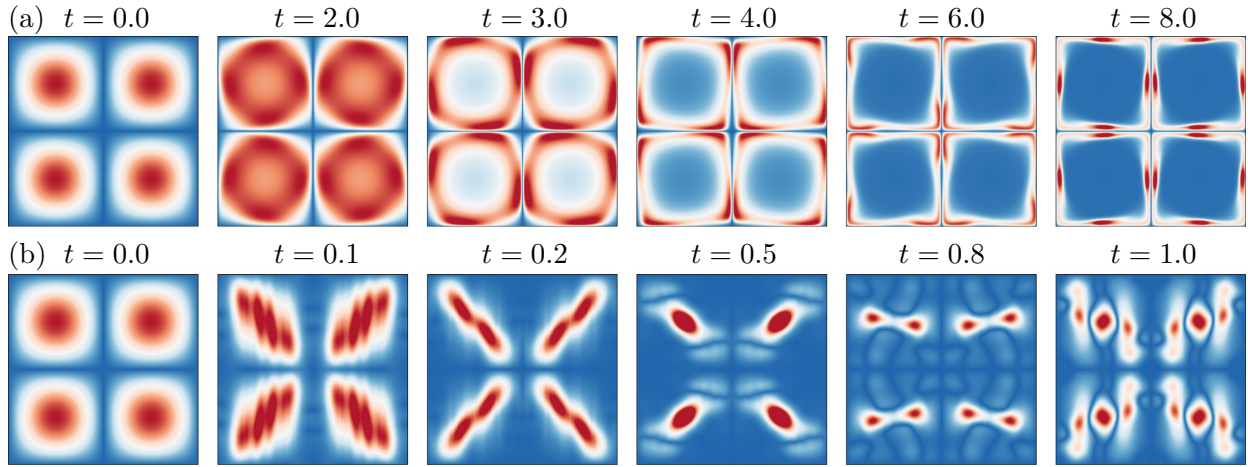
\begin{figure}
    \centering
    \begin{overpic}[width=\linewidth]{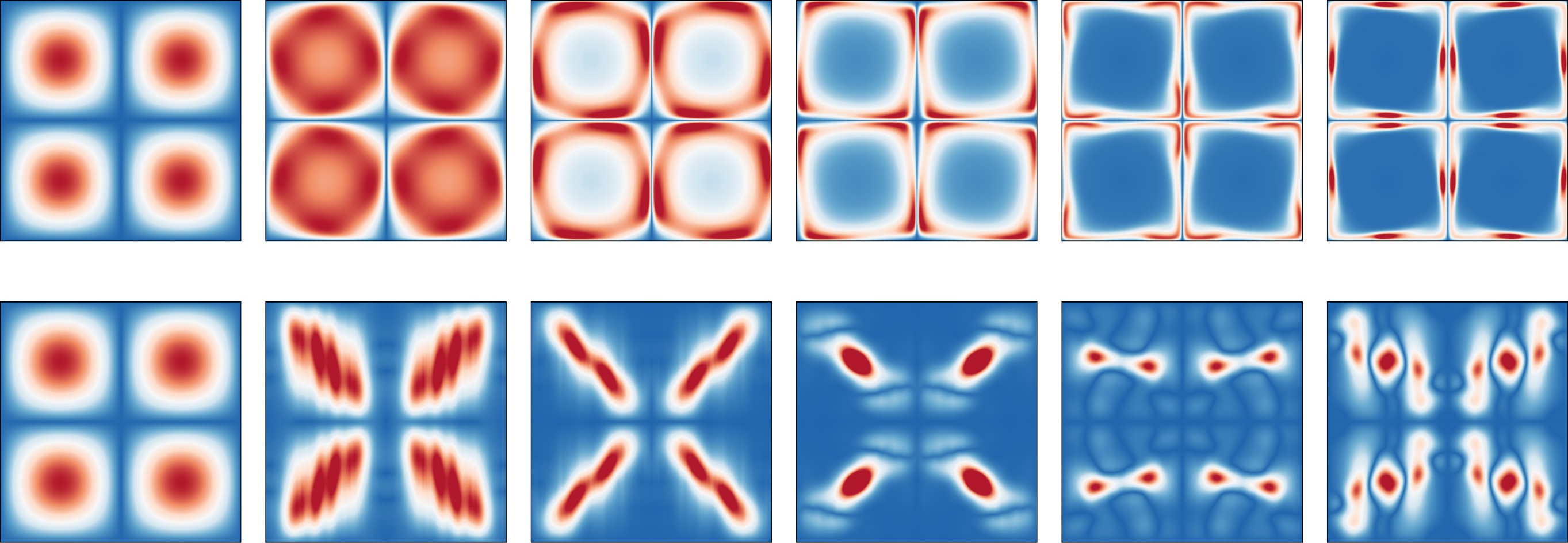}
        \begin{small}
            \put(0,35.4) {(a)}
            \put(4.5,35.4) {$t=0.0$}
            \put(21.5,35.4) {$t=2.0$}
            \put(38.3,35.4) {$t=3.0$}
            \put(55.2,35.4) {$t=4.0$}
            \put(72,35.4) {$t=6.0$}
            \put(89.1,35.4) {$t=8.0$}
            \put(0,16.25) {(b)}
            \put(4.5,16.25) {$t=0.0$}
            \put(21.5,16.25) {$t=0.1$}
            \put(38.3,16.25) {$t=0.2$}
            \put(55.2,16.25) {$t=0.5$}
            \put(72,16.25) {$t=0.8$}
            \put(89.1,16.25) {$t=1.0$}
        \end{small}
    \end{overpic}
    \caption{Evolution of $|\boldsymbol\omega|$ on the $x$--$y$ plane at $z=\pi$ for TG vortices in the (a) INSF with $\Rey=1000$ and (b) ISF with $\hbar=1$. The contour is color-coded by $0 \le|\boldsymbol\omega|/|\vec{\omega}|_{\max}\le 1$ from blue to red.}
    \label{fig:TG_ContourVor_z=pi}
\end{figure}

We apply the TG initial condition in Eq.~\eqref{eq:TG_vel} to the INSE and IHSE.
The construction of the initial wave function is detailed in Appendix~\ref{app:TG}.
%The velocity field reconstructed by the wave function introduces negligible numerical errors compared to the theoretical one.
%
The evolutions of the TG vortex in the ISF and the incompressible NS flow (INSF) are compared in Fig.~\ref{fig:TG_ContourVor_z=pi} using the contour of the vorticity magnitude $|\boldsymbol \omega|$ with $\boldsymbol \omega\equiv\bn\times\boldsymbol u$.
In the INSF, the initial blob-like vortices are stretched into sheet-like structures and move towards symmetry planes at $t=4$.
%, then a pair of vortices in the opposite direction undergoes a strong shearing effect, as shown in Fig.~\ref{fig:TG_ContourVor_z=pi}(a).
%
In the ISF, the vortices undergo strongly oscillating shearing motion, and they break up into smaller-scale vortices at $t=1$.
%Interestingly, the initial spherical vortex is stretched toward the four corners of the periodic cube, like a butterfly spreading its wings, which is particularly evident in Fig.~\ref{fig:TG_ContourVor_z=pi}(b) at $t=1.0$.
%
%Thus, the vortex stretching in the ISF seems to be much stronger than that in the INSF, and
The characteristic time scale of vortex dynamics in the ISF is much smaller than that in the INSF. % due to the LLF
%The mechanism of this phenomenon is to be studied in the future.

\begin{figure}
    \centering
    \begin{overpic}[width=\linewidth]{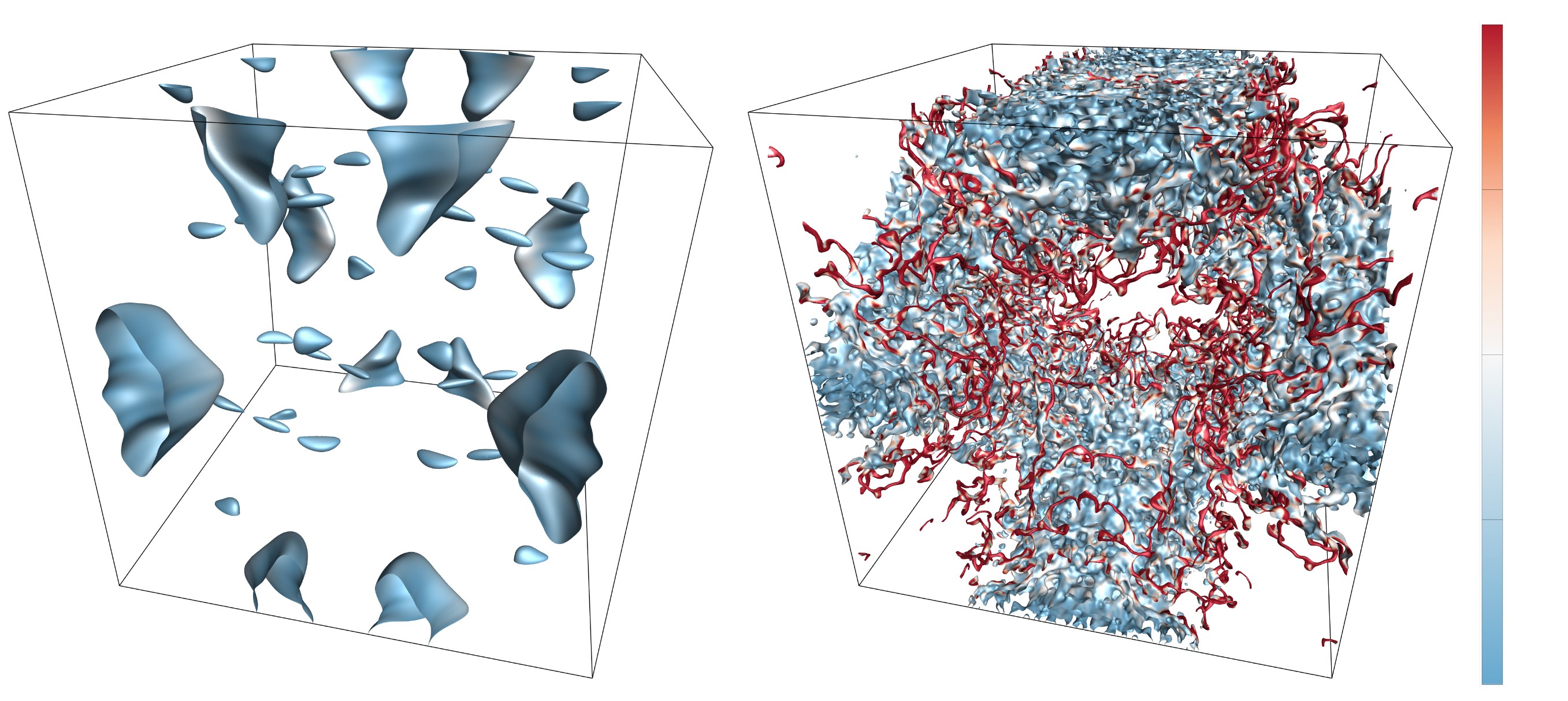}
        \begin{small}
            \put(0,41) {(a)}
            \put(48,41) {(b)}
            \begin{footnotesize}
                \put(93.6,44) {$|\vec{\omega}|$}
                \put(96.5,0.4) {$0$}
                \put(96.5,10.9) {$2.5$}
                \put(96.5,21.4) {$5$}
                \put(96.5,32) {$7.5$}
                \put(96.5,42.4) {$10$}
            \end{footnotesize}
        \end{small}
    \end{overpic}
    \caption{Isosurfaces of $s_1=-0.9$ at $t=1$ for TG vortices in the ISF with (a) $\hbar=1$ and (b) $\hbar=0.1$. Note that the isosurface of $s_1$ is a vortex surface~\cite{Yang2010_On, Hao2019_Tracking} consisting of vortex lines, and the initial isosurface is a vortex column shown in Fig.~\ref{fig:TG_VortexSurface}(a). The isosurfaces are color-coded by $|\boldsymbol\omega|$.}
    \label{fig:TG_VortexSurface_IHSE}
\end{figure}

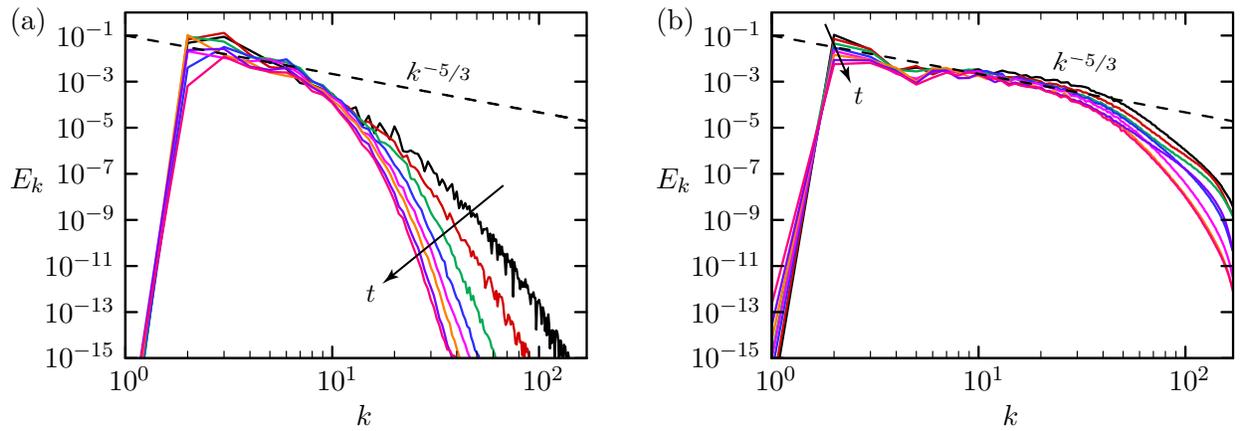
\begin{figure}
    \centering
    \begin{overpic}[width=0.4\linewidth]{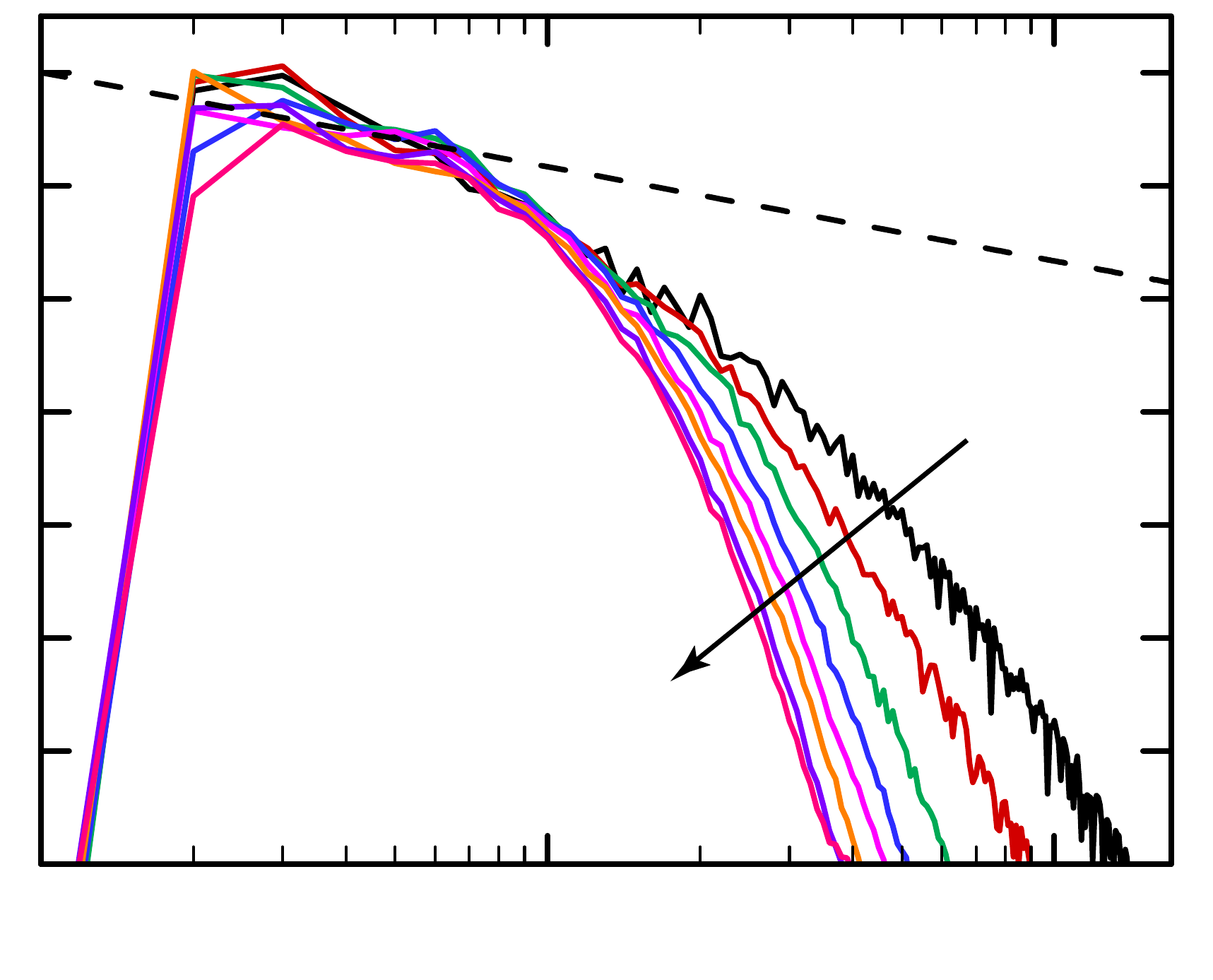}
        \begin{small}
            % x-axis
            \put(0,3) {$10^{0}$}
            \put(42,3) {$10^1$}
            \put(83.6,3) {$10^2$}
            \put(50,-4) {$k$}
            % y-axis
            \put(-13,7.8) {$10^{-15}$}
            \put(-13,16.8) {$10^{-13}$}
            \put(-13,26.2) {$10^{-11}$}
            \put(-10.6,35.5) {$10^{-9}$}
            \put(-10.6,44.9) {$10^{-7}$}
            \put(-10.6,54.2) {$10^{-5}$}
            \put(-10.6,63.5) {$10^{-3}$}
            \put(-10.6,72.7) {$10^{-1}$}
            \put(-20,44) {$E_k$}
            \put(-20,76) {(a)}
            % legend & mark
            \begin{footnotesize}
                \put(52,21) {$t$}
                \put(60,66) {\begin{rotate}{-10} $k^{-5/3}$ \end{rotate}}
            \end{footnotesize}
        \end{small}
    \end{overpic}
    \hspace{50pt}
    \begin{overpic}[width=0.4\linewidth]{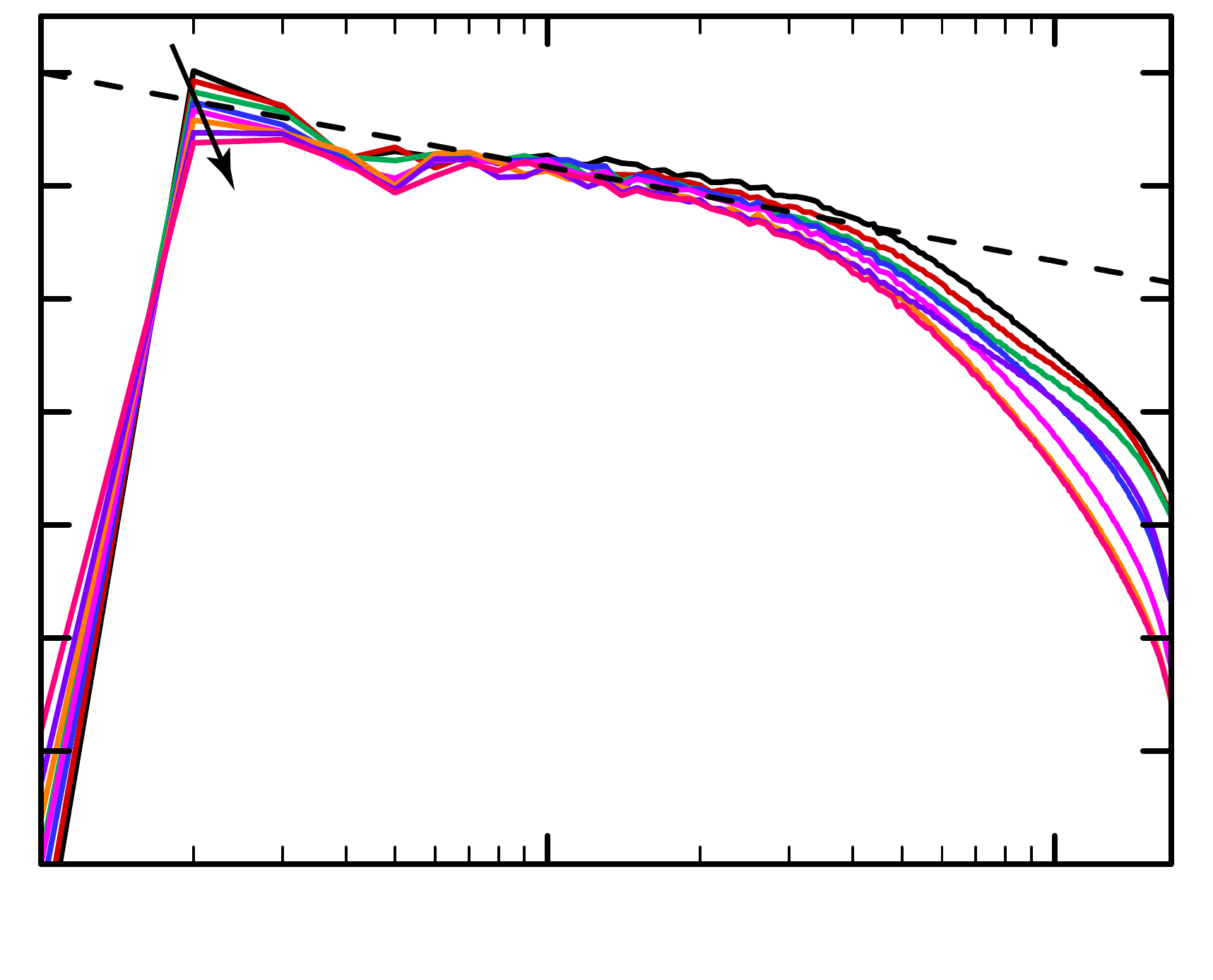}
        \begin{small}
            % x-axis
            \put(0,3) {$10^{0}$}
            \put(42,3) {$10^1$}
            \put(83.6,3) {$10^2$}
            \put(50,-4) {$k$}
            % y-axis
            \put(-13,7.8) {$10^{-15}$}
            \put(-13,16.8) {$10^{-13}$}
            \put(-13,26.2) {$10^{-11}$}
            \put(-10.6,35.5) {$10^{-9}$}
            \put(-10.6,44.9) {$10^{-7}$}
            \put(-10.6,54.2) {$10^{-5}$}
            \put(-10.6,63.5) {$10^{-3}$}
            \put(-10.6,72.7) {$10^{-1}$}
            \put(-20,44) {$E_k$}
            \put(-20,76) {(b)}
            % legend & mark
            \begin{footnotesize}
                \put(20,61) {$t$}
                \put(60,68) {\begin{rotate}{-10} $k^{-5/3}$ \end{rotate}}
            \end{footnotesize}
        \end{small}
    \end{overpic}
    \hspace{-45pt}
    \caption{Evolution of the energy spectra for TG vortices in the ISF at $t=1 \sim 8$ with (a) $\hbar=1$ and (b) $\hbar=0.1$.}
    \label{fig:TG_EnergySpectrum_ISF}
\end{figure}

The effect of the parameter $\hbar$ in the ISF is similar to the kinetic viscosity $\nu$ in the INSF.
Figure~\ref{fig:TG_VortexSurface_IHSE} shows similar large-scale vortical structures with $\hbar=1$ and 0.1, whereas much more small-scale tube-like structures emerge for smaller $\hbar=0.1$.
%Overall, the smaller $\hbar$ is, the smaller the scale of the vortex structures after evolving from the same initial condition.
In general, the length scale of vortices is proportional to $\hbar$ via the vorticity Clebsch mapping~\cite{Chern2016_Schrodinger, Chern2017_Fluid, Tao2021_Construction}, and the flow stability depends on the value of $\hbar$.
%This can be explained by the vorticity Clebsch mapping~\cite{Chern2016_Schrodinger, Chern2017_Fluid, Tao2021_Construction}, which maps a vortex tube to a patch of $\mathbb{S}^2$ so that the vorticity flux is proportional to the area of the patch by a factor of $\hbar$.
%Besides, similar to $\nu$ in the INSF, the parameter $\hbar$ is closely related to the flow stability and thus determines whether the flow transits to turbulence.
In Fig.~\ref{fig:TG_VortexSurface_IHSE}, the flow with smooth large-scale structures does not have a transition for $\hbar=1$, whereas the flow breaks down into turbulence with numerous chaotic vortex tubes for $\hbar=0.1$.
%Fig.~\ref{fig:TG_VortexSurface_IHSE}(c) shows the complex and chaotic vortex tube system of turbulent ISF very similar to the classical viscous turbulence.
% However, the ISF seems to be full of axial vortices represented by vortex tubes and lacks the typical coiled-up sheet vortex structures of the INSF.
%
The energy spectrum $E_k(k)$ of a turbulent ISF in Fig.~\ref{fig:TG_EnergySpectrum_ISF} exhibits a $-5/3$ scaling law in the inertial range as in classical turbulence~\cite{Pope2000_Turbulent}.
As $\hbar$ decreases, the inertial range broadens with a more pronounced $-5/3$ scaling.
%This property is very analogous to the $\nu$-dependent nature of the energy spectrum in classical turbulence.
% Besides, Fig.~\ref{fig:TG_Energy} shows the total kinetic energy curves for different $\hbar$ in the ISF and compares them with the classical viscous flow with $\Rey=1000$.
% The ISF is not strictly dissipative due to the presence of the LLF, but there is an overall fluctuating downward trend in energy.
% As $\hbar$ decreases, the energy fluctuations become smoother in contrast to the vortex dynamics, which becomes more intense.
% When $\hbar=1$, the energy decays continuously thereafter except for a brief rise at the beginning of the evolution, but the concavity of the curve is different from that of INSF.

\subsection{Decaying HIT}
%For the second comparative study, we focus on the decaying homogeneous isotropic turbulence (HIT).
%The primary difficulty is how to construct an appropriate initial condition $\vec{\psi}(\vec{x},t=0)$ for simulating HIT in the IHSE.
%The velocity field corresponding to this initial wave function needs to be random, non-divergent and needs to satisfy a certain energy spectrum.
%So, we briefly describe the process of constructing an initial field that meets the above conditions.

We construct an initial $\vec{\psi}(\vec{x},t=0)$, corresponding to a random divergence-free velocity, for simulating HIT in the ISF.
First, a normalized Gaussian-random wave function
\begin{equation}\label{eq:HIT_psi0}
    \begin{split}
        \vec{\psi}^* =\ & \frac{1}{\sqrt{-2\ln(r_1r_3)}}\left(\sqrt{-2\ln r_1}\cos(2\pi r_2) + \sqrt{-2\ln r_1}\sin(2\pi r_2)\vec{i} \right.
        \\
        &+ \left.\sqrt{-2\ln r_3}\cos(2\pi r_4)\vec{j} + \sqrt{-2\ln r_3}\sin(2\pi r_4)\vec{k} \right)
    \end{split}
\end{equation}
was generated, where $r_1$, $r_2$, $r_3$, and $r_4$ are independently generated real random numbers satisfying the uniform distribution within $[0,1]$.
Second, a divergence-free projection $\vec{\psi}^{**}=\ee^{-\ii q/\hbar}\vec{\psi}^*$ was applied, where $q$ is solved from $\nabla^2q = \hbar(\nabla^2\widebar{\vec{\psi}^{*}}\vec{i}	\vec{\psi}^{*} - \widebar{\vec{\psi}^{*}}\vec{i}\nabla^2	\vec{{\psi}}^{*})/2$.
Third, $\vec{\psi}^{**}$ was evolved using the IHSE for a time period $t_0=5$ to smooth the noisy initial $\vec{\psi}^{**}$.
Finally, $\vec{\psi}(\vec{x},t=0) = \vec{\psi}^{**}(\vec{x},t_0=5)$ and its corresponding velocity were taken as the initial conditions of IHSE and INSE, respectively.

% Fig.~\ref{fig:HIT_Ek0} shows the energy spectrum $E_k(k,t=0)$ of the velocity field corresponding to the wave function constructed in the above way.
% Surprisingly, although we cannot specify a designated energy spectrum $E_k(k,t=0)\sim k^\sigma \ee^{-\sigma(k/k_p)^2/2}$ for the Gaussian random field as the initial field of the HIT, as in INSF~\cite{Andre1977_Influence}, the velocity field constructed in the above way meets a good model spectrum $E_k(k,t=0)\sim k\ee^{-k/5}$ [it is not meaningful for an intermediate state. its scaling can be anyone.].
% The difference between this and the energy spectrum commonly used in INSF is only that the high wave number energy is slightly higher.
% [figure 13 can be removed.]
% To visualize the initial field we constructed, Fig.~\ref{fig:HIT_initial_s1=0`9} shows a family of vortex surfaces using an isosurface of $s_1 =0.9$ with several integrated vortex lines.
% The typical lamellar and axial vortex structures in classical viscous flows can be observed.
% From the distribution of coloring, we find the vorticity is concentrated near some fine vortex tubes, similar to classical turbulence.
% Note that the three components of the spin vector are almost indistinguishable for visualization in HIT, perhaps due to the nature of homogeneous isotropy, so the isosurfaces of the other two components $s_2,s_3$ are not shown.

\begin{figure}
    \centering
    \begin{overpic}[width=\linewidth]{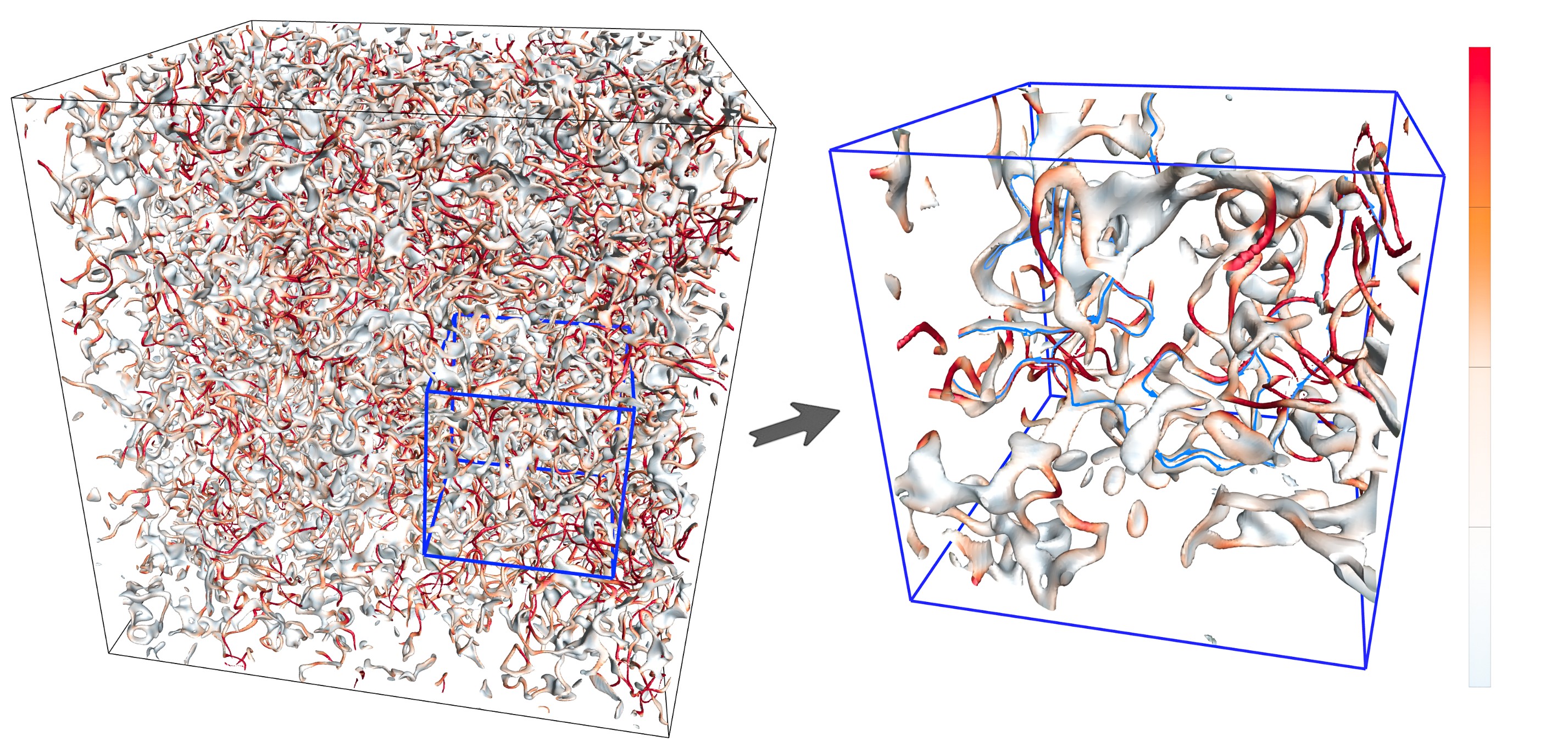}
        \begin{small}
            \put(0,45) {(a)}
            \put(52,45) {(b)}
            % legend
            \begin{footnotesize}
                \put(92.8,46) {$|\vec{\omega}|$}
                \put(96,3.2) {$0$}
                \put(96,13.375) {$1.25$}
                \put(96,23.55) {$2.5$}
                \put(96,33.725) {$3.75$}
                \put(96,43.9) {$5$}
            \end{footnotesize}
        \end{small}
    \end{overpic}
    \caption{Visualization of the tangled vortex tubes for the decaying HIT in the ISF. (a) Isosurface of $s_1s_2s_3=0.18$ color-coded by $|\boldsymbol\omega|$ at $t=5$. (b) Close-up view of the region marked by the blue box in (a). Some vortex lines (blue) are integrated and plotted on the isosurface in (b).}
    \label{fig:HIT_ISF_VortexSurface}
\end{figure}
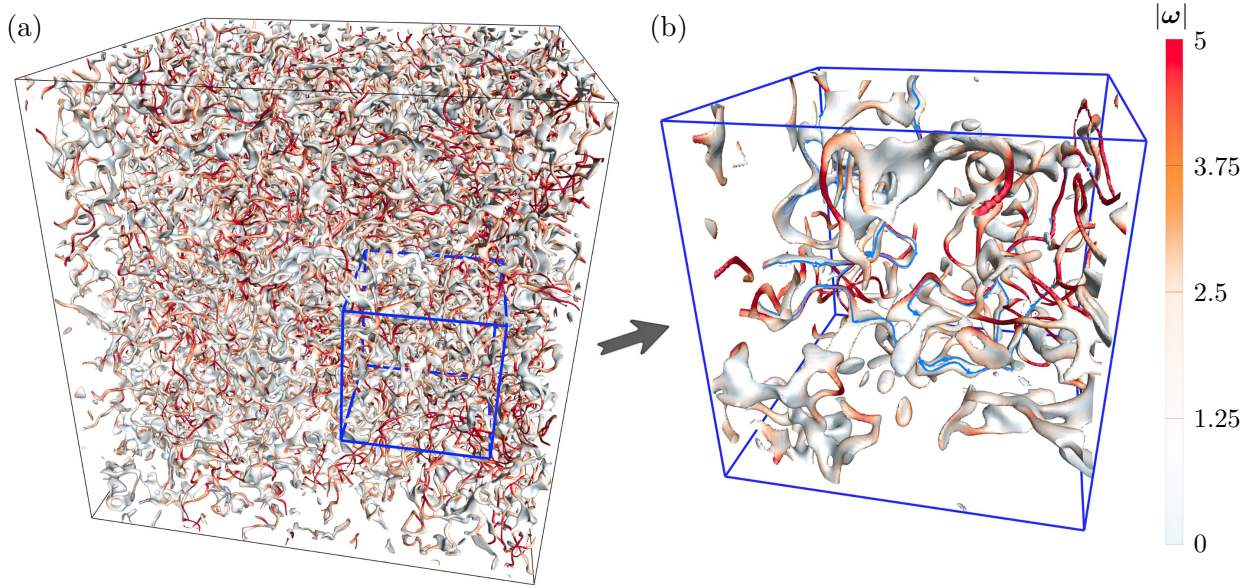

The vortex surface~\cite{Yang2010_On} in the fully developed turbulent ISF at $t=5$ is visualized in Fig.~\ref{fig:HIT_ISF_VortexSurface} using the isosurface of $s_1s_2s_3=0.18$.
We observe a network of entangled vortex tubes and sheets, which can be mapped to closed curves, the intersection of $s_1^2+s_2^2+s_3^2=1$ and $s_1s_2s_3=0.18$, on the unit sphere $\mathbb{S}^2$ (or the Bloch sphere) via the vorticity Clebsch mapping~\cite{Chern2016_Schrodinger, Chern2017_Fluid, Tao2021_Construction}.
The geometry of vortex surfaces in the ISF is in between the vortex filaments in quantum turbulence~\cite{Vahala2011_Unitary, Madeira2020_Quantum, Muller2021_Intermittency} and the tangle of spiral vortex tubes and sheets in classical turbulence~\cite{She1990_Intermittent, Cardesa2017_The, Xiong2019_Identifying}.
%Specifically, the vortex lines in the turbulent ISF are very similar to the vortex filaments in quantum turbulence, which are random, closed and knotted.
Therefore, the turbulent ISF manifests the features of both quantum and classical turbulent flows.
%be considered as a fusion
%However, there is an obvious difference between the turbulent ISF and practical turbulence, that the ISF appears to lack some of the typical vortex structures formed by the coiling of large-scale lamellar vortices.

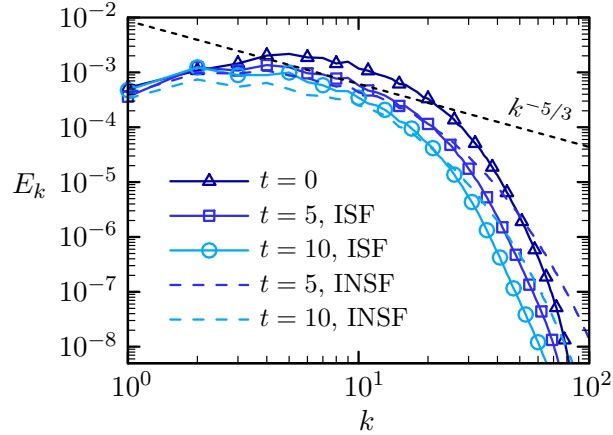
\begin{figure}
    \centering
    \begin{overpic}[width=0.4\linewidth]{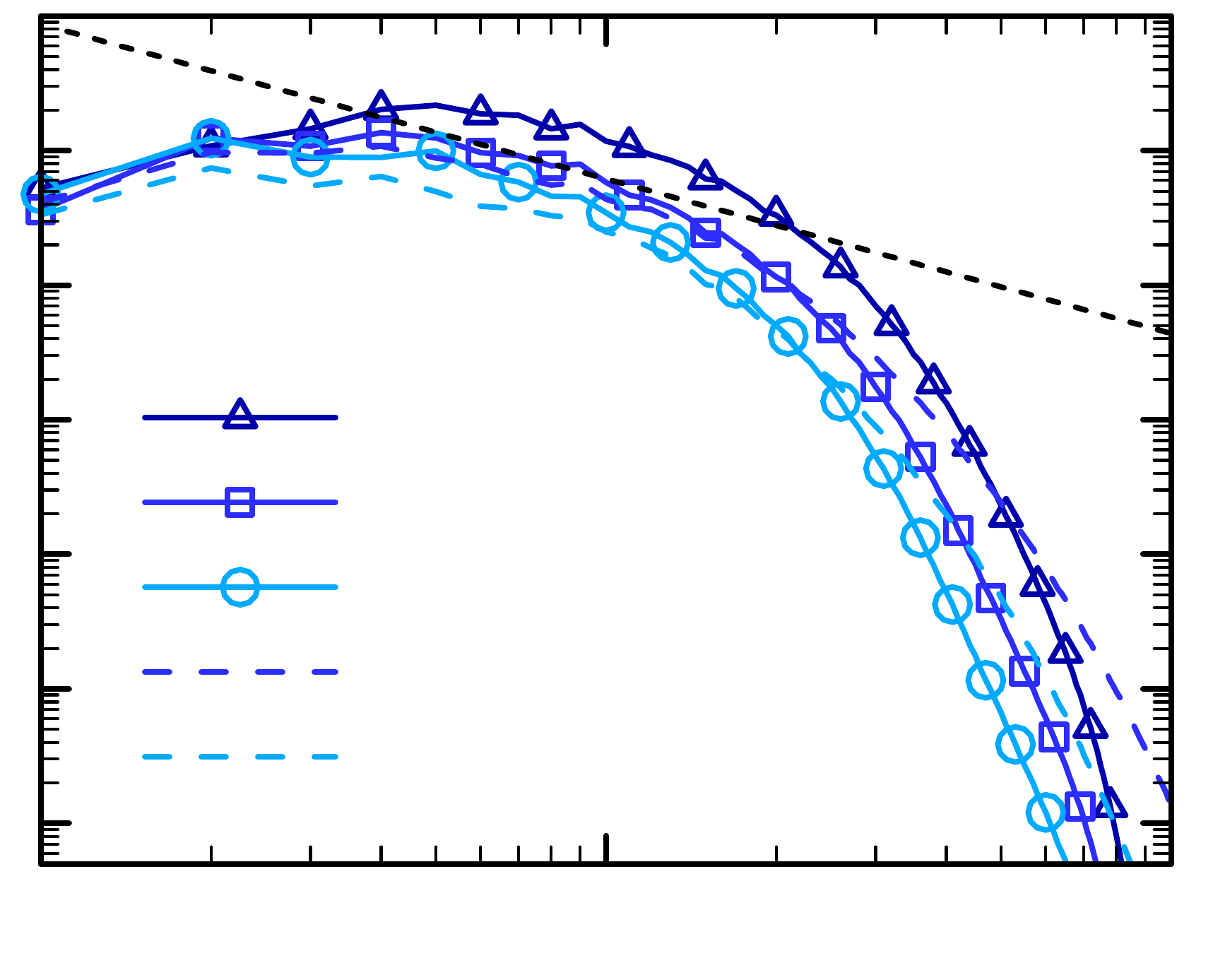}
        \begin{small}
            % x-axis
            \put(0,3) {$10^{0}$}
            \put(47,3) {$10^1$}
            \put(93,3) {$10^2$}
            \put(50,-4) {$k$}
            % y-axis
            \put(-10.6,11) {$10^{-8}$}
            \put(-10.6,22.2) {$10^{-7}$}
            \put(-10.6,33.3) {$10^{-6}$}
            \put(-10.6,44.2) {$10^{-5}$}
            \put(-10.6,55.4) {$10^{-4}$}
            \put(-10.6,66.5) {$10^{-3}$}
            \put(-10.6,77.6) {$10^{-2}$}
            \put(-20,44) {$E_k$}
            % legend & mark
            \begin{footnotesize}
                \put(30,45) {$t=0$}
                \put(30,38) {$t=5$, ISF}
                \put(30,31) {$t=10$, ISF}
                \put(30,24) {$t=5$, INSF}
                \put(30,17) {$t=10$, INSF}
                \put(80,60) {\begin{rotate}{-15} $k^{-5/3}$ \end{rotate}}
            \end{footnotesize}
        \end{small}
    \end{overpic}
    \caption{Evolution of the energy spectra for the decaying HIT in the ISF with $\hbar=0.1$ and the INSF with $\Rey=2000$.}
    \label{fig:HIT_Ek}
\end{figure}

Figure~\ref{fig:HIT_Ek} shows the evolution of $E_k(k)$ for the decaying HIT in the ISF with $\hbar=0.1$ and the INSF with $\nu=0.0005$ (or $\Rey=2000$ for unity length and velocity scales).
The energy spectrum in the ISF exhibits the $-5/3$ scaling law in the inertial range as in classical turbulence~\cite{Pope2000_Turbulent}, and it decays with time due to energy dissipation.
In addition, the total kinetic energy decays with time in the turbulent ISF as in the classical HIT (not shown).
%Figure~\ref{fig:HIT_Energy} shows the decay of the total kinetic energy $E$ with time in the ISF with $\hbar=0.1$ and in the classical HIT with $\nu=0.001$ and 0.0005.
%For the turbulent ISF at $\hbar=0.1$, we roughly estimate that the energy dissipation rate is roughly equivalent to an INSF at $\nu=$ 0.0001--0.0002.
%As discussed in Section~\ref{sec:TG_Flow}, $\hbar$ in the ISF plays a similar role of $\nu$ in the INSF.
%, i.e., changing the rate of energy dissipation.
%Since it is impossible to construct an identical random initial field when changing the value of $\hbar$, here we do not show profiles with different $\hbar$.

\section{Quantum algorithm for the ISF}\label{sec:QA_ISF}

\subsection{Prediction-correction approach} % and flowchart
We develop a quantum algorithm for simulating the ISF. % with $\rho_0=1$.
As sketched in Fig.~\ref{fig:SystemConv}, the algorithm can be executed on a quantum processor with measurements only at the end of the simulation,
%The information transfer between classical and quantum hardware takes place only at the start (state preparation) and end (measurement) of the simulation.
so it does not involve frequent information exchanges between classical and quantum hardware as in existing hybrid quantum-classical methods \cite{Liu2022_Application, Pfeffer2022_Hybrid, Lapworth2022_A, Chen2022_Quantum}.
This quantum algorithm can have significant advantages over classical and hybrid ones in terms of computational speedup, memory saving, and reduction of noises introduced by measurements.
%In addition, the present quantum algorithm can save memory exponentially due to the state superposition, compared with the classical one.

\begin{figure}
    \centering
    \includegraphics[]{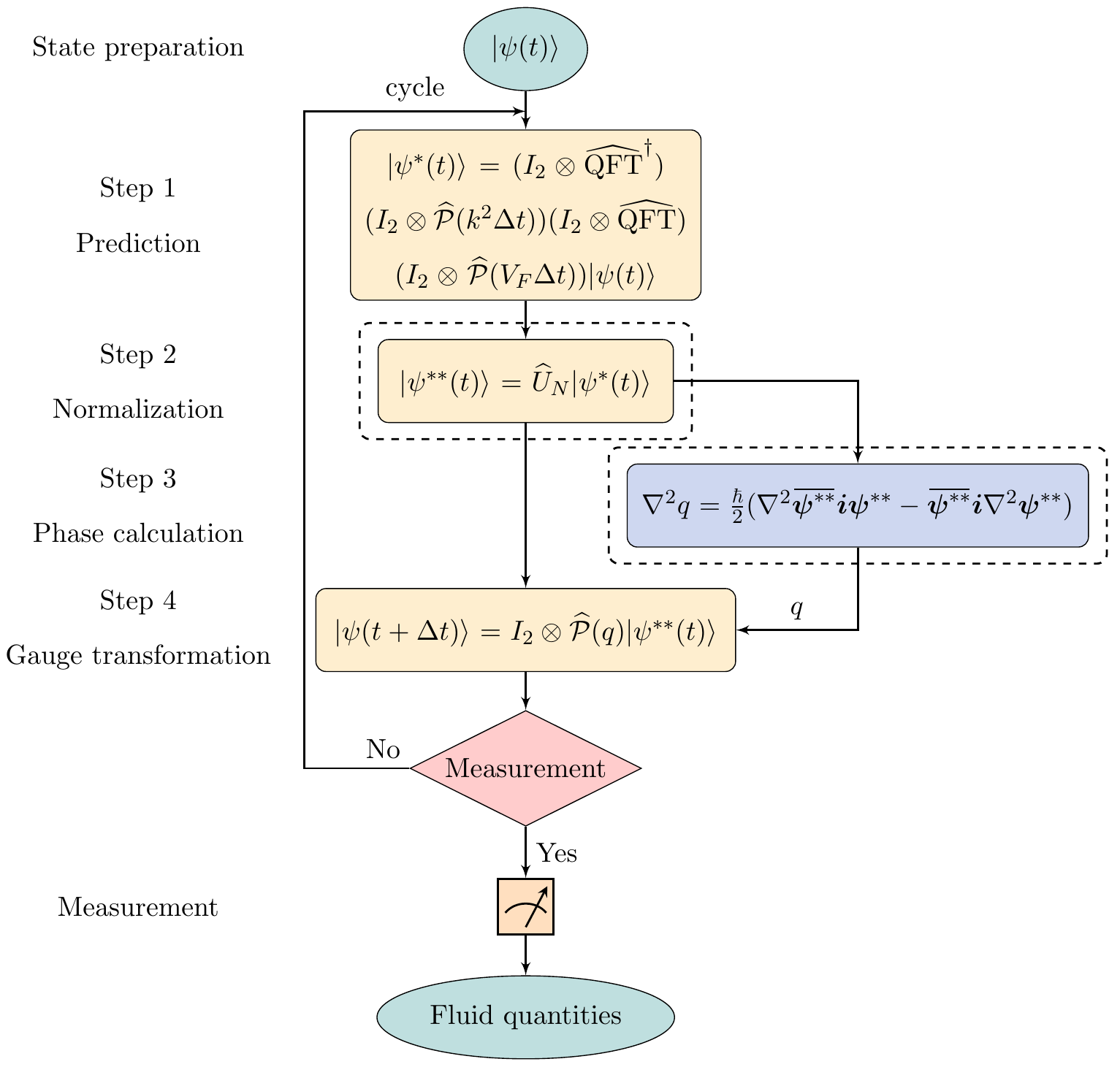}
    \caption{Flowchart of the quantum algorithm for simulating the ISF. The dashed boxes highlight the bottlenecks of quantum speedup in the present algorithm.}
    \label{fig:Process_QCISF}
\end{figure}

We apply a prediction-correction approach to bypass handling the nonlinear potential in Eq.~\eqref{eq:potential} in the IHSE.
As in classical algorithms~\cite{Patankar1972_A,Kim1985_Application,Issa1986_The} for simulating incompressible flows, the pressure is not solved using the pressure-Poisson equation
\begin{equation}\label{eq:Poisson-p}
    \nabla^2 p
    = \bn\cdot\left(\vec{u}\times\vec{\omega} - \frac{\hbar^2}{4}\bn\vec{s}\cdot\nabla^2\vec{s}\right) - \nabla^2\left(V_F + \frac{|\vec{u}|^2}{2}\right),
\end{equation}
because the RHS in Eq.~\eqref{eq:Poisson-p} is difficult to encode on a quantum computer.
First, we perform a prediction to obtain a temporary wave function using Eq.~\eqref{eq:ISF_SchEq} with ignoring $p - \hbar^2|\bn\vec{s}|^2/8$.
Second, we apply a divergence-free projection of the temporary wave function.
%In this way, we bypass solving Eq.~\eqref{eq:Poisson-p} and the strongly nonlinear term $-\frac{\hbar^2}{8}|\bn\vec{s}|^2$.
The flowchart of this quantum algorithm the ISF is illustrated in Fig.~\ref{fig:Process_QCISF}.
Next, we elaborate each step in the algorithm in Fig.~\ref{fig:Process_QCISF} using a 1D problem, and it is straightforward to extend the algorithm to 3D problems.

\subsection{Quantum encoding of the IHSE}
In an $n+1$-qubit quantum register, the state of a ``Pauli particle"~\cite{Bjorken1964_Relativistic, Davydov1965_Quantum, Messiah1968_Quantum}, whose motion is governed by the IHSE~\eqref{eq:ISF_SchEq} with $\rho=1$, can be encoded as follows.
We use $n$ qubits with state vectors $\ket{j_0},\ket{j_1},\cdots,\ket{j_{n-1}}$ to encode a particle location.
The last qubit $\ket{j_{n}} = \ket{s}$ stores the spin state of the particle.
The state of each qubit $j_0,j_1,\cdots,j_{n-1},j_n$ takes the value 0 or 1.
The domain $-d\le x\le d$ for the particle location is discretized into $2^n$ segments with the spacing $\Delta x=2d/2^n$.
These segments can be represented by the computational basis $\ket{x_j}=\ket{j_{n-1}j_{n-2}\cdots j_0}$ in the Hilbert space $\mathbb{C}^{2^n}$ with the shorthand $\ket{j_{n-1}j_{n-2}\cdots j_0}\equiv \ket{j_{n-1}}\otimes\ket{j_{n-2}}\otimes\cdots\otimes\ket{j_0}$.

In this way, the quaternionic wave function $\vec{\psi}(x,t)$ is approximated by the state vector
\begin{equation}
    \ket{\psi} = \frac{1}{\mathcal{N}}\sum_{s=0}^1\sum_{j_{n-1}=0}^1\cdots \sum_{j_0=0}^1\psi_s(x_{j},t)\ket{s}\otimes\ket{j_{n-1}j_{n-2}\cdots j_0}
\end{equation}
with $\psi_0\equiv a+\ii b$, $\psi_1\equiv c+\ii d$, $x_j\equiv -d+(j+\frac{1}{2})\Delta x$, $j=\sum_{i=0}^{n-1}j_i2^i$, and
\begin{equation}\label{eq:N_define}
    \mathcal{N}
    = \sqrt{\sum_{s=0}^{1}\sum_{j=0}^{2^n-1}|\psi_{s}(x_j,t)|^2}
    = \sqrt{2^n\rho}
    = \sqrt{2^n}.
\end{equation}
Hence, the wave function is reconstructed by
\begin{equation}
    \vec{\psi}(x_j,t) = \sqrt{2^n}\left(\Real{\langle x_j0 \,|\, \psi \rangle} + \Imag{\langle x_j0 \,|\, \psi \rangle}\vec{i} + \Real{\langle x_j1 \,|\, \psi \rangle}\vec{j} + \Imag{\langle x_j1 \,|\, \psi \rangle}\vec{k} \right).
\end{equation}
%where 0 and 1 are [...].

\subsection{Quantum algorithm for solving the IHSE}\label{sec:QA_IHSE}
\subsubsection*{Step 1: Prediction}
\begin{figure}
    \centering
    \includegraphics[width=\textwidth]{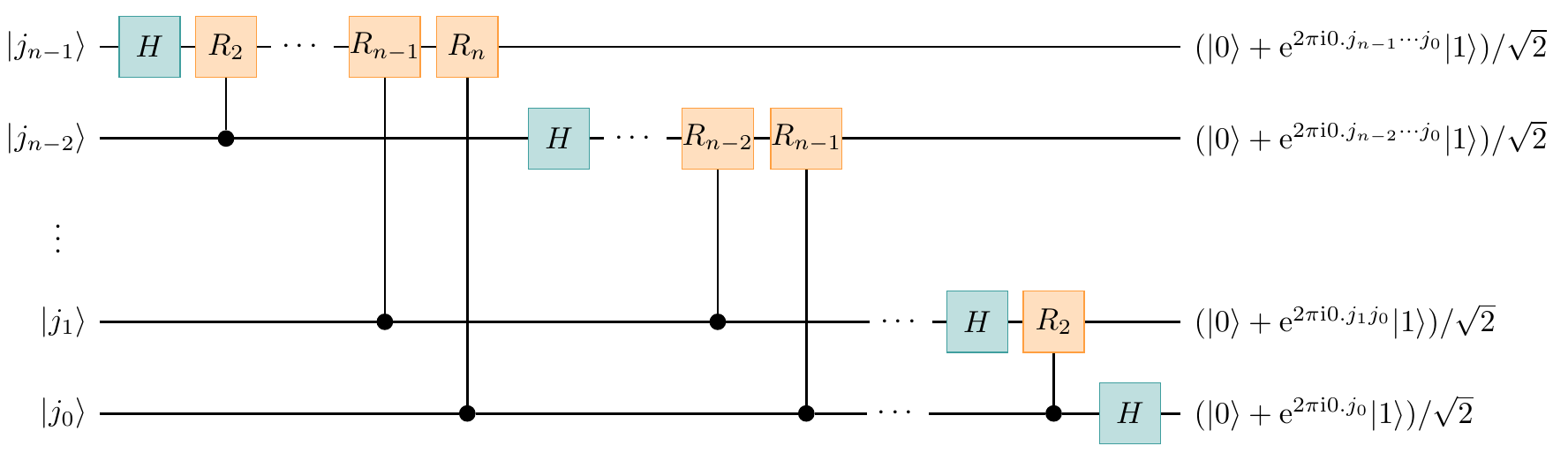}
    \caption{Quantum circuit for the QFT in Eq.~\eqref{eq:QFT} with $n$ qubits, where the SWAP gates for reversing the order of qubits in the end of the QFT are not shown.}
    \label{fig:QFT_circuit}
\end{figure}

In step 1, we treat the particle governed by the IHSE as a free Pauli particle. %without an external potential.
Namely,  Eq.~\eqref{eq:ISF_SchEq} becomes $\vec{i}\hbar\frac{\p}{\p t}\vec{\psi}(x,t) = \left(-\frac{\hbar^2}{2}\p_x^2 + V_F(x) \right)\vec{\psi}(x,t)$.
The motion of such a particle is described by a temporary solution~\cite{Benenti2008_Quantum, Georgescu2014_Quantum, Ostrowski2017_Quantum, Bogdanov2021_Solution}
\begin{equation}\label{eq:psi_star}
    \ket{\psi^*(t)} = \left(I_2\otimes\widehat{\mathrm{QFT}}^\dag\right)
    \left(I_2\otimes\widehat{\mathcal{P}}(k^2\Delta t)\right)
    \left(I_2\otimes\widehat{\mathrm{QFT}}\right)
    \left(I_2\otimes\widehat{\mathcal{P}}(V_F\Delta t)\right)\ket{\psi(t)}
\end{equation}
with the $2^1\times 2^1$ identity matrix $I_2$, the quantum Fourier transform (QFT)~\cite{Coppersmith1994_An, Jozsa1998_Quantum, Weinstein2001_Implementation}
\begin{equation}\label{eq:QFT}
    \widehat{\mathrm{QFT}}:~
    \ket{j}
    \to \frac{1}{\sqrt{2^n}}\sum_{k=0}^{2^n-1}\ee^{2\pi\ii\frac{jk}{2^n}}\ket{k},
\end{equation}
the Hermitian transpose $\dag$, and the diagonal unitary transformation $\widehat{\mathcal{P}}(f)\equiv \ee^{-\ii f/\hbar}$.
The QFT can be implemented by $\Od{n^2}$ quantum gates in Fig.~\ref{fig:QFT_circuit}, which achieves an exponential acceleration compared to $\Od{n2^n}$ operations of the fast Fourier transform.
Equation \eqref{eq:psi_star} is an approximation based on the second-order Trotter decomposition~\cite{Nielsen2010_Quantum}
\begin{equation}
    \ee^{-\ii(H_0+V_F)\Delta t/\hbar}
    = \ee^{-\ii H_0\Delta t/\hbar}\ee^{-\ii V_F(x)\Delta t/\hbar} + \Od{\Delta t^2},
\end{equation}
because the kinetic energy $H_0=|\widehat{\vec{p}}|^2/2$ and $V_F$ are not commute, i.e., $[H_0,V_F]\ne 0$.
The time stepping $\Delta t$ should be small enough to ensure accuracy.

An efficient quantum implementation of $\widehat{\mathcal{P}}(f)$ in Eq.~\eqref{eq:psi_star} is important.
The variable
\begin{equation}
    x_j = -d + \left(j+\frac{1}{2}\right)\Delta x
    = c_0\sum_{i=0}^{n-1}\left(j_i2^i + c_1\right)
\end{equation}
in $V(x)$ is discretized with constants $c_0=\Delta x$ and $c_1=(-d+\Delta x/2)/(n\Delta x)$, and thus
\begin{equation}
    \widehat{\mathcal{P}}(V_F\Delta t):~
    \ket{j} \to \ee^{-\ii V_F\left(c_0\sum_{i=0}^{n-1}(j_i2^i + c_1)\right)\Delta t/\hbar}\ket{j}
\end{equation}
is computable.

\begin{figure}
    \centering
    \includegraphics{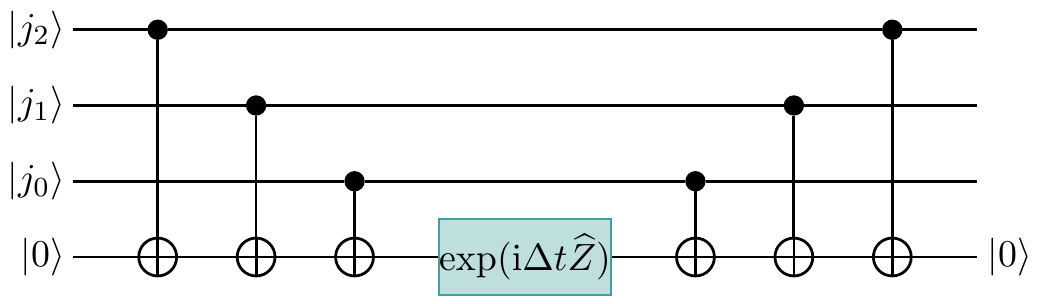}
    \caption{Quantum circuit for calculating $\widehat{H}=\exp(\ii\Delta t \widehat{Z})$ with an ancilla qubit and  $\widehat{Z}=\widehat{Z}_2\otimes\widehat{Z}_1\otimes\widehat{Z}_0$.}
    \label{fig:PhaseShift_Z}
\end{figure}

\begin{figure}
    \centering
    \includegraphics[width=\linewidth]{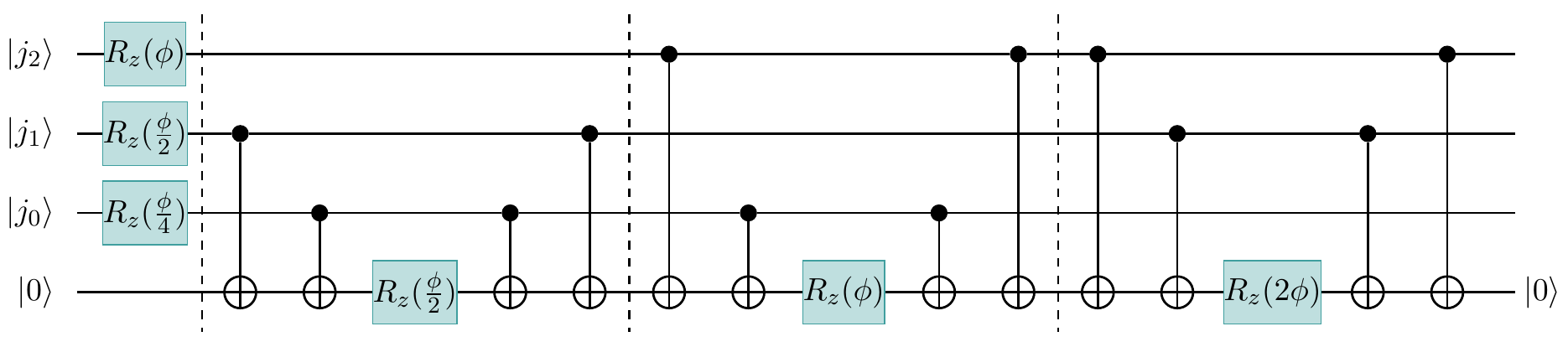}
    \caption{Quantum circuit for calculating $\widehat{\mathcal{P}}(k^2\Delta t)$ in Eq.~\eqref{eq:P_k2t_ex} with $n=3$ qubits and an ancilla qubit.}
    \label{fig:PhaseTransEx}
\end{figure}

Similarly, we express the $n$-bit number $k=\sum_{j=0}^{n-1}k_{j}2^{j}$ with $k_0,k_1,\cdots,k_{n-1}\in \{0,1\}$ in the momentum operator $\widehat{\mathcal{P}}(k^2\Delta t)$ in Eq.~\eqref{eq:psi_star}.
Using the wavenumber expressed by~\cite{Benenti2008_Quantum, Rodrigues2018_Validation}
\begin{equation}
    k = \sqrt{\frac{1}{2^{2n-3}}\frac{\phi\hbar}{\Delta t}}\left(1 + \sum_{j=0}^{n-1} 2^{j}\widehat{Z}_j \right),
\end{equation}
we obtain
\begin{equation}\label{eq:P_k2t}
    \widehat{\mathcal{P}}(k^2\Delta t)
    = \exp\left(\frac{\ii\phi}{2^{2n-3}} \right)\prod_{\ell=0}^{n-1}\exp\left( \frac{\ii\phi}{2^{2n-\ell-4}}\widehat{Z}_\ell\right)\prod_{\substack{i,j=0 \\ i>j}}^{n-1}\exp\left(\frac{\ii\phi}{2^{2n-i-j}}\widehat{Z}_i\otimes\widehat{Z}_j \right).
\end{equation}
Here, $\widehat{Z}_j$ denotes a phase-shift gate at the $j$-th qubit and $\phi=\Delta t/\hbar$ is a small phase shift on a time step.
The quantum circuit for calculating a Hamiltonian with the form of $\widehat{H}=\exp(\ii\Delta t \widehat{Z}_2\otimes\widehat{Z}_1\otimes\widehat{Z}_0)$ with a given $\Delta t$ is shown in Fig.~\ref{fig:PhaseShift_Z}, using an ancilla qubit~\cite{Nielsen2010_Quantum}.
Taking, e.g., $n=3$ qubits and ignoring the global phase $\ee^{\ii\phi/8}$, Eq.~\eqref{eq:P_k2t} becomes
\begin{equation}\label{eq:P_k2t_ex}
    \widehat{P}(k^2\Delta t)
    = \exp\left[\ii\phi\left(\widehat{Z}_2 + \frac12\widehat{Z}_1 + \frac14\widehat{Z}_0 + 2\widehat{Z}_2\otimes\widehat{Z}_1 + \widehat{Z}_2\otimes\widehat{Z}_0 + \frac12\widehat{Z}_1\otimes\widehat{Z}_0 \right) \right],
\end{equation}
which can be realized by a quantum circuit in Fig.~\ref{fig:PhaseTransEx}.
Thus, only $\Od{n^2}$ quantum gates are sufficient to calculate the momentum operator in Eq.~\eqref{eq:P_k2t}.

\subsubsection*{Step 2: Normalization}
Since the temporary solution $\vec{\psi}^*(x,t)$ obtained from Eq.~\eqref{eq:psi_star} in step 1 is not necessarily on $\mathbb{S}^3$, i.e., $\widebar{\vec{\psi}^*}\vec{\psi}^*\ne 1$, it is normalized as
\begin{equation}\label{eq:UN}
    \ket{\psi^{**}(t)} = \widehat{U}_N\ket{\psi^*(t)}
\end{equation}
in step 2, with an unitary operation $\widehat{U}_N$ to obtain $\widebar{\vec{\psi}^{**}}\vec{\psi}^{**}=1$.

\begin{figure}
    \centering
    \includegraphics{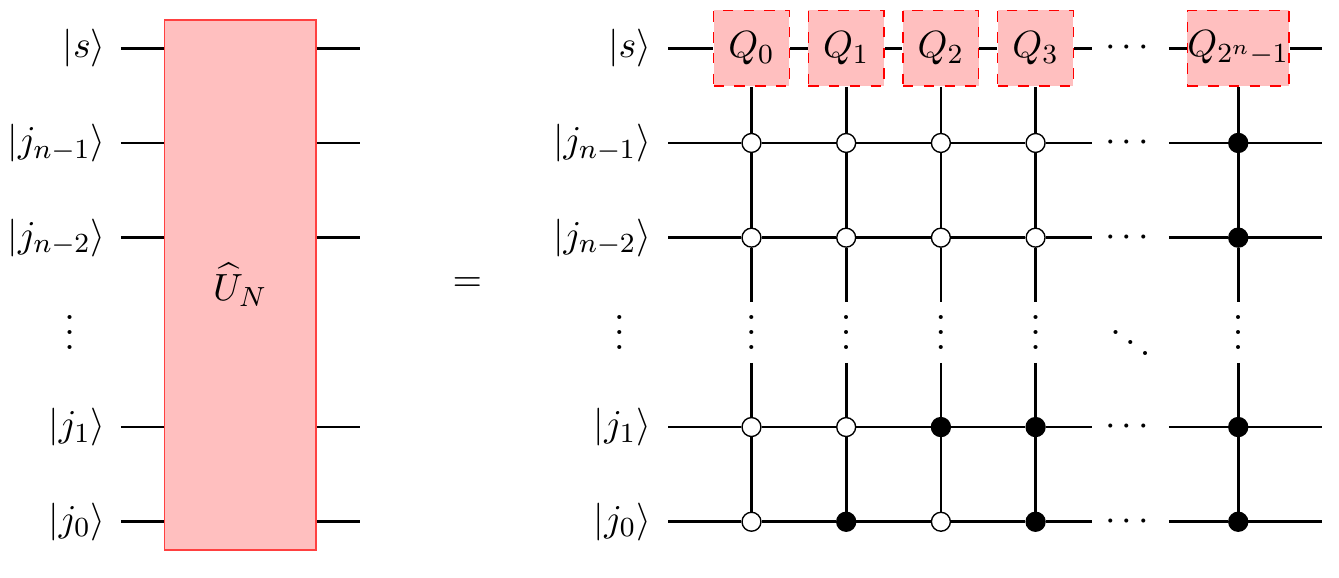}
    \caption{Conceptual quantum circuit for calculating $\widehat{U}_N$ in Eq.~\eqref{eq:U_N1}. The dashed boxes mark that  $Q_i,~i=1,2,\cdots,2^n-1$ may be non-unitary, so they are not realizable quantum gates. Hollow and solid circles denote control qubits with values 0 and 1, respectively. The operation $Q_j$ on the target qubit $\ket{s}$ is active only when the control qubits are in the state $\ket{j}=\ket{j_{n-1}j_{n-2}\cdots j_0}$ with $j=\sum_{i=0}^{n-1}j_i2^i$.}
    \label{fig:UN}
\end{figure}

This normalization step appears to be difficult to implement on a quantum computer.
As illustrated in Fig.~\ref{fig:UN}, we can only conceptually decompose the operator
\begin{equation}\label{eq:U_N1}
    \widehat{U}_N
    = Q_{2^n-1}Q_{2^n-2}\cdots Q_1Q_0
    = \prod_{i=0}^{2^n-1}Q_{2^n-1-i},
\end{equation}
with scale transformations
\begin{equation}\label{eq:Qi}
    Q_i:~ \frac{1}{\sqrt{2^n}}\sum_{s=0}^1\sum_{j=0}^{2^n-1} \psi_s(x_j,t)\ket{s}\otimes\ket{j}
    \to
    \frac{1}{\sqrt{2^n}}\sum_{s=0}^1\sum_{j=0}^{2^n-1}\frac{\psi_s(x_j,t)}{\sqrt{|\psi_0(x_j,t)|^2 + |\psi_1(x_j,t)|^2}}\ket{s}\otimes\ket{j}.
\end{equation}
Here, each $Q_i$ may be non-unitary with $Q_i^\dag Q_i\ne I$ and thus it is not realizable using a quantum gate, whereas their product Eq.~\eqref{eq:U_N1} is unitary.
%Equation~\eqref{eq:Qi} is only used to explain the capability of $\widehat{U}_N$, as it is likely that it cannot be implemented on a quantum computer.
Therefore, an effective quantum algorithm for calculating $\widehat{U}_N$ with the complexity $\Od{\mathrm{poly}(n)}$ remains an open problem.

\subsubsection*{Step 3: Phase calculation}
After step 2, the divergence of the velocity $u^{**}=\hbar(\p_x\widebar{\vec{\psi}^{**}}\vec{i}\vec{\psi}^{**}-\widebar{\vec{\psi}^{**}}\vec{i}\p_x\vec{\psi}^{**})/2$ can be non-zero.
%\begin{equation}
%    \p_x u^{**}
%    = \frac{\hbar}{2}(\p_x^2\widebar{\vec{\psi}^{**}}\vec{i}\vec{\psi}^{**}
%    - \widebar{\vec{\psi}^{**}}\vec{i}\p_x^2 \vec{{\psi}}^{**})
%    \ne 0.
%\end{equation}
%
A divergence-free projection of $u^{**}$, as a gauge transformation $\vec{u}\to\vec{u}-\bn q$, is applied, where the phase $q$ is solved from a Poisson equation
\begin{equation}\label{eq:Poi_psiss}
    \p_x^2q = \frac{\hbar}{2}\left(\p_x^2\widebar{\vec{\psi}^{**}}\vec{i}	\vec{\psi}^{**} - \widebar{\vec{\psi}^{**}}\vec{i}\p_x^2	\vec{{\psi}}^{**}\right).
\end{equation}
This projection corresponds to the gauge transformation $\vec{\psi}\to \ee^{-\ii q/\hbar}\vec{\psi}$ for the wave function~\cite{Yang2021_Clebsch}.

The encoding of the RHS of Eq.~\eqref{eq:Poi_psiss} without affecting the quantum state $\ket{\psi^{**}}$ appears to be challenging.
%If it can be successfully encoded, the subsequent solving process is relatively easy.
A solution of this issue admits an efficient quantum algorithm~\cite{Cao2013_Quantum, Arrazola2019_Quantum, Childs2020_Quantum, Liu2021_Variational, Childs2021_High} to solve Eq.~\eqref{eq:Poi_psiss} using $\Od{\mathrm{poly}(n)}$ basic quantum gates.

\subsubsection*{Step 4: Gauge transformation}
\begin{figure}
    \centering
    \includegraphics[]{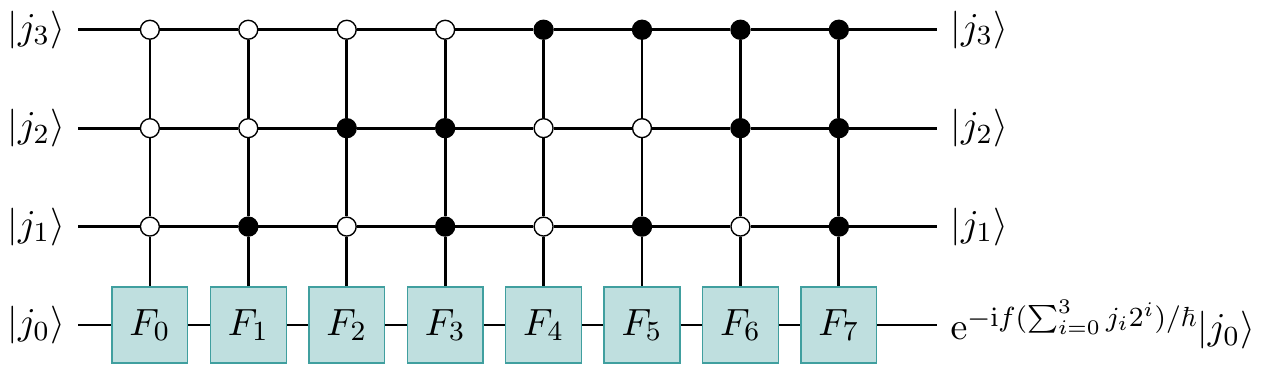}
    \caption{Quantum circuit for calculating $\widehat{\mathcal{P}}(f(x))$ in Eq.~\eqref{eq:GDUT} with $n=4$ qubits.
    Hollow and solid circles denote control qubits with values 0 and 1, respectively. The operation $F_j$ on the target qubit $\ket{j_0}$ is active only when the control qubits are in the state $\ket{j} = \ket{j_{3}j_{2} j_1}$ with $j=\sum_{i=0}^2j_{i+1}2^i$.}
%are set to the correct series of 0 and 1
    \label{fig:GDUT}
\end{figure}

In the final step, we take a gauge transformation
\begin{equation}\label{eq:GaugeTrans}
    \ket{\psi(t+\Delta t)}=I_2\otimes\widehat{\mathcal{P}}(q)\ket{\psi^{**}(t)}
\end{equation}
from the temporary state $\ket{\psi^{**}}$ at $t$ to the state $\ket{\psi}$ at $t+\Delta t$.
The diagonal unitary transformation
\begin{equation}\label{eq:GDUT}
    \widehat{\mathcal{P}}(f(x)):\quad
    \ket{x} \to \ee^{-\ii f(x)/\hbar}\ket{x}
\end{equation}
of a function $f(x)$ can be implemented by $\Od{2^n}$ generalized controlled-phase shift gates.
They apply the single qubit gate $F_j$ to a target qubit $\ket{j_0}$ only when the other $n-1$ controlled qubits are in the state $\ket{j}=\ket{j_{n-1}j_{n-2}\cdots j_1}$ \cite{Benenti2008_Quantum}.
An example with $n=4$ qubits is shown in Fig.~\ref{fig:GDUT}.

However, such an implementation is inefficient because its complexity scales exponentially with $n$.
A more efficient quantum circuit can be designed for a specific form of $f(x)$, e.g., only $\Od{n^2}$ basic quantum gates are used for $f(k)=k^2\Delta t$ in Eq.~\eqref{eq:psi_star}.
In general, if $f(x)$ has a specific form, e.g., $f(x)$ for the harmonic oscillator, square well, and quantum tunneling, Eq.~\eqref{eq:GDUT} can be calculated with the complexity $\Od{\mathrm{poly}(n)}$.
%The operation $\widehat{\mathcal{P}}(q)$ implemented by $\Od{2^n}$ generalized controlled-phase shift gates in Eq.~\eqref{eq:GaugeTrans} may be possible to implement more efficiently in the future.

\subsubsection*{Algorithm complexity}
\begin{figure}
    \centering
    \includegraphics{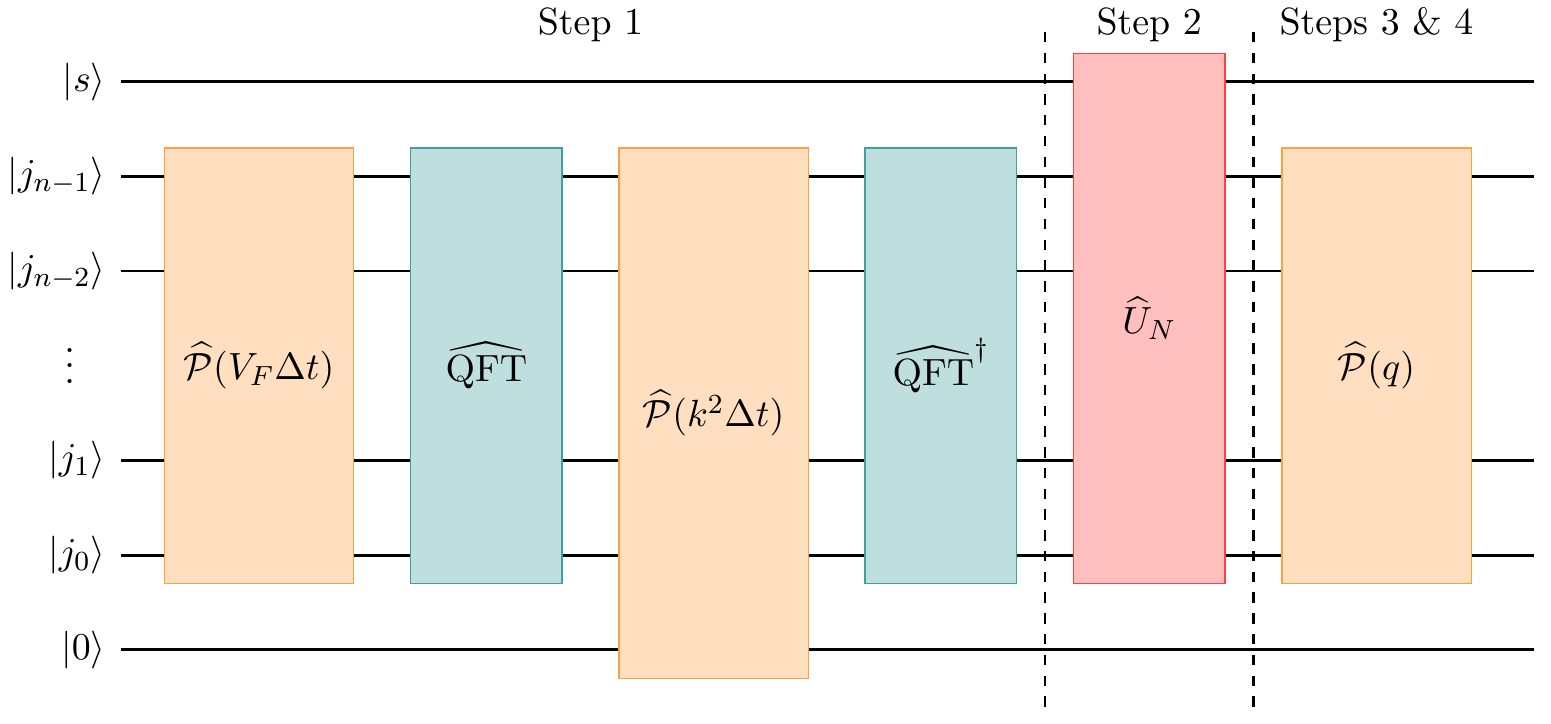}
    \caption{Overall quantum circuit for solving the IHSE, from time $t$ to $t+\Delta t$. The bottom qubit is an auxiliary one for implementing the momentum operator.}
    \label{fig:ISF_circuit}
\end{figure}

We estimate the total complexity of the quantum algorithm for solving the IHSE in Eq.~\eqref{eq:ISF_SchEq}.
The overall quantum circuit with the prediction, normalization, phase calculation, and gauge transformation is illustrated in Fig.~\ref{fig:ISF_circuit}.
The prediction is a standard simulation of a potential-free Pauli particle, using only $\Od{n^2}$ basic quantum gates.
The normalization is unconventional in quantum computing. The upper and lower bounds of operations are $\Od{2^n}$ and $\Od{\mathrm{poly}(n)}$, respectively.
%In addition, the quantum encoding of the RHS of Eq.~\eqref{eq:Poi_psiss} without affecting the quantum state $\ket{\psi^{**}}$ is still an open problem.
The gauge transformation, currently, can only be realized through the generic diagonal unitary transformation in Fig.~\ref{fig:GDUT}, using $\Od{2^n}$ basic quantum gates.

The complexities of each step and the entire algorithm are summarized in Table~\ref{tab:QC_Compx}.
The present quantum algorithm can achieve exponential speedup in steps 1 and 3, and possible
\begin{equation}
    S_1 = \frac{n2^n+\mathrm{poly}(2^n)}{\mathrm{poly}(n)}
\end{equation}
speedup overall.
The bottlenecks for the computational efficiency in steps 2 and 4 need to be tackled in the future work.

%The above discussion only considers the space complexity and do not including the time evolution.
Besides the spatial complexity, the temporal and spatial steps are related by the Courant--Friedrichs--Lewy (CFL) condition in both the quantum and classical implementations.
This implies that the number of time iterations is $N_t=\Od{N}$, where $N=2^n$ is the total number of grid points.
Thus, the numerical error $\epsilon$ decreases polynomially with the number of grid points.
Assuming that the unitary operators are smooth enough and the norm of the exponential operators are bounded by one~\cite{Wiebe2010_Higher, Fillion2017_Algorithm}, the error after $N_t$ iterations scales as $\epsilon\sim N_t\Delta t^2\sim N^{-1}$ for the second-order Trotter decomposition.
Given an error tolerance, the upper and lower bounds of the number of gates scale as $N_{\mathrm{gate}} \sim \epsilon^{-1}\mathrm{poly}(\log_2\epsilon^{-1})$ and $N_{\mathrm{gate}}\sim \epsilon^{-2}$, respectively, for a quantum algorithm.
In the classical algorithm, the number of operations scales as $N_{\mathrm{op}}\sim \epsilon^{-1}(\epsilon^{-1}\log_2\epsilon^{-1} + \mathrm{poly}(\epsilon^{-1}))$.
Therefore, even if the CFL condition constrains the time stepping, the quantum algorithm has a possible exponential speedup
\begin{equation}
    S_2 = \frac{\epsilon^{-1}(\epsilon^{-1}\log_2\epsilon^{-1} + \mathrm{poly}(\epsilon^{-1}))}{\epsilon^{-1}\mathrm{poly}(\log_2\epsilon^{-1})}
    = \frac{\epsilon^{-1}\log_2\epsilon^{-1} + \mathrm{poly}(\epsilon^{-1})}{\mathrm{poly}(\log_2\epsilon^{-1})}.
\end{equation}

\begin{table}
\caption{Breakdown of algorithm complexities of quantum and classical algorithms.}
\begin{ruledtabular}
    \renewcommand\arraystretch{1}
    \begin{tabular}{lccccc}
    & Step 1 & Step 2 & Step 3 & Step 4 & Total \\
    \hline
    Quantum & $\Od{n^2}$ & $\Od{\mathrm{poly}(n)}$--$\Od{2^n}$ & $\Od{\mathrm{poly}(n)}$ & $\Od{\mathrm{poly}(n)}$--$\Od{2^n}$ & $\Od{\mathrm{poly}(n)}$--$\Od{2^n}$ \\
    Classical & $\Od{n2^n}$ & $\Od{2^n}$ & $\Od{\mathrm{poly}(2^n)}$ & $\Od{2^n}$ & $\Od{n2^n+\mathrm{poly}(2^n)}$ \\
    \end{tabular}
\end{ruledtabular}
\label{tab:QC_Compx}
\end{table}

\subsection{Qiskit implementation} % of a simple arithmetic example}
We provide a simple 1D example to go through the entire algorithm, which is implemented on a quantum computer with exponential speedup.
The initial wave function is
\begin{equation}\label{eq:Simple1}
    \vec{\psi}(x,t=0) = \frac{\sqrt{2}}{2}\left(\cos\frac{x}{\hbar} + \sin\frac{x}{\hbar}\vec{i} + \cos\frac{x}{\hbar}\vec{j} + \sin\frac{x}{\hbar}\vec{k}\right),
\end{equation}
and the corresponding spin vector is $\vec{s}(x,0)=(0,0,1)$.
The solution of this steady ISF satisfies a Helmholtz equation
\begin{equation}
    \p_x^2\vec{\psi} - \left(\frac{2}{\hbar^2}p - \frac{1}{4}|\p_x\vec{s}|^2 \right)\vec{\psi}
    = \vec{0}.
\end{equation}
This simplified IHSE with $V_F=0$ with the initial condition in Eq.~\eqref{eq:Simple1} has a steady solution $u=1$, where the nonlinear potential in Eq.~\eqref{eq:ISF_SchEq} is simplified to $p=-1/2$.
%\begin{equation}
%    V = p - \frac{\hbar^2}{8}|\bn\vec{s}|^2
%    = -\frac{1}{2}.
%\end{equation}
Thus, we only need to perform steps 1 and 4 as
\begin{equation}\label{eq:Simple2}
    \ket{\psi(t+\Delta t)}
    = \left(I_2\otimes\widehat{\mathcal{P}}(-\frac{\Delta t}{2})\right)
    \left(I_2\otimes\widehat{\mathrm{QFT}}^\dag\right)
    \left(I_2\otimes\widehat{\mathcal{P}}(k^2\Delta t)\right)
    \left(I_2\otimes\widehat{\mathrm{QFT}}\right) \ket{\psi(t)}.
\end{equation}
The complexity for calculating a time step of Eq.~\eqref{eq:Simple2} is $\Od{n^2}$.
Compared to $\Od{\mathrm{poly}(2^nn)}$ for the classical algorithm, the exponential quantum speedup is achieved.

We validate the algorithm in Eq.~\eqref{eq:Simple2} using IBM's Qiskit~\cite{Qiskit}.
The Qiskit is an open-source software development kit for quantum computers at the level of pulses, circuits, and application modules.
We used a simulator with the quantum assembly language (QASM) on a classical computer, which mimics a quantum computer by adding small noises to the result~\cite{Koch2019_Introduction}.
The QASM also operates by running the quantum circuit multiple times and storing the number of times when an outcome occurs, similar to the procedure on a practical quantum computer.

The reconstruction of the probability distribution $|\vec{\psi}(x,t)|^2$ with a small statistical error needs to repeat the quantum simulation a large number of times.
After an outcome $x_j$ is obtained $M_j$ times in $M$ runs, $|\vec{\psi}(x_j,t)|^2 \approx \mathcal{N}^2M_j/M$ is estimated, with a normalization factor $\mathcal{N}$ in Eq.~\eqref{eq:N_define}.
Moreover, it is also possible to reconstruct the entire $\vec{\psi}(x_j,t)$ using a Ramsey-type quantum interferometry method~\cite{Gardiner1997_Quantum, Benenti2008_Quantum}, the quantum-state tomography~\cite{Smithey1993_Measurement, Breitenbach1997_Measurement, James2001_Measurement}, or the direct weak tomography~\cite{Vallone2016_Strong}.

\begin{figure}
    \centering
    \begin{overpic}[width=0.8\linewidth]{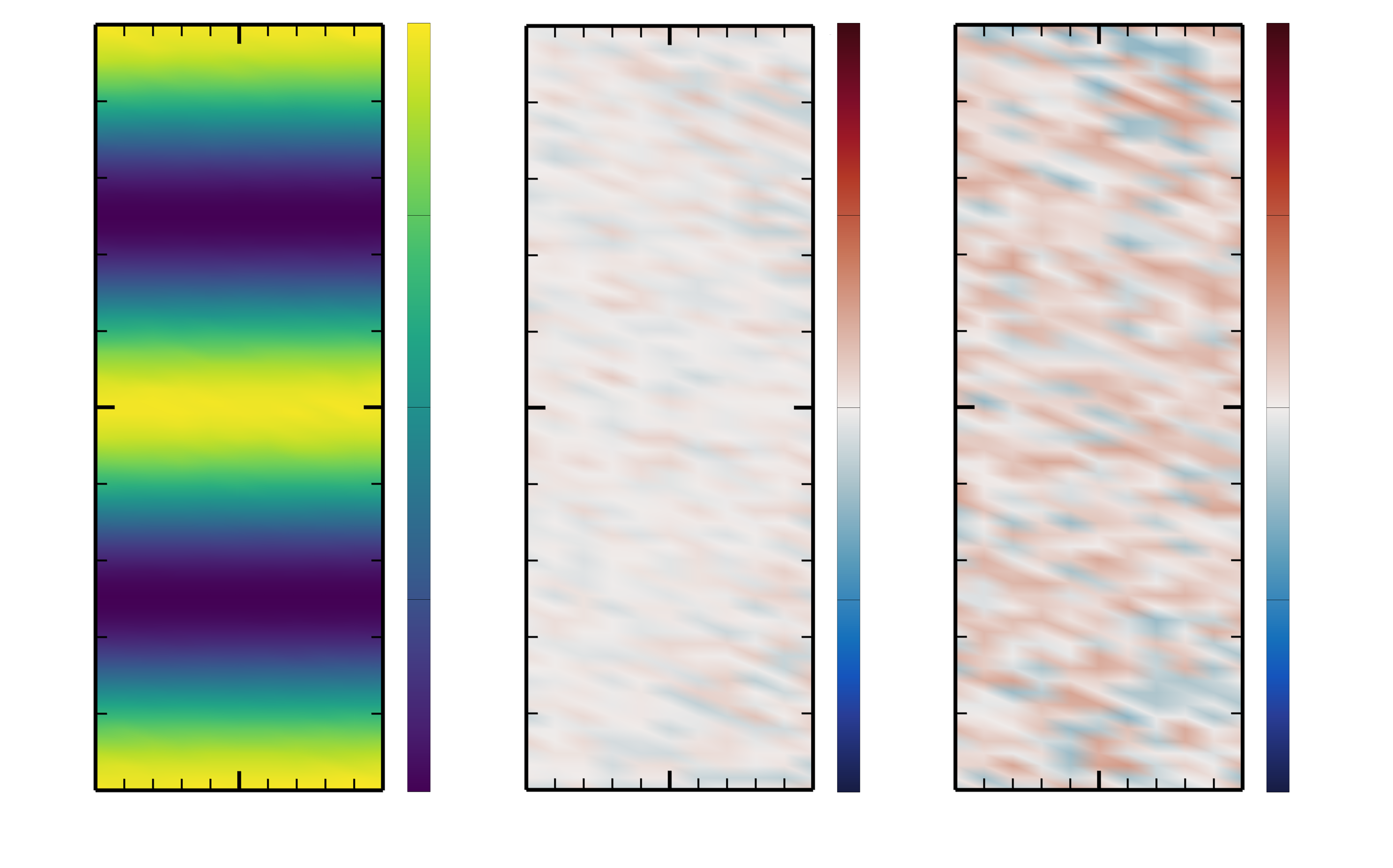}
        \begin{small}
            \put(6.7,61.5) {(a)}
            \put(37.5,61.5) {(b)}
            \put(68.4,61.5) {(c)}
            \put(13,61.5) {$2\Re{\psi_0}^2$}
            \put(47.3,61.5) {$\rho$}
            \put(78.3,61.5) {$u$}
            \put(4.5,60) {$\pi$}
            \put(4.5,32.2) {$0$}
            \put(2,5) {$-\pi$}
            \put(1,32.35) {$x$}
            \put(6.2,3) {$0$}
            \put(15.3,3) {$0.5$}
            \put(26.9,3) {$1$}
            \put(16.8,0) {$t$}
            \put(37.1,3) {$0$}
            \put(46.3,3) {$0.5$}
            \put(57.8,3) {$1$}
            \put(47.7,0) {$t$}
            \put(68,3) {$0$}
            \put(77.1,3) {$0.5$}
            \put(88.7,3) {$1$}
            \put(78.6,0) {$t$}
            % colorbar
            \begin{footnotesize}
                \put(31.5,59.8) {$1$}
                \put(31.5,46.15) {$0.75$}
                \put(31.5,32.3) {$0.5$}
                \put(31.5,18.4) {$0.25$}
                \put(31.5,4.8) {$0$}
                \put(62.3,59.8) {$1.2$}
                \put(62.3,46.15) {$1.1$}
                \put(62.3,32.3) {$1$}
                \put(62.3,18.4) {$0.9$}
                \put(62.3,4.8) {$0.8$}
                \put(93.1,59.8) {$1.2$}
                \put(93.1,46.15) {$1.1$}
                \put(93.1,32.3) {$1$}
                \put(93.1,18.4) {$0.9$}
                \put(93.1,4.8) {$0.8$}
            \end{footnotesize}
        \end{small}
    \end{overpic}
    \caption{Contours of (a) $2\Re{\psi_0(x,t)}^2$, (b) $\rho(x,t)=|\vec{\psi}(x,t)|^2$, and (c) $u(x,t)=\hbar(\p_x\widebar{\vec{\psi}}\vec{i\psi} - \widebar{\vec{\psi}}\vec{i}\p_x\vec{\psi})/2$ with $n=6$ qubits after $M=10^6$ runs. The $t$-axis is divided into 10 time steps within $0\le t\le 1$, and the $x$-axis is discretized into $2^n=64$ grid points within $-\pi\le x\le \pi$.} %Each run is followed by a projective measurement to reconstruct the wave function.
    \label{fig:QC_Example}
\end{figure}
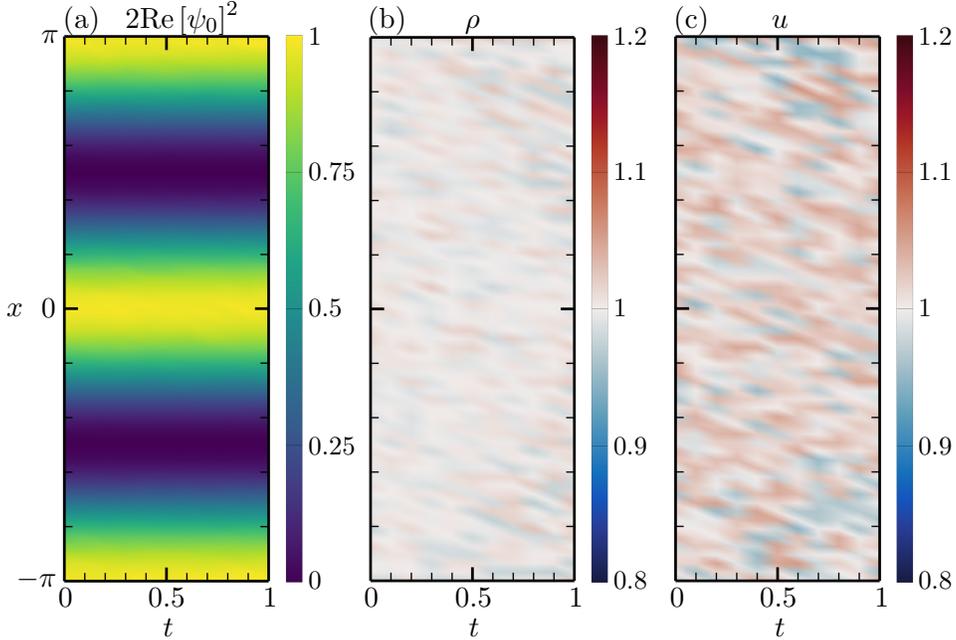

In Fig.~\ref{fig:QC_Example}(a), the result of the simple example described in Eq.~\eqref{eq:Simple2} from the Qiskit simulation with $M=10^6$ runs agrees with the theoretical distribution $\Re{\psi_0(x_j,t)}^2=\cos^2(x_j/\hbar)/2$.
%Note that $\psi_1(x_j,t)=\psi_0(x_j,t)$ in this example, so we just show $\psi_0$.
Figures~\ref{fig:QC_Example}(b, c) show that the reconstructed mass density and velocity have slight deviations from the theoretical values $\rho=1$ and $u=1$ due to the statistical errors and the noises introduced by the QASM simulator.

Moreover, we performed a hybrid quantum-classical simulation of a 2D unsteady TG ISF in Appendix~\ref{app:hybrid_2DTG} to demonstrate the capability of simulating high-dimensional ISFs illustrated in Section~\ref{sec:ISF_DNS}.
The obstacles (steps 2--4) in the quantum algorithm were tentatively treated on the classical computer.

\section{Conclusions}\label{sec:conclusion}

We develop a framework for the quantum computing of fluid dynamics based on the HSE with a generalized Madelung transform.
The SF, a flow with finite vorticity and dissipation, is governed by the HSE in Eq.~\eqref{eq:HSE} of a two-component wave function or by the continuity and momentum equations in Eqs.~\eqref{eq:contEq} and \eqref{eq:Du/Dt_SF}.
Since the Hamiltonian of the SF is Hermitian, we are able to obtain $\vec{\psi}(\vec{x},t)$ from an initial wave function in the quantum computing of the HSE (see Fig.~\ref{fig:SystemConv}).
%The new framework of HSE may lead to immeasurable application potential in the intersection of quantum computing and CFD.

In particular, we develop a prediction-correction quantum algorithm for the ISF, a constant-density incompressible SF governed by the IHSE~\eqref{eq:ISF_SchEq}.
This algorithm can be executed on a quantum processor with measurements only at the end of the simulation (see Fig.~\ref{fig:Process_QCISF}). Thus, it does not involve frequent information exchanges as in existing hybrid quantum-classical methods,
% \cite{Liu2022_Application, Pfeffer2022_Hybrid, Lapworth2022_A, Chen2022_Quantum}
which brings a significant advantage in the computational speedup over classical methods and in the reduction of noises introduced by measurements.
%
%In addition, the present quantum algorithm can save memory exponentially due to the state superposition, compared with the classical one.

We estimate the complexity of the quantum algorithm for solving the IHSE. %in Eq.~\eqref{eq:ISF_SchEq}.
The overall quantum circuit contains four steps of the prediction, normalization, phase calculation, and gauge transformation in the algorithm (see Fig.~\ref{fig:ISF_circuit}).
The breakdown of the algorithm complexities is summarized in Table~\ref{tab:QC_Compx}.
The present quantum algorithm can achieve exponential speedup in the steps of prediction and phase calculation, and possible $\Od{(n2^n+\mathrm{poly}(2^n))/\mathrm{poly}(n)}$ speedup overall.

The quantum algorithm is implemented using IBM's Qiskit for a simple 1D flow. The result agrees with the theoretical solution with finite noises, and demonstrates an exponential speedup on a quantum computer.

Note that the HSE without a viscous term and with an external LLF term is different from the NSE, but the SF resembles the viscous flow in terms of the similar flow statistics and structures.
We use the TG vortex and decaying HIT to demonstrate the similarities between the ISF and the viscous flow.  %using the DNS on a classical computer.
The role of the parameter $\hbar$ in the HSE is similar to the kinetic viscosity.
The flow stability depends on the value of $\hbar$ in the TG vortex, and the inertial range with the $-5/3$ scaling broadens with decreasing $\hbar$ in the HIT.
The tangle of vortex tubes and sheets is observed in the turbulent ISF as in the classical turbulent flow.
%
%Meanwhile, there is an obvious difference between the turbulent ISF and practical turbulence, that the ISF appears to lack some of the typical vortex structures formed by the coiling of large-scale lamellar vortices.

With the development of hardware and algorithms for quantum computing, the HSE framework, involving the quantum unitary evolution and characterizing 3D turbulent statistics and structures, can be promising in CFD applications.
In the future work, the bottlenecks for the efficient quantum algorithm will be tackled in the steps of normalization and gauge transformation.
Moreover, the difference between the SF and real flows can be reduced by introducing further modifications and models in the HSE.

\begin{acknowledgments}
The authors thank S. Xiong, Y. Shi, and C. Yang for their helpful discussions. Numerical simulations and visualizations were carried out on the TH-2A supercomputer in Guangzhou, China. This work has been supported in part by the National Natural Science Foundation of China (grant nos~11925201 and 11988102), the National Key R\&D Program of China (No.~2020YFE0204200), and the Xplore Prize.
\end{acknowledgments}

\appendix

\section{Momentum equation for the SF}\label{app:derive_SF}
We derive the momentum equation for the SF. With
\begin{equation}\label{eq:Du/Dt0}
    \frac{\DD\vec{u}}{\DD t}
    %= \frac{\DD}{\DD t}\frac{\vec{J}}{\rho}
    = \frac{1}{\rho}\frac{\DD\vec{J}}{\DD t} + (\bn\cdot\vec{u})\vec{u}
\end{equation}
and $\vec{J}=\hbar\Real[(\bn\widebar{\vec{\psi}})\vec{i\psi}]$, we have
\begin{equation}\label{eq:DjDt1}
    \frac{\DD\vec{J}}{\DD t}
    = \hbar\Real\left[\frac{\DD\bn\widebar{\vec{\psi}}}{\DD t}\vec{i\psi} + \bn\widebar{\vec{\psi}}\vec{i}\frac{\DD\vec{\psi}}{\DD t}\right].
\end{equation}
Substituting the vector identity
\begin{equation}
    \frac{\DD\bn\widebar{\vec{\psi}}}{\DD t}
    %= \bn\frac{\p\widebar{\vec{\psi}}}{\p t} + \bn(\vec{u}\cdot\bn\widebar{\vec{\psi}}) - \bn\vec{u}\cdot\bn\widebar{\vec{\psi}}
    = \bn\frac{\DD\widebar{\vec{\psi}}}{\DD t} - \bn\vec{u}\cdot\bn\widebar{\vec{\psi}}
\end{equation}
into Eq.~\eqref{eq:DjDt1} yields
\begin{equation}\label{eq:DjDt2}
    \frac{\DD\vec{J}}{\DD t}
    = \hbar\Real\left[\bn\frac{\DD\widebar{\vec{\psi}}}{\DD t}\vec{i\psi} + \bn\widebar{\vec{\psi}}\vec{i}\frac{\DD\vec{\psi}}{\DD t}\right] - \rho\bn\frac{|\vec{u}|^2}{2}.
\end{equation}

Next, we derive $\DD\vec{\psi}/\DD t$. The spin vector in Eq.~\eqref{eq:SpinVector} can be expanded as
\begin{equation}\label{eq:SpinVector1}
    \vec{s}
    = (a^2+b^2-c^2-d^2)\vec{i} + 2(bc-ad)\vec{j} + 2(ac+bd)\vec{k},
\end{equation}
which is a pure quaternion and has
\begin{equation}\label{eq:Grad_s}
    \bn\vec{s} = \bn\widebar{\vec{\psi}}\vec{i}\vec{\psi} + \widebar{\vec{\psi}}\vec{i}\bn\vec{\psi}.
\end{equation}
Substituting Eq.~\eqref{eq:Grad_s} into Eq.~\eqref{eq:j1} yields
\begin{equation}\label{eq:Grad_psi0}
    \widebar{\vec{\psi}}\vec{i}\bn\vec{\psi}
    = \frac{1}{2}\bn\vec{s} - \frac{1}{\hbar}\vec{J}.
\end{equation}
%Left multiplying $\vec{i}\vec{\psi}$ to Eq.~\eqref{eq:Grad_psi0} gives
From Eqs.~\eqref{eq:Grad_psi0} and \eqref{eq:IHSE_Vel}, we derive
%\begin{equation}\label{eq:Grad_psi}
%    \bn\vec{\psi}
%    = \vec{i\psi}\left(\frac{\vec{u}}{\hbar} - \frac{1}{2\rho}\bn\vec{s}\right).
%\end{equation}
%Taking the divergence of Eq.~\eqref{eq:Grad_psi} yields
\begin{equation}\label{eq:Lap_psi}
    \nabla^2\vec{\psi}
    = \frac{2\vec{i}}{\hbar}\vec{u}\cdot\bn\vec{\psi} + \frac{1}{\hbar}(\bn\cdot\vec{u})\vec{i\psi} + \frac{1}{\rho}|\bn\vec{\psi}|^2\vec{\psi} + \frac{\vec{i\psi}}{2\rho^2}\bn\rho\cdot\bn\vec{s} - \frac{\vec{i\psi}}{2\rho}\nabla^2\vec{s}.
\end{equation}
%From Eq.~\eqref{eq:Lap_psi},
Then, we obtain the convective term of the wave function
\begin{equation}\label{eq:convect_psi1}
    \vec{u}\cdot\bn\vec{\psi}
    = -\frac{\hbar}{2}\vec{i}\nabla^2\vec{\psi} - \frac12(\bn\cdot\vec{u})\vec{\psi} + \frac{\hbar}{4\rho}\left(2|\bn\vec{\psi}|^2\vec{i\psi} - \frac{\vec{\psi}}{\rho}\bn\rho\cdot\bn\vec{s} + \vec{\psi}\nabla^2\vec{s}\right).
\end{equation}

Combining Eqs.~\eqref{eq:SchEq1} and \eqref{eq:convect_psi1}, we have
\begin{equation}\label{eq:Dpsi/Dt}
    \frac{\DD\vec{\psi}}{\DD t} = - \frac12(\bn\cdot\vec{u})\vec{\psi} + \frac{\hbar}{4\rho}\left(2|\bn\vec{\psi}|^2\vec{i\psi} - \frac{\vec{\psi}}{\rho}\bn\rho\cdot\bn\vec{s} + \vec{\psi}\nabla^2\vec{s}\right) - \frac{V}{\hbar}\vec{i\psi}.
\end{equation}
%Taking the complex conjugate of Eq.~\eqref{eq:Dpsi/Dt} yields
%\begin{equation}\label{eq:Dpsiconj/Dt}
%    \frac{\DD\widebar{\vec{\psi}}}{\DD t}
%    = - \frac{1}{2}(\bn\cdot\vec{u})\widebar{\vec{\psi}}
%    - \frac{\hbar}{4\rho}\left(2|\bn\vec{\psi}|^2\widebar{\vec{\psi}}\vec{i} - \frac{1}{\rho}\bn\rho\cdot\bn\vec{s}\widebar{\vec{\psi}} + (\nabla^2\vec{s})\widebar{\vec{\psi}} \right) + \frac{V}{\hbar}\widebar{\vec{\psi}}\vec{i}.
%\end{equation}
Substituting Eqs.~\eqref{eq:Dpsi/Dt} with its complex conjugate into Eq.~\eqref{eq:DjDt2} yields
\begin{equation}\label{eq:Dj/Dt2}
    \frac{\DD\vec{J}}{\DD t}
    = -\rho(\bn\cdot\vec{u})\vec{u} - \hbar\bn(\vec{\sigma}\cdot\vec{s}) + \hbar\bn\vec{s}\cdot\vec{\sigma} - \rho\bn V + \frac{{\hbar}^2}{2}\rho\bn\frac{|\bn\vec{\psi}|^2}{\rho}
    - \rho\bn\frac{|\vec{u}|^2}{2}
\end{equation}
with $\vec{\sigma}= -\hbar\bn\cdot\left(\bn\vec{s} /\rho\right)/4$.
Substituting Eq.~\eqref{eq:Dj/Dt2} into Eq.~\eqref{eq:Du/Dt0}, we obtain
\begin{equation}\label{eq:DuDt_SF0}
    \frac{\DD\vec{u}}{\DD t}
    = \bn\left(\frac{{\hbar}^2}{8\rho^2}|\bn\vec{s}|^2 - V \right) - \frac{\hbar}{\rho}\bn(\vec{\sigma}\cdot\vec{s}) + \frac{\hbar}{\rho}\bn\vec{s}\cdot\vec{\sigma}.
\end{equation}

Taking the nonlinear potential in Eq.~\eqref{eq:potential} and the equation of state in Eq.~\eqref{eq:p_SF}, and substituting them into Eq.~\eqref{eq:DuDt_SF0}, we obtain the momentum equation~\eqref{eq:Du/Dt_SF} for the SF.
%\begin{equation}\label{eq:p_SF_app}
%    p = -\frac{{\hbar}^2}{4}\vec{s}\cdot\left[\bn\cdot\left(\frac{1}{\rho}\bn\vec{s}\right)\right],
%\end{equation}
%the momentum equation
%~\eqref{eq:DuDt_SF0} finally reads
%\begin{equation}\label{eq:Du/Dt_SF_app}
%    \frac{\DD\vec{u}}{\DD t} = -\frac{1}{\rho}\bn p - \bn V_F
%    - \frac{{\hbar}^2}{4\rho}\bn\vec{s}\cdot\left[\bn\cdot\left(\frac{1}{\rho}\bn\vec{s}\right)\right].
%\end{equation}
%with the potential $V_F$ of conservative body forces (e.g., gravity or buoyancy) which can be arbitrarily set artificially.
In addition, combining Eqs.~\eqref{eq:SpinVector} and \eqref{eq:Dpsi/Dt} gives the transport equation of the spin vector
\begin{equation}
    \frac{\DD\vec{s}}{\DD t}
    = -(\bn\cdot\vec{u})\vec{s} + \frac{\hbar}{2\rho}\vec{s}\times\nabla^2\vec{s}.
\end{equation}

\section{Physical meaning of the LLF}\label{app:LLF}
We discuss the physical meaning of the LLF
\begin{equation}\label{eq:LL-force}
    \vec{F}_{\mathrm{LL}} \equiv -\frac{\hbar^2}{4\rho_0^2}\bn\vec{s}\cdot\nabla^2\vec{s}
\end{equation}
in the momentum equation~\eqref{eq:Du/Dt_ISF1} for the ISF.
Without loss of generality, we set $\rho_0=1$ here.
The transport equation of the spin vector in Eq.~\eqref{eq:SpinVector} reads
\begin{equation}\label{eq:ps/pt_ISF}
    \frac{\p\vec{s}}{\p t}
    = \frac{\hbar}{2}\vec{s}\times\nabla^2 \vec{s} - \hbar\vec{s}\times\vec{m},
\end{equation}
where $\vec{m} \equiv \bn\widebar{\vec{\psi}}\cdot\vec{i}\bn\vec{\psi}$ is a pure quaternion and can be expanded as
\begin{equation}
    \vec{m}
    = (|\bn a|^2 + |\bn b|^2 - |\bn c|^2 - |\bn d|^2)\vec{i}
    + 2(\bn b\cdot\bn c - \bn a\cdot\bn d)\vec{j}
    + 2(\bn a\cdot\bn c + \bn b\cdot\bn d)\vec{k}.
\end{equation}
Note that Eq.~\eqref{eq:ps/pt_ISF} is very similar to the Landau--Lifshitz--Gilbert equation \cite{Gilbert1955_A}, which is a quasi-linear equation describing the evolution of the magnetization vector in ferromagnetic materials \cite{Landau1935_On}.
%Therefore, the quantum system governed by the IHSE~\eqref{eq:ISF_SchEq} may share some properties of ferromagnetic crystals.

Then, we analyze the properties of the LLF in Eq.~\eqref{eq:LL-force}.
Take
\begin{equation}\label{eq:App_psi}
    \vec{\psi} = \cos\theta\cos\frac{\phi_1}{\hbar}
    + \vec{i}\cos\theta\sin\frac{\phi_1}{\hbar}
    + \vec{j}\sin\theta\cos\frac{\phi_2}{\hbar}
    + \vec{k}\sin\theta\sin\frac{\phi_2}{\hbar}
\end{equation}
with real-valued functions $\theta=\theta(x,y,z)$ and $\phi_\alpha=\phi_\alpha(x,y,z),~\alpha=1,2$, which satisfies $|\vec{\psi}|^2=1$.
The corresponding velocity and vorticity can be expressed as
\begin{equation}\label{eq:VelVor_SpCleb}
    \vec{u} = \cos^2\theta\bn(\dphi) + \bn\phi_2, \quad
    \vec{\omega} = \frac{1}{2}\bn(\cos2\theta)\times\bn(\dphi)
\end{equation}
with $\dphi\equiv\phi_1 - \phi_2$.
The spin vector reads
\begin{equation}\label{eq:S_1}
    s_1 = \cos2\theta, \quad
    s_2 = \sin 2\theta\sin\frac{\Delta\phi}{\hbar}, \quad
    s_3 = \sin 2\theta\cos\frac{\Delta\phi}{\hbar}.
\end{equation}
Moreover, the incompressibility condition imposes a constraint
\begin{equation}
    \bn(\cos^2\theta)\cdot\bn(\dphi) + \cos^2\theta\nabla^2(\dphi) + \nabla^2\phi_2
    = 0.
\end{equation}

From the gradient and Laplacian of Eq.~\eqref{eq:S_1},
%\begin{equation}\label{eq:Grad_APP}
%    \begin{split}
%        \bn s_1
%        =\ & -2\sin2\theta\bn\theta, \\
%        \bn s_2
%        =\ & 2\cos2\theta\sin\frac{\dphi}{\hbar}\bn\theta + \frac{1}{\hbar}\sin2\theta\cos\frac{\dphi}{\hbar}\bn(\dphi), \\
%        \bn s_3
%        =\ & 2\cos2\theta\cos\frac{\dphi}{\hbar}\bn\theta - \frac{1}{\hbar}\sin2\theta\sin\frac{\dphi}{\hbar}\bn(\dphi).
%    \end{split}
%\end{equation}
%We then take the divergence of the above three equations, which gives
%\begin{equation}\label{eq:Lap_APP}
%    \begin{split}
%        \nabla^2 s_1
%        =\ & -2\sin2\theta\nabla^2\theta - 4\cos2\theta|\bn\theta|^2, \\
%        \nabla^2 s_2
%        =\ & 2\cos2\theta\sin\frac{\dphi}{\hbar}\nabla^2\theta - 4\sin2\theta\sin\frac{\dphi}{\hbar}|\bn\theta|^2 + \frac{4}{\hbar}\cos2\theta\cos\frac{\dphi}{\hbar}\bn(\dphi)\cdot\bn\theta
%        \\
%        &+ \frac{1}{\hbar}\sin2\theta\cos\frac{\dphi}{\hbar}\nabla^2(\dphi) - \frac{1}{{\hbar}^2}\sin2\theta\sin\frac{\dphi}{\hbar}|\bn(\dphi)|^2,
%        \\
%        \nabla^2 s_3
%        =\ & 2\cos2\theta\cos\frac{\dphi}{\hbar}\nabla^2\theta - 4\sin2\theta\cos\frac{\dphi}{\hbar}|\bn\theta|^2 - \frac{4}{\hbar}\cos2\theta\sin\frac{\dphi}{\hbar}\bn(\dphi)\cdot\bn\theta
%        \\
%        &- \frac{1}{\hbar}\sin2\theta\sin\frac{\dphi}{\hbar}\nabla^2(\dphi) - \frac{1}{{\hbar}^2}\sin2\theta\cos\frac{\dphi}{\hbar}|\bn(\dphi)|^2.
%    \end{split}
%\end{equation}
%Combining Eqs.~\eqref{eq:Grad_APP} and \eqref{eq:Lap_APP},
we obtain
\begin{equation}\label{eq:F_LL-APP}
    \begin{split}
        \vec{F}_{\mathrm{LL}}
        =\ & \left[ \frac{1}{2}\sin2\theta\cos2\theta|\bn(\dphi)|^2 - {\hbar}^2\nabla^2\theta \right]\bn\theta
        \\
        &- \left[\sin2\theta\cos2\theta\bn(\dphi)\cdot\bn\theta + \frac14\sin^22\theta\nabla^2(\dphi)\right]\bn(\dphi)
    \end{split}
\end{equation}
after some algebra.
%
%Note that three terms on the RHS of Eq.~\eqref{eq:F_LL-APP} are independent of $\hbar$, and only one is proportional to $\hbar^2$.
%When $\theta$ is a harmonic function, i.e., $\nabla^2\theta=0$, $\vec{F}_{\mathrm{LL}}$ is completely independent of $\hbar$.
%Hence, the LLF is not proportional to $\hbar^2$ as shown by the form in Eq.~\eqref{eq:LL-force}.
%Based on the dependence of $\hbar$,
We decompose $\vec{F}_{\mathrm{LL}}=\vec{F}_{E}+\vec{F}_D$, where
\begin{equation}
    \vec{F}_E
    \equiv \frac{1}{2}\sin2\theta\cos2\theta|\bn(\dphi)|^2\bn\theta
    - \left[\sin2\theta\cos2\theta\bn(\dphi)\cdot\bn\theta + \frac14\sin^22\theta\nabla^2(\dphi)\right]\bn(\dphi),
\end{equation}
independent on $\hbar$, behaves as an external body force to stir the flow, and
\begin{equation}
    \vec{F}_D \equiv -\hbar^2\nabla^2\theta\bn\theta
\end{equation}
is similar to a viscous term to dissipate the flow with an effective viscosity correlated to $\hbar^2$.

\section{Wave function for the initial TG field}\label{app:TG}
We convert the TG initial condition \cite{Taylor1937_Mechanism}
\begin{equation}\label{eq:TG_vel}
    \vec{u} = (\sin x\cos y\cos z, -\cos x\sin y\cos z, 0)
\end{equation}
into the form of the wave function.
For the wave function in Eq.~\eqref{eq:App_psi},
%with the corresponding velocity and vorticity~\eqref{eq:VelVor_SpCleb} and spin vector~\eqref{eq:S_1},
we determine the real-valued functions $\theta$ and $\phi_\alpha$ for the TG initial condition.
Following the form of the Clebsch potentials \cite{Clebsch1859_Ueber} for the TG initial condition \cite{Nore1997_Decaying},  we let
\begin{equation}\label{eq:Let1_TG}
    \theta = \frac{1}{2}\arccos\left(\cos x\,|\!\cos z|^{1/2} \right), \quad
    \dphi = 4\cos y\,|\!\cos z|^{1/2}\,\mathrm{sgn}(\cos z).
\end{equation}
Substituting Eqs.~\eqref{eq:TG_vel} and \eqref{eq:Let1_TG} into Eq.~\eqref{eq:VelVor_SpCleb} yields
\begin{equation}
    \phi_2 = -\cos x\cos y\cos z - 2\cos y\,|\!\cos z|^{1/2}\,\mathrm{sgn}(\cos z).
\end{equation}

\begin{figure}
    \centering
    \begin{overpic}[width=0.6\linewidth]{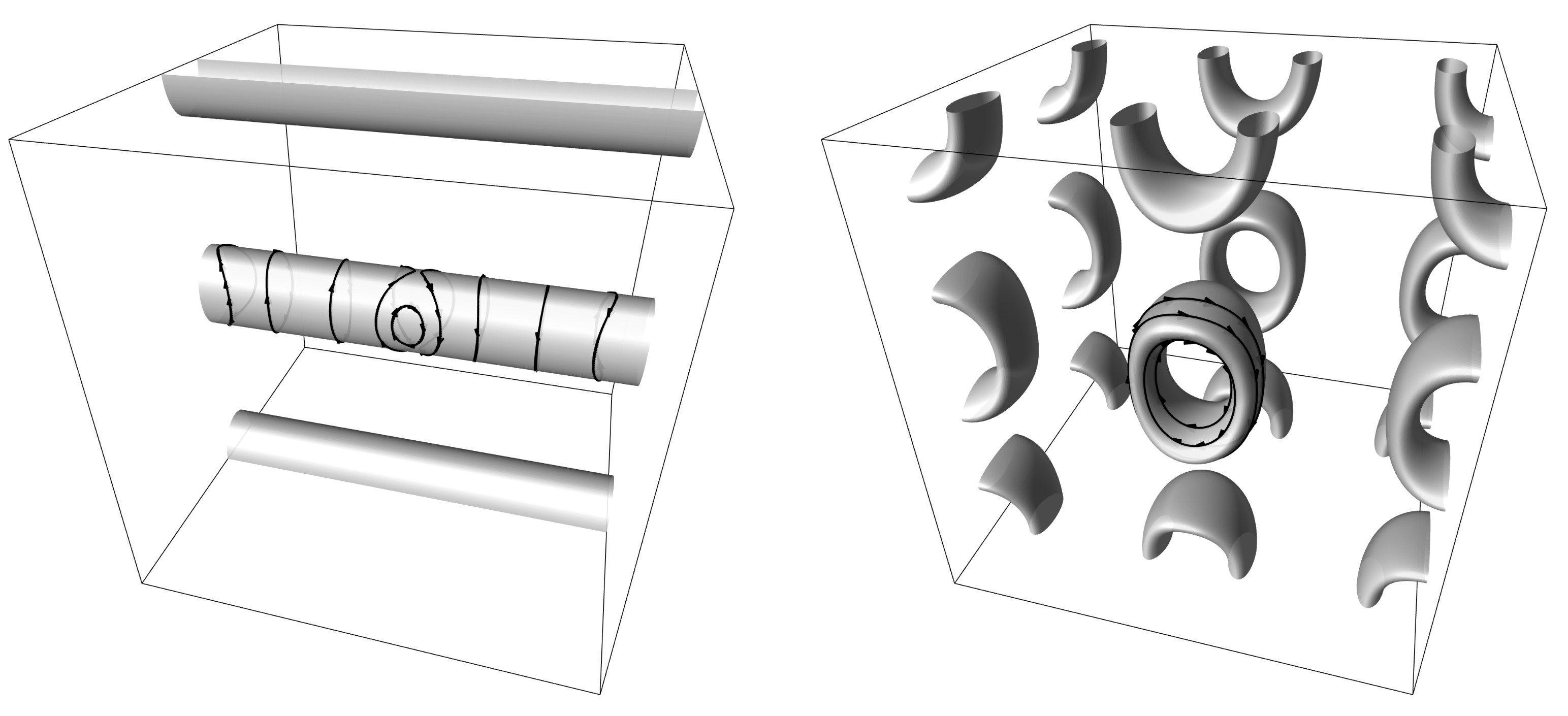}
        \begin{small}
            \put(0,42) {(a)}
            \put(52,42) {(b)}
        \end{small}
    \end{overpic}
    \caption{Isosurfaces of (a) $s_1=-0.9$ and (b) $s_3=-0.9$ in the TG initial field. Some vortex lines are integrated and plotted on the isosurfaces.}
    \label{fig:TG_VortexSurface}
\end{figure}

Thus, we obtain the wave function
\begin{equation}\label{eq:TG_WF}
    \begin{dcases}
    a = \cos\left(\frac{1}{2}\arccos\left(\cos x\,|\!\cos z|^{1/2} \right) \right)\cos\left(\frac{\cos y\,[2\,|\!\cos z|^{1/2}\,\mathrm{sgn}(\cos z) - \cos x\cos z]}{\hbar} \right), \\
    b = \cos\left(\frac{1}{2}\arccos\left(\cos x\,|\!\cos z|^{1/2} \right) \right)\sin\left(\frac{\cos y\,[2\,|\!\cos z|^{1/2}\,\mathrm{sgn}(\cos z) - \cos x\cos z]}{\hbar} \right), \\
    c = \sin\left(\frac{1}{2}\arccos\left(\cos x\,|\!\cos z|^{1/2} \right) \right)\cos\left(\frac{\cos y\,[2\,|\!\cos z|^{1/2}\,\mathrm{sgn}(\cos z) + \cos x\cos z]}{\hbar} \right), \\
    d = -\sin\left(\frac{1}{2}\arccos\left(\cos x\,|\!\cos z|^{1/2} \right) \right)\sin\left(\frac{\cos y\,[2\,|\!\cos z|^{1/2}\,\mathrm{sgn}(\cos z) + \cos x\cos z]}{\hbar} \right)
    \end{dcases}
\end{equation}
and the spin vector
\begin{equation}\label{eq:TG_SV}
    \begin{dcases}
    s_1 = \cos x\,|\!\cos z|^{1/2}, \\
    s_2 = \sin\left(\arccos\left(\cos x\,|\!\cos z|^{1/2} \right) \right)\sin\left(\frac{4\cos y\,|\!\cos z|^{1/2}\,\mathrm{sgn}(\cos z)}{\hbar} \right), \\
    s_3 = \sin\left(\arccos\left(\cos x\,|\!\cos z|^{1/2} \right) \right)\cos\left(\frac{4\cos y\,|\!\cos z|^{1/2}\,\mathrm{sgn}(\cos z)}{\hbar} \right).
    \end{dcases}
\end{equation}
for the TG initial field.
Typical vortex surfaces consisting of ring-like vortex lines for the TG initial condition are obtained from Eq.~\eqref{eq:TG_SV} and plotted in Fig.~\ref{fig:TG_VortexSurface}.

\section{Hybrid simulation of the 2D TG ISF}\label{app:hybrid_2DTG} %quantum-classical

We perform a hybrid quantum-classical simulation of a 2D unsteady TG ISF to demonstrate the capability of solving the IHSE~\eqref{eq:ISF_SchEq} with the quantum simulator Qiskit~\cite{Qiskit}.
The obstacles (steps 2--4) in the present quantum algorithm are tentatively treated on the classical computer.
According to Appendix~\ref{app:TG}, the wave function for the 2D TG initial condition $\boldsymbol u=(\sin x\cos y,-\cos x\sin y)$ is
\begin{equation}
    \begin{dcases}
        a = \cos\left(H(x) \right)\cos\frac{\cos y(2 - \cos x)}{\hbar}, \\
        b = \cos\left(H(x) \right)\sin\frac{\cos y(2 - \cos x)}{\hbar}, \\
        c = \sin\left(H(x) \right)\cos\frac{\cos y(2 + \cos x)}{\hbar}, \\
        d = -\sin\left(H(x) \right)\sin\frac{\cos y(2 + \cos x)}{\hbar}.
    \end{dcases}
    ~\text{with}~
    H(x) = \begin{dcases}
    \frac{x}{2}, & 0\le x\le \pi, \\
    \pi - \frac{x}{2}, & \pi<x\le 2\pi.
    \end{dcases}
\end{equation}

The hybrid quantum-classical simulation is governed by Eq.~\eqref{eq:ISF_SchEq} with $\rho_0=1$.
In each time step, we first implement step 1 in the algorithm in Section~\ref{sec:QA_IHSE} using $n=10$ qubits, where $n$ depends on the computational resource.
We set five qubits in each of the $x$- and $y$-directions, equivalent to using $2^5\times 2^5=32^2$ grid points.
Then, we measure the state vector $\ket{\psi^*(t)}$ in Eq.~\eqref{eq:psi_star} to obtain the corresponding $a^*$, $b^*$, $c^*$, and $d^*$.
Note that the algorithm for the 2D or higher-dimensional problem is essentially the same as that for the 1D problem in Section~\ref{sec:QA_ISF}.
Finally, we implement steps 2--4 in Section~\ref{sec:QA_IHSE} using $32^2$ grid points using the methods for the classical computer, and then prepare the state vector corresponding to the final output wave function.
The above process is iterated for time marching.
Since step 1 dominates the computational complexity in the classical simulation (see Table~\ref{tab:QC_Compx}), the hybrid simulation itself can be a temporal method for simulating 3D ISFs.
%if the efficiency of complex quantum state preparation and measurement can be addressed.

\begin{figure}
    \centering
    \begin{overpic}[width=\textwidth]{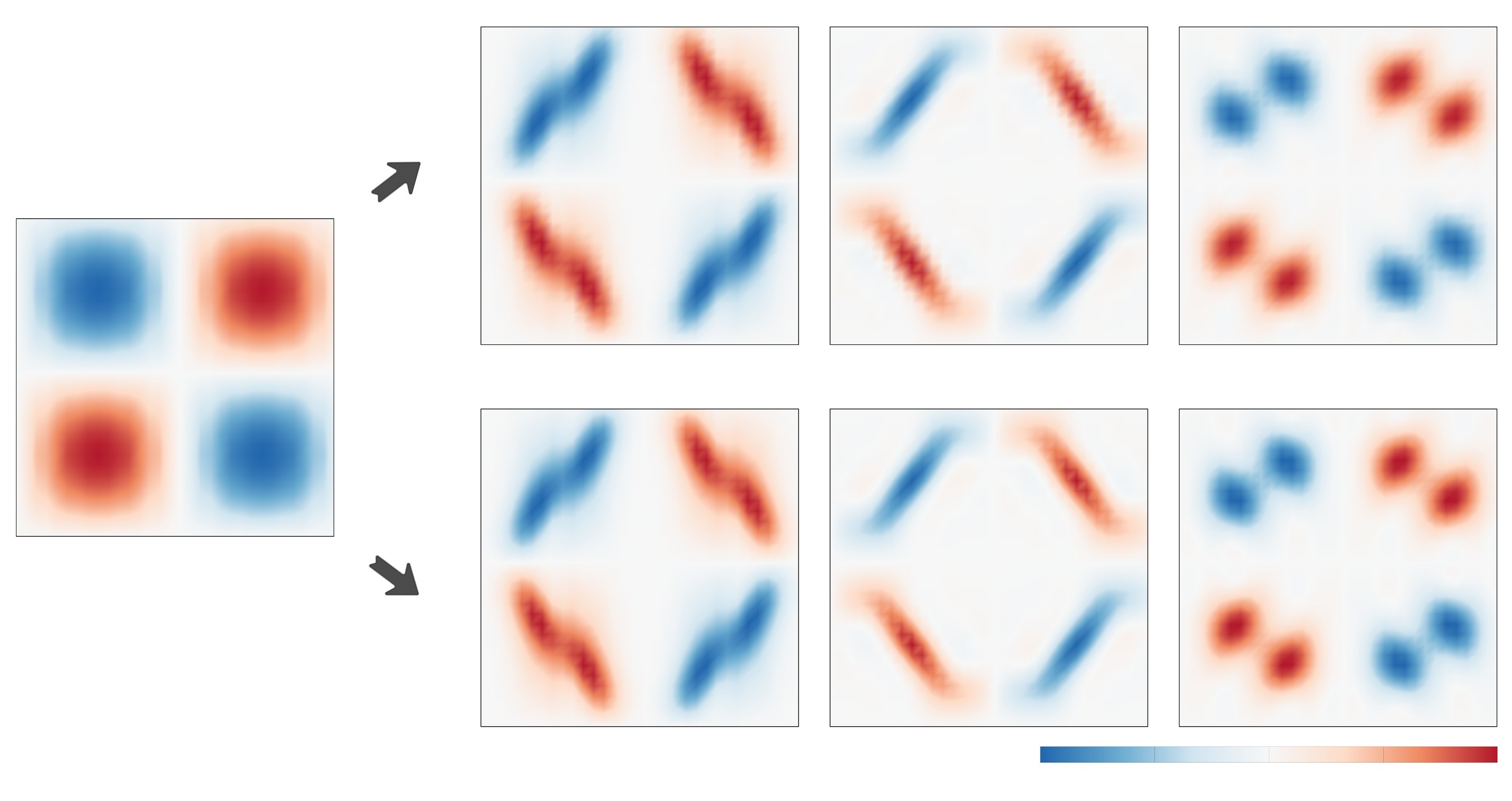}
        \begin{small}
            \put(1,39.3) {(a)}
            \put(9,39.3) {$t=0$}
            \put(27,49.5) {(b)}
            \put(39,52) {$t=0.2$}
            \put(62,52) {$t=0.5$}
            \put(86.3,52) {$t=1$}
            \put(27,24.3) {(c)}
            \put(39,26.8) {$t=0.2$}
            \put(62,26.8) {$t=0.5$}
            \put(86.3,26.8) {$t=1$}
            \put(22,43) {hybrid}
            \put(22,11.5) {classical}
            % legend:
            \begin{footnotesize}
                \put(58.5,2.6) {$\omega/|\omega|_{\max}$}
                \put(67.5,0.7) {$-1$}
                \put(74,0.7) {$-0.5$}
                \put(83.4,0.7) {$0$}
                \put(90.2,0.7) {$0.5$}
                \put(98.5,0.7) {$1$}
            \end{footnotesize}
        \end{small}
    \end{overpic}
    \caption{Comparison of the instantaneous contours of the normalized vorticity in the hybrid quantum-classical simulation and classical simulation of the 2D TG ISF with $\hbar=1$ and $n=10$ qubits.}
    \label{fig:2D-TG_vorz-Contour}
\end{figure}

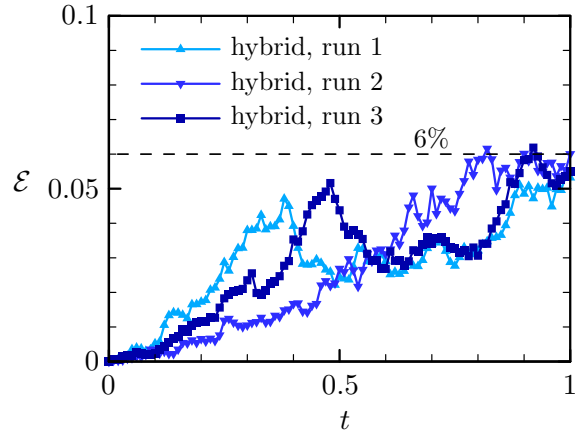
\begin{figure}
    \centering
    \begin{overpic}[width=0.4\textwidth]{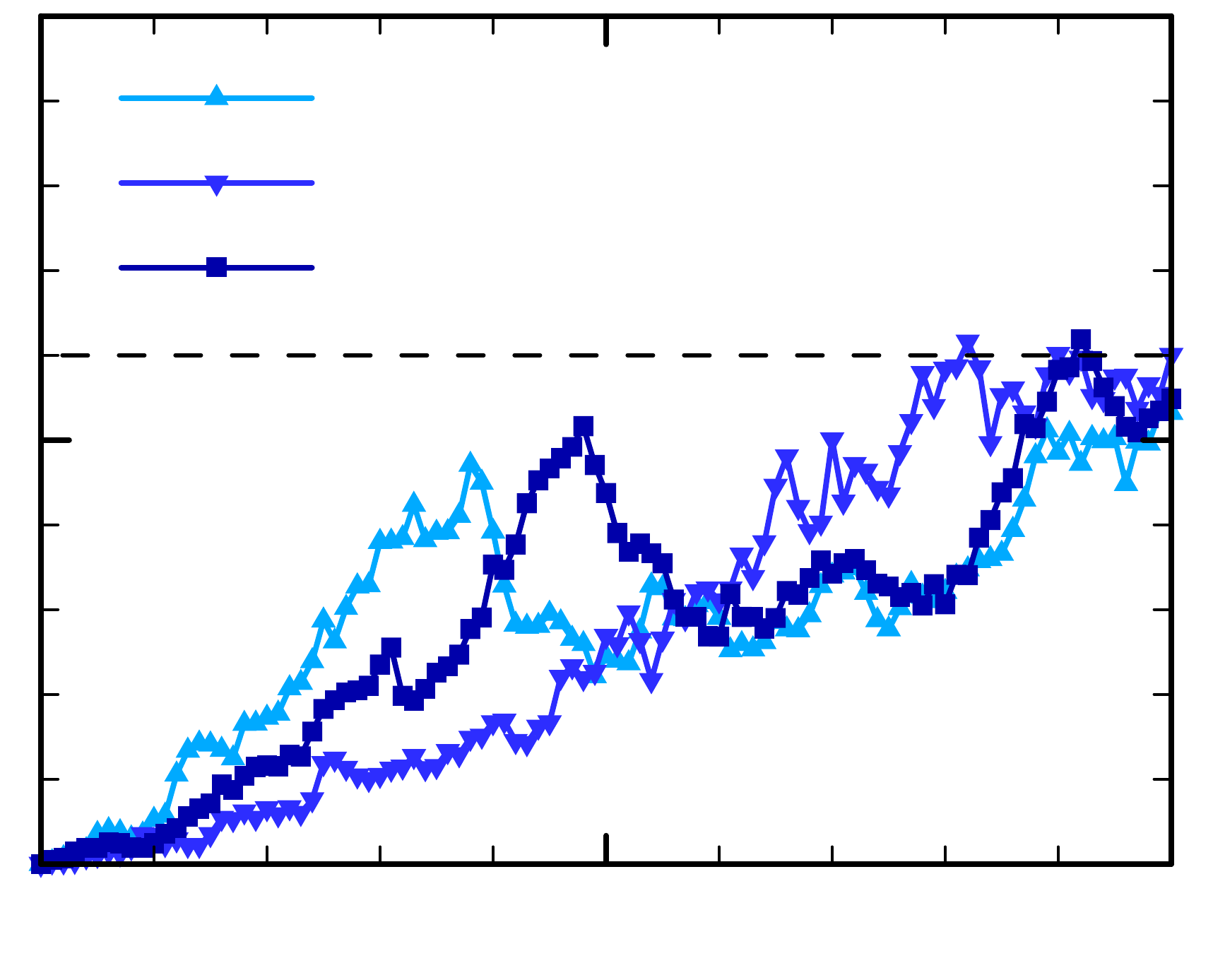}
        \begin{small}
            % x axis:
            \put(1.8,3) {$0$}
            \put(46.4,3) {$0.5$}
            \put(95.1,3) {$1$}
            \put(50,-4) {$t$}
            % y axis:
            \put(-1.5,7.8) {$0$}
            \put(-9,42.7) {$0.05$}
            \put(-6.2,77.8) {$0.1$}
            \put(-16,44) {$\mathcal{E}$}
            % legend:
            \begin{footnotesize}
                \put(28,71.5) {hybrid, run 1}
                \put(28,64.5) {hybrid, run 2}
                \put(28,57.5) {hybrid, run 3}
                \put(65,53) {$6\%$}
            \end{footnotesize}
        \end{small}
    \end{overpic}
    \caption{Discrepancies $\mathcal{E}(t)\equiv  \||\omega|_{\mathrm{hybrid}} - |\omega|_{\mathrm{classical}}\|_2^2$ of three independent hybrid quantum-classical runs from the classical simulation result of the 2D TG ISF with $\hbar=1$ and $n=10$ qubits.}
    %The error is defined as $\mathcal{E}(t)\equiv  \|\omega_z^{\mathrm{hybrid}}(t) - \omega_z^{\mathrm{classical}}(t)\|_2^2$, which is smaller than $6\%$ when $t<1$ using $n=10$ qubits.
    \label{fig:error}
\end{figure}

Figure~\ref{fig:2D-TG_vorz-Contour} shows evolution of the contour of $\omega/|\omega|_{\max}$ in the 2D TG ISF with $\hbar=1$, where $\omega$ denotes the $z$-component of $\boldsymbol\omega$ and $|\omega|_{\max}$ denotes the instantaneous maximum $|\omega|$ in the computational domain.
The hybrid simulation result shows a good agreement with the classical one, and only has slight oscillations due to the statistical error and noises introduced by the QASM simulator.
%, slight oscillation seems to be more obvious in the hybrid quantum-classical case.
The discrepancy $\mathcal{E}(t)\equiv  \||\omega|_{\mathrm{hybrid}} - |\omega|_{\mathrm{classical}}\|_2^2$ between the two simulations is smaller than $6\%$ at $t<1$ for three independent runs in Fig.~\ref{fig:error}.
In addition, the total kinetic energy and enstrophy are almost identical for the two simulations (not shown). %$E\equiv \int_{\mathcal{L}^2}|\vec{u}|^2/2\dif A$ $\varOmega\equiv \int_{\mathcal{L}^2}|\vec{\omega}|^2/2\dif A$
%In addition, we perform three hybrid quantum-classical simulations independently and find that the statistical error is quite notable.
%This may be due to the fact that the reconstruction of the velocity field through the wave function requires derivative operations, which amplifies the statistical errors and the noises.

\begin{thebibliography}{120}%
\makeatletter
\providecommand \@ifxundefined [1]{%
 \@ifx{#1\undefined}
}%
\providecommand \@ifnum [1]{%
 \ifnum #1\expandafter \@firstoftwo
 \else \expandafter \@secondoftwo
 \fi
}%
\providecommand \@ifx [1]{%
 \ifx #1\expandafter \@firstoftwo
 \else \expandafter \@secondoftwo
 \fi
}%
\providecommand \natexlab [1]{#1}%
\providecommand \enquote  [1]{``#1''}%
\providecommand \bibnamefont  [1]{#1}%
\providecommand \bibfnamefont [1]{#1}%
\providecommand \citenamefont [1]{#1}%
\providecommand \href@noop [0]{\@secondoftwo}%
\providecommand \href [0]{\begingroup \@sanitize@url \@href}%
\providecommand \@href[1]{\@@startlink{#1}\@@href}%
\providecommand \@@href[1]{\endgroup#1\@@endlink}%
\providecommand \@sanitize@url [0]{\catcode `\\12\catcode `\$12\catcode
  `\&12\catcode `\#12\catcode `\^12\catcode `\_12\catcode `\%12\relax}%
\providecommand \@@startlink[1]{}%
\providecommand \@@endlink[0]{}%
\providecommand \url  [0]{\begingroup\@sanitize@url \@url }%
\providecommand \@url [1]{\endgroup\@href {#1}{\urlprefix }}%
\providecommand \urlprefix  [0]{URL }%
\providecommand \Eprint [0]{\href }%
\providecommand \doibase [0]{https://doi.org/}%
\providecommand \selectlanguage [0]{\@gobble}%
\providecommand \bibinfo  [0]{\@secondoftwo}%
\providecommand \bibfield  [0]{\@secondoftwo}%
\providecommand \translation [1]{[#1]}%
\providecommand \BibitemOpen [0]{}%
\providecommand \bibitemStop [0]{}%
\providecommand \bibitemNoStop [0]{.\EOS\space}%
\providecommand \EOS [0]{\spacefactor3000\relax}%
\providecommand \BibitemShut  [1]{\csname bibitem#1\endcsname}%
\let\auto@bib@innerbib\@empty
%</preamble>
\bibitem [{\citenamefont {Feynman}(1982)}]{Feynman1982_Simulating}%
  \BibitemOpen
  \bibfield  {author} {\bibinfo {author} {\bibfnamefont {R.~P.}\ \bibnamefont
  {Feynman}},\ }\bibfield  {title} {\bibinfo {title} {Simulating physics with
  computers},\ }\href@noop {} {\bibfield  {journal} {\bibinfo  {journal} {Int.
  J. Theor. Phys.}\ }\textbf {\bibinfo {volume} {21}},\ \bibinfo {pages} {467}
  (\bibinfo {year} {1982})}\BibitemShut {NoStop}%
\bibitem [{\citenamefont {Nielsen}\ and\ \citenamefont
  {Chuang}(2010)}]{Nielsen2010_Quantum}%
  \BibitemOpen
  \bibfield  {author} {\bibinfo {author} {\bibfnamefont {M.~A.}\ \bibnamefont
  {Nielsen}}\ and\ \bibinfo {author} {\bibfnamefont {I.~L.}\ \bibnamefont
  {Chuang}},\ }\href@noop {} {\emph {\bibinfo {title} {Quantum Computation and
  Quantum Information}}},\ \bibinfo {edition} {10th}\ ed.\ (\bibinfo
  {publisher} {Cambridge University Press},\ \bibinfo {address} {New York},\
  \bibinfo {year} {2010})\BibitemShut {NoStop}%
\bibitem [{\citenamefont {Sleator}\ and\ \citenamefont
  {Weinfurter}(1995)}]{Sleator1995_Realizable}%
  \BibitemOpen
  \bibfield  {author} {\bibinfo {author} {\bibfnamefont {T.}~\bibnamefont
  {Sleator}}\ and\ \bibinfo {author} {\bibfnamefont {H.}~\bibnamefont
  {Weinfurter}},\ }\bibfield  {title} {\bibinfo {title} {Realizable universal
  quantum logic gates},\ }\href@noop {} {\bibfield  {journal} {\bibinfo
  {journal} {Phys. Rev. Lett.}\ }\textbf {\bibinfo {volume} {74}},\ \bibinfo
  {pages} {4087} (\bibinfo {year} {1995})}\BibitemShut {NoStop}%
\bibitem [{\citenamefont {Makhlin}\ \emph {et~al.}(2001)\citenamefont
  {Makhlin}, \citenamefont {Sch\"on},\ and\ \citenamefont
  {Shnirman}}]{Makhlin2001_Quantum}%
  \BibitemOpen
  \bibfield  {author} {\bibinfo {author} {\bibfnamefont {Y.}~\bibnamefont
  {Makhlin}}, \bibinfo {author} {\bibfnamefont {G.}~\bibnamefont {Sch\"on}},\
  and\ \bibinfo {author} {\bibfnamefont {A.}~\bibnamefont {Shnirman}},\
  }\bibfield  {title} {\bibinfo {title} {{Quantum-state engineering with
  Josephson-junction devices}},\ }\href@noop {} {\bibfield  {journal} {\bibinfo
   {journal} {Rev. Mod. Phys.}\ }\textbf {\bibinfo {volume} {73}},\ \bibinfo
  {pages} {357} (\bibinfo {year} {2001})}\BibitemShut {NoStop}%
\bibitem [{\citenamefont {Kok}\ \emph {et~al.}(2007)\citenamefont {Kok},
  \citenamefont {Munro}, \citenamefont {Nemoto}, \citenamefont {Ralph},
  \citenamefont {Dowling},\ and\ \citenamefont {Milburn}}]{Kok2007_Linear}%
  \BibitemOpen
  \bibfield  {author} {\bibinfo {author} {\bibfnamefont {P.}~\bibnamefont
  {Kok}}, \bibinfo {author} {\bibfnamefont {W.~J.}\ \bibnamefont {Munro}},
  \bibinfo {author} {\bibfnamefont {K.}~\bibnamefont {Nemoto}}, \bibinfo
  {author} {\bibfnamefont {T.~C.}\ \bibnamefont {Ralph}}, \bibinfo {author}
  {\bibfnamefont {J.~P.}\ \bibnamefont {Dowling}},\ and\ \bibinfo {author}
  {\bibfnamefont {G.~J.}\ \bibnamefont {Milburn}},\ }\bibfield  {title}
  {\bibinfo {title} {Linear optical quantum computing with photonic qubits},\
  }\href@noop {} {\bibfield  {journal} {\bibinfo  {journal} {Rev. Mod. Phys.}\
  }\textbf {\bibinfo {volume} {79}},\ \bibinfo {pages} {135} (\bibinfo {year}
  {2007})}\BibitemShut {NoStop}%
\bibitem [{\citenamefont {Saffman}\ \emph {et~al.}(2010)\citenamefont
  {Saffman}, \citenamefont {Walker},\ and\ \citenamefont
  {M\o{}lmer}}]{Saffman2010_Quantum}%
  \BibitemOpen
  \bibfield  {author} {\bibinfo {author} {\bibfnamefont {M.}~\bibnamefont
  {Saffman}}, \bibinfo {author} {\bibfnamefont {T.~G.}\ \bibnamefont
  {Walker}},\ and\ \bibinfo {author} {\bibfnamefont {K.}~\bibnamefont
  {M\o{}lmer}},\ }\bibfield  {title} {\bibinfo {title} {{Quantum information
  with Rydberg atoms}},\ }\href@noop {} {\bibfield  {journal} {\bibinfo
  {journal} {Rev. Mod. Phys.}\ }\textbf {\bibinfo {volume} {82}},\ \bibinfo
  {pages} {2313} (\bibinfo {year} {2010})}\BibitemShut {NoStop}%
\bibitem [{\citenamefont {Leibfried}\ \emph {et~al.}(2011)\citenamefont
  {Leibfried}, \citenamefont {Ospelkaus}, \citenamefont {Warring},
  \citenamefont {Colombe}, \citenamefont {Brown}, \citenamefont {Amini},\ and\
  \citenamefont {Wineland}}]{Leibfried2011_Microwave}%
  \BibitemOpen
  \bibfield  {author} {\bibinfo {author} {\bibfnamefont {D.}~\bibnamefont
  {Leibfried}}, \bibinfo {author} {\bibfnamefont {C.}~\bibnamefont
  {Ospelkaus}}, \bibinfo {author} {\bibfnamefont {U.}~\bibnamefont {Warring}},
  \bibinfo {author} {\bibfnamefont {Y.}~\bibnamefont {Colombe}}, \bibinfo
  {author} {\bibfnamefont {K.}~\bibnamefont {Brown}}, \bibinfo {author}
  {\bibfnamefont {J.}~\bibnamefont {Amini}},\ and\ \bibinfo {author}
  {\bibfnamefont {D.}~\bibnamefont {Wineland}},\ }\bibfield  {title} {\bibinfo
  {title} {Microwave quantum logic gates for trapped ions},\ }\href@noop {}
  {\bibfield  {journal} {\bibinfo  {journal} {Nature}\ }\textbf {\bibinfo
  {volume} {476}},\ \bibinfo {pages} {181} (\bibinfo {year}
  {2011})}\BibitemShut {NoStop}%
\bibitem [{\citenamefont {Zhang}\ \emph {et~al.}(2021)\citenamefont {Zhang},
  \citenamefont {Feng}, \citenamefont {Li}, \citenamefont {Chen}, \citenamefont
  {Zhang}, \citenamefont {He}, \citenamefont {Guo}, \citenamefont {Guo},
  \citenamefont {Ren},\ and\ \citenamefont {Dai}}]{Zhang2021_Supercompact}%
  \BibitemOpen
  \bibfield  {author} {\bibinfo {author} {\bibfnamefont {M.}~\bibnamefont
  {Zhang}}, \bibinfo {author} {\bibfnamefont {L.}~\bibnamefont {Feng}},
  \bibinfo {author} {\bibfnamefont {M.}~\bibnamefont {Li}}, \bibinfo {author}
  {\bibfnamefont {Y.}~\bibnamefont {Chen}}, \bibinfo {author} {\bibfnamefont
  {L.}~\bibnamefont {Zhang}}, \bibinfo {author} {\bibfnamefont
  {D.}~\bibnamefont {He}}, \bibinfo {author} {\bibfnamefont {G.}~\bibnamefont
  {Guo}}, \bibinfo {author} {\bibfnamefont {G.}~\bibnamefont {Guo}}, \bibinfo
  {author} {\bibfnamefont {X.}~\bibnamefont {Ren}},\ and\ \bibinfo {author}
  {\bibfnamefont {D.}~\bibnamefont {Dai}},\ }\bibfield  {title} {\bibinfo
  {title} {Supercompact photonic quantum logic gate on a silicon chip},\
  }\href@noop {} {\bibfield  {journal} {\bibinfo  {journal} {Phys. Rev. Lett.}\
  }\textbf {\bibinfo {volume} {126}},\ \bibinfo {pages} {130501} (\bibinfo
  {year} {2021})}\BibitemShut {NoStop}%
\bibitem [{\citenamefont {Shor}(1994)}]{Shor1994_Algorithms}%
  \BibitemOpen
  \bibfield  {author} {\bibinfo {author} {\bibfnamefont {P.~W.}\ \bibnamefont
  {Shor}},\ }\bibfield  {title} {\bibinfo {title} {Algorithms for quantum
  computation: discrete logarithms and factoring},\ }in\ \href@noop {} {\emph
  {\bibinfo {booktitle} {Proceedings 35th Annual Symposium on Foundations of
  Computer Science}}}\ (\bibinfo {year} {1994})\ pp.\ \bibinfo {pages}
  {124--134}\BibitemShut {NoStop}%
\bibitem [{\citenamefont {Ekert}\ and\ \citenamefont
  {Jozsa}(1996)}]{Ekert1996_Quantum}%
  \BibitemOpen
  \bibfield  {author} {\bibinfo {author} {\bibfnamefont {A.}~\bibnamefont
  {Ekert}}\ and\ \bibinfo {author} {\bibfnamefont {R.}~\bibnamefont {Jozsa}},\
  }\bibfield  {title} {\bibinfo {title} {{Quantum computation and Shor's
  factoring algorithm}},\ }\href@noop {} {\bibfield  {journal} {\bibinfo
  {journal} {Rev. Mod. Phys.}\ }\textbf {\bibinfo {volume} {68}},\ \bibinfo
  {pages} {733} (\bibinfo {year} {1996})}\BibitemShut {NoStop}%
\bibitem [{\citenamefont {Shor}(1997)}]{Shor1997_polynomial}%
  \BibitemOpen
  \bibfield  {author} {\bibinfo {author} {\bibfnamefont {P.~W.}\ \bibnamefont
  {Shor}},\ }\bibfield  {title} {\bibinfo {title} {{Polynomial-time algorithms
  for prime factorization and discrete logarithms on a quantum computer}},\
  }\href@noop {} {\bibfield  {journal} {\bibinfo  {journal} {SIAM J. Comput.}\
  }\textbf {\bibinfo {volume} {26}},\ \bibinfo {pages} {1484} (\bibinfo {year}
  {1997})}\BibitemShut {NoStop}%
\bibitem [{\citenamefont {Grover}(1996)}]{Grover1996_A}%
  \BibitemOpen
  \bibfield  {author} {\bibinfo {author} {\bibfnamefont {L.~K.}\ \bibnamefont
  {Grover}},\ }\bibfield  {title} {\bibinfo {title} {A fast quantum mechanical
  algorithm for database search},\ }in\ \href@noop {} {\emph {\bibinfo
  {booktitle} {Proceedings of the 28th Annual ACM Symposium on Theory of
  Computing}}}\ (\bibinfo  {publisher} {ACM},\ \bibinfo {year} {1996})\ pp.\
  \bibinfo {pages} {212--219}\BibitemShut {NoStop}%
\bibitem [{\citenamefont {Harrow}\ \emph {et~al.}(2009)\citenamefont {Harrow},
  \citenamefont {Hassidim},\ and\ \citenamefont {Lloyd}}]{Harrow2009_Quantum}%
  \BibitemOpen
  \bibfield  {author} {\bibinfo {author} {\bibfnamefont {A.~W.}\ \bibnamefont
  {Harrow}}, \bibinfo {author} {\bibfnamefont {A.}~\bibnamefont {Hassidim}},\
  and\ \bibinfo {author} {\bibfnamefont {S.}~\bibnamefont {Lloyd}},\ }\bibfield
   {title} {\bibinfo {title} {Quantum algorithm for linear systems of
  equations},\ }\href@noop {} {\bibfield  {journal} {\bibinfo  {journal} {Phys.
  Rev. Lett.}\ }\textbf {\bibinfo {volume} {103}},\ \bibinfo {pages} {150502}
  (\bibinfo {year} {2009})}\BibitemShut {NoStop}%
\bibitem [{\citenamefont {Reiher}\ \emph {et~al.}(2017)\citenamefont {Reiher},
  \citenamefont {Wiebe}, \citenamefont {Svore}, \citenamefont {Wecker},\ and\
  \citenamefont {Troyer}}]{Reiher2017_Elucidating}%
  \BibitemOpen
  \bibfield  {author} {\bibinfo {author} {\bibfnamefont {M.}~\bibnamefont
  {Reiher}}, \bibinfo {author} {\bibfnamefont {N.}~\bibnamefont {Wiebe}},
  \bibinfo {author} {\bibfnamefont {K.~M.}\ \bibnamefont {Svore}}, \bibinfo
  {author} {\bibfnamefont {D.}~\bibnamefont {Wecker}},\ and\ \bibinfo {author}
  {\bibfnamefont {M.}~\bibnamefont {Troyer}},\ }\bibfield  {title} {\bibinfo
  {title} {Elucidating reaction mechanisms on quantum computers},\ }\href@noop
  {} {\bibfield  {journal} {\bibinfo  {journal} {Proc. Natl. Acad. Sci. U. S.
  A.}\ }\textbf {\bibinfo {volume} {114}},\ \bibinfo {pages} {7555} (\bibinfo
  {year} {2017})}\BibitemShut {NoStop}%
\bibitem [{\citenamefont {Schuld}\ and\ \citenamefont
  {Killoran}(2019)}]{Schuld2019_Quantum}%
  \BibitemOpen
  \bibfield  {author} {\bibinfo {author} {\bibfnamefont {M.}~\bibnamefont
  {Schuld}}\ and\ \bibinfo {author} {\bibfnamefont {N.}~\bibnamefont
  {Killoran}},\ }\bibfield  {title} {\bibinfo {title} {{Quantum machine
  learning in feature Hilbert spaces}},\ }\href@noop {} {\bibfield  {journal}
  {\bibinfo  {journal} {Phys. Rev. Lett.}\ }\textbf {\bibinfo {volume} {122}},\
  \bibinfo {pages} {040504} (\bibinfo {year} {2019})}\BibitemShut {NoStop}%
\bibitem [{\citenamefont {McArdle}\ \emph {et~al.}(2020)\citenamefont
  {McArdle}, \citenamefont {Endo}, \citenamefont {Aspuru-Guzik}, \citenamefont
  {Benjamin},\ and\ \citenamefont {Yuan}}]{McArdle2020_Quantum}%
  \BibitemOpen
  \bibfield  {author} {\bibinfo {author} {\bibfnamefont {S.}~\bibnamefont
  {McArdle}}, \bibinfo {author} {\bibfnamefont {S.}~\bibnamefont {Endo}},
  \bibinfo {author} {\bibfnamefont {A.}~\bibnamefont {Aspuru-Guzik}}, \bibinfo
  {author} {\bibfnamefont {S.~C.}\ \bibnamefont {Benjamin}},\ and\ \bibinfo
  {author} {\bibfnamefont {X.}~\bibnamefont {Yuan}},\ }\bibfield  {title}
  {\bibinfo {title} {Quantum computational chemistry},\ }\href@noop {}
  {\bibfield  {journal} {\bibinfo  {journal} {Rev. Mod. Phys.}\ }\textbf
  {\bibinfo {volume} {92}},\ \bibinfo {pages} {015003} (\bibinfo {year}
  {2020})}\BibitemShut {NoStop}%
\bibitem [{\citenamefont {Bharti}\ \emph {et~al.}(2022)\citenamefont {Bharti},
  \citenamefont {Cervera-Lierta}, \citenamefont {Kyaw}, \citenamefont {Haug},
  \citenamefont {Alperin-Lea}, \citenamefont {Anand}, \citenamefont {Degroote},
  \citenamefont {Heimonen}, \citenamefont {Kottmann}, \citenamefont {Menke},
  \citenamefont {Mok}, \citenamefont {Sim}, \citenamefont {Kwek},\ and\
  \citenamefont {Aspuru-Guzik}}]{Bharti2022_Noisy}%
  \BibitemOpen
  \bibfield  {author} {\bibinfo {author} {\bibfnamefont {K.}~\bibnamefont
  {Bharti}}, \bibinfo {author} {\bibfnamefont {A.}~\bibnamefont
  {Cervera-Lierta}}, \bibinfo {author} {\bibfnamefont {T.~H.}\ \bibnamefont
  {Kyaw}}, \bibinfo {author} {\bibfnamefont {T.}~\bibnamefont {Haug}}, \bibinfo
  {author} {\bibfnamefont {S.}~\bibnamefont {Alperin-Lea}}, \bibinfo {author}
  {\bibfnamefont {A.}~\bibnamefont {Anand}}, \bibinfo {author} {\bibfnamefont
  {M.}~\bibnamefont {Degroote}}, \bibinfo {author} {\bibfnamefont
  {H.}~\bibnamefont {Heimonen}}, \bibinfo {author} {\bibfnamefont {J.~S.}\
  \bibnamefont {Kottmann}}, \bibinfo {author} {\bibfnamefont {T.}~\bibnamefont
  {Menke}}, \bibinfo {author} {\bibfnamefont {W.-K.}\ \bibnamefont {Mok}},
  \bibinfo {author} {\bibfnamefont {S.}~\bibnamefont {Sim}}, \bibinfo {author}
  {\bibfnamefont {L.-C.}\ \bibnamefont {Kwek}},\ and\ \bibinfo {author}
  {\bibfnamefont {A.}~\bibnamefont {Aspuru-Guzik}},\ }\bibfield  {title}
  {\bibinfo {title} {Noisy intermediate-scale quantum algorithms},\ }\href@noop
  {} {\bibfield  {journal} {\bibinfo  {journal} {Rev. Mod. Phys.}\ }\textbf
  {\bibinfo {volume} {94}},\ \bibinfo {pages} {015004} (\bibinfo {year}
  {2022})}\BibitemShut {NoStop}%
\bibitem [{\citenamefont {Somaroo}\ \emph {et~al.}(1999)\citenamefont
  {Somaroo}, \citenamefont {Tseng}, \citenamefont {Havel}, \citenamefont
  {Laflamme},\ and\ \citenamefont {Cory}}]{Somaroo1999_Quantum}%
  \BibitemOpen
  \bibfield  {author} {\bibinfo {author} {\bibfnamefont {S.}~\bibnamefont
  {Somaroo}}, \bibinfo {author} {\bibfnamefont {C.~H.}\ \bibnamefont {Tseng}},
  \bibinfo {author} {\bibfnamefont {T.~F.}\ \bibnamefont {Havel}}, \bibinfo
  {author} {\bibfnamefont {R.}~\bibnamefont {Laflamme}},\ and\ \bibinfo
  {author} {\bibfnamefont {D.~G.}\ \bibnamefont {Cory}},\ }\bibfield  {title}
  {\bibinfo {title} {Quantum simulations on a quantum computer},\ }\href@noop
  {} {\bibfield  {journal} {\bibinfo  {journal} {Phys. Rev. Lett.}\ }\textbf
  {\bibinfo {volume} {82}},\ \bibinfo {pages} {5381} (\bibinfo {year}
  {1999})}\BibitemShut {NoStop}%
\bibitem [{\citenamefont {Georgescu}\ \emph {et~al.}(2014)\citenamefont
  {Georgescu}, \citenamefont {Ashhab},\ and\ \citenamefont
  {Nori}}]{Georgescu2014_Quantum}%
  \BibitemOpen
  \bibfield  {author} {\bibinfo {author} {\bibfnamefont {I.~M.}\ \bibnamefont
  {Georgescu}}, \bibinfo {author} {\bibfnamefont {S.}~\bibnamefont {Ashhab}},\
  and\ \bibinfo {author} {\bibfnamefont {F.}~\bibnamefont {Nori}},\ }\bibfield
  {title} {\bibinfo {title} {Quantum simulation},\ }\href@noop {} {\bibfield
  {journal} {\bibinfo  {journal} {Rev. Mod. Phys.}\ }\textbf {\bibinfo {volume}
  {86}},\ \bibinfo {pages} {153} (\bibinfo {year} {2014})}\BibitemShut
  {NoStop}%
\bibitem [{\citenamefont {Ju}\ \emph {et~al.}(2014)\citenamefont {Ju},
  \citenamefont {Lei}, \citenamefont {Xu}, \citenamefont {Culcer},
  \citenamefont {Zhang},\ and\ \citenamefont {Du}}]{Ju2014_NV}%
  \BibitemOpen
  \bibfield  {author} {\bibinfo {author} {\bibfnamefont {C.}~\bibnamefont
  {Ju}}, \bibinfo {author} {\bibfnamefont {C.}~\bibnamefont {Lei}}, \bibinfo
  {author} {\bibfnamefont {X.}~\bibnamefont {Xu}}, \bibinfo {author}
  {\bibfnamefont {D.}~\bibnamefont {Culcer}}, \bibinfo {author} {\bibfnamefont
  {Z.}~\bibnamefont {Zhang}},\ and\ \bibinfo {author} {\bibfnamefont
  {J.}~\bibnamefont {Du}},\ }\bibfield  {title} {\bibinfo {title}
  {{NV-center-based digital quantum simulation of a quantum phase transition in
  topological insulators}},\ }\href@noop {} {\bibfield  {journal} {\bibinfo
  {journal} {Phys. Rev. B}\ }\textbf {\bibinfo {volume} {89}},\ \bibinfo
  {pages} {045432} (\bibinfo {year} {2014})}\BibitemShut {NoStop}%
\bibitem [{\citenamefont {Zhong}\ \emph {et~al.}(2020)\citenamefont {Zhong},
  \citenamefont {Wang}, \citenamefont {Deng}, \citenamefont {Chen},
  \citenamefont {Peng}, \citenamefont {i~Jun~Zhang}, \citenamefont {Li},
  \citenamefont {Li}, \citenamefont {Jiang}, \citenamefont {Gan}, \citenamefont
  {Yang}, \citenamefont {You}, \citenamefont {Wang}, \citenamefont {Li},
  \citenamefont {Liu}, \citenamefont {Lu},\ and\ \citenamefont
  {Pan}}]{Zhong2020_Quantum}%
  \BibitemOpen
  \bibfield  {author} {\bibinfo {author} {\bibfnamefont {H.-S.}\ \bibnamefont
  {Zhong}}, \bibinfo {author} {\bibfnamefont {H.}~\bibnamefont {Wang}},
  \bibinfo {author} {\bibfnamefont {Y.-H.}\ \bibnamefont {Deng}}, \bibinfo
  {author} {\bibfnamefont {M.-C.}\ \bibnamefont {Chen}}, \bibinfo {author}
  {\bibfnamefont {L.-C.}\ \bibnamefont {Peng}}, \bibinfo {author}
  {\bibfnamefont {Y.-H.~L.}\ \bibnamefont {i~Jun~Zhang}}, \bibinfo {author}
  {\bibfnamefont {H.}~\bibnamefont {Li}}, \bibinfo {author} {\bibfnamefont
  {Y.}~\bibnamefont {Li}}, \bibinfo {author} {\bibfnamefont {X.}~\bibnamefont
  {Jiang}}, \bibinfo {author} {\bibfnamefont {L.}~\bibnamefont {Gan}}, \bibinfo
  {author} {\bibfnamefont {G.}~\bibnamefont {Yang}}, \bibinfo {author}
  {\bibfnamefont {L.}~\bibnamefont {You}}, \bibinfo {author} {\bibfnamefont
  {Z.}~\bibnamefont {Wang}}, \bibinfo {author} {\bibfnamefont {L.}~\bibnamefont
  {Li}}, \bibinfo {author} {\bibfnamefont {N.-L.}\ \bibnamefont {Liu}},
  \bibinfo {author} {\bibfnamefont {C.-Y.}\ \bibnamefont {Lu}},\ and\ \bibinfo
  {author} {\bibfnamefont {J.-W.}\ \bibnamefont {Pan}},\ }\bibfield  {title}
  {\bibinfo {title} {Quantum computational advantage using photons},\
  }\href@noop {} {\bibfield  {journal} {\bibinfo  {journal} {Science}\ }\textbf
  {\bibinfo {volume} {370}},\ \bibinfo {pages} {1460} (\bibinfo {year}
  {2020})}\BibitemShut {NoStop}%
\bibitem [{\citenamefont {Yuan}(2020)}]{Yuan2020_A}%
  \BibitemOpen
  \bibfield  {author} {\bibinfo {author} {\bibfnamefont {X.}~\bibnamefont
  {Yuan}},\ }\bibfield  {title} {\bibinfo {title} {A quantum-computing
  advantage for chemistry},\ }\href@noop {} {\bibfield  {journal} {\bibinfo
  {journal} {Science}\ }\textbf {\bibinfo {volume} {369}},\ \bibinfo {pages}
  {1054} (\bibinfo {year} {2020})}\BibitemShut {NoStop}%
\bibitem [{\citenamefont {Han}\ \emph {et~al.}(2021)\citenamefont {Han},
  \citenamefont {Cai}, \citenamefont {Hu}, \citenamefont {Mu}, \citenamefont
  {Ma}, \citenamefont {Xu}, \citenamefont {Wang}, \citenamefont {Wang},
  \citenamefont {Song}, \citenamefont {Zou},\ and\ \citenamefont
  {Sun}}]{Han2021_Experimental}%
  \BibitemOpen
  \bibfield  {author} {\bibinfo {author} {\bibfnamefont {J.}~\bibnamefont
  {Han}}, \bibinfo {author} {\bibfnamefont {W.}~\bibnamefont {Cai}}, \bibinfo
  {author} {\bibfnamefont {L.}~\bibnamefont {Hu}}, \bibinfo {author}
  {\bibfnamefont {X.}~\bibnamefont {Mu}}, \bibinfo {author} {\bibfnamefont
  {Y.}~\bibnamefont {Ma}}, \bibinfo {author} {\bibfnamefont {Y.}~\bibnamefont
  {Xu}}, \bibinfo {author} {\bibfnamefont {W.}~\bibnamefont {Wang}}, \bibinfo
  {author} {\bibfnamefont {H.}~\bibnamefont {Wang}}, \bibinfo {author}
  {\bibfnamefont {Y.~P.}\ \bibnamefont {Song}}, \bibinfo {author}
  {\bibfnamefont {C.-L.}\ \bibnamefont {Zou}},\ and\ \bibinfo {author}
  {\bibfnamefont {L.}~\bibnamefont {Sun}},\ }\bibfield  {title} {\bibinfo
  {title} {{Experimental simulation of open quantum system dynamics via
  Trotterization}},\ }\href@noop {} {\bibfield  {journal} {\bibinfo  {journal}
  {Phys. Rev. Lett.}\ }\textbf {\bibinfo {volume} {127}},\ \bibinfo {pages}
  {020504} (\bibinfo {year} {2021})}\BibitemShut {NoStop}%
\bibitem [{\citenamefont {Monroe}\ \emph {et~al.}(2021)\citenamefont {Monroe},
  \citenamefont {Campbell}, \citenamefont {Duan}, \citenamefont {Gong},
  \citenamefont {Gorshkov}, \citenamefont {Hess}, \citenamefont {Islam},
  \citenamefont {Kim}, \citenamefont {Linke}, \citenamefont {Pagano},
  \citenamefont {Richerme}, \citenamefont {Senko},\ and\ \citenamefont
  {Yao}}]{Monroe2021_Programmable}%
  \BibitemOpen
  \bibfield  {author} {\bibinfo {author} {\bibfnamefont {C.}~\bibnamefont
  {Monroe}}, \bibinfo {author} {\bibfnamefont {W.~C.}\ \bibnamefont
  {Campbell}}, \bibinfo {author} {\bibfnamefont {L.-M.}\ \bibnamefont {Duan}},
  \bibinfo {author} {\bibfnamefont {Z.-X.}\ \bibnamefont {Gong}}, \bibinfo
  {author} {\bibfnamefont {A.~V.}\ \bibnamefont {Gorshkov}}, \bibinfo {author}
  {\bibfnamefont {P.~W.}\ \bibnamefont {Hess}}, \bibinfo {author}
  {\bibfnamefont {R.}~\bibnamefont {Islam}}, \bibinfo {author} {\bibfnamefont
  {K.}~\bibnamefont {Kim}}, \bibinfo {author} {\bibfnamefont {N.~M.}\
  \bibnamefont {Linke}}, \bibinfo {author} {\bibfnamefont {G.}~\bibnamefont
  {Pagano}}, \bibinfo {author} {\bibfnamefont {P.}~\bibnamefont {Richerme}},
  \bibinfo {author} {\bibfnamefont {C.}~\bibnamefont {Senko}},\ and\ \bibinfo
  {author} {\bibfnamefont {N.~Y.}\ \bibnamefont {Yao}},\ }\bibfield  {title}
  {\bibinfo {title} {Programmable quantum simulations of spin systems with
  trapped ions},\ }\href@noop {} {\bibfield  {journal} {\bibinfo  {journal}
  {Rev. Mod. Phys.}\ }\textbf {\bibinfo {volume} {93}},\ \bibinfo {pages}
  {025001} (\bibinfo {year} {2021})}\BibitemShut {NoStop}%
\bibitem [{\citenamefont {Bharadwaj}\ and\ \citenamefont
  {Sreenivasan}(2020)}]{Bharadwaj2020_Quantum}%
  \BibitemOpen
  \bibfield  {author} {\bibinfo {author} {\bibfnamefont {S.~S.}\ \bibnamefont
  {Bharadwaj}}\ and\ \bibinfo {author} {\bibfnamefont {K.~R.}\ \bibnamefont
  {Sreenivasan}},\ }\bibfield  {title} {\bibinfo {title} {Quantum computation
  of fluid dynamics},\ }\href@noop {} {\bibfield  {journal} {\bibinfo
  {journal} {Indian Acad. Sci. Conf. Ser.}\ }\textbf {\bibinfo {volume} {3}},\
  \bibinfo {pages} {77} (\bibinfo {year} {2020})}\BibitemShut {NoStop}%
\bibitem [{\citenamefont {Dodin}\ and\ \citenamefont
  {Startsev}(2021)}]{Dodin2021_On}%
  \BibitemOpen
  \bibfield  {author} {\bibinfo {author} {\bibfnamefont {I.~Y.}\ \bibnamefont
  {Dodin}}\ and\ \bibinfo {author} {\bibfnamefont {E.~A.}\ \bibnamefont
  {Startsev}},\ }\bibfield  {title} {\bibinfo {title} {On applications of
  quantum computing to plasma simulations},\ }\href@noop {} {\bibfield
  {journal} {\bibinfo  {journal} {Phys. Plasmas}\ }\textbf {\bibinfo {volume}
  {28}},\ \bibinfo {pages} {092101} (\bibinfo {year} {2021})}\BibitemShut
  {NoStop}%
\bibitem [{\citenamefont {Giannakis}\ \emph {et~al.}(2022)\citenamefont
  {Giannakis}, \citenamefont {Ourmazd}, \citenamefont {Pfeffer}, \citenamefont
  {Schumacher},\ and\ \citenamefont {Slawinska}}]{Giannakis2022_Embedding}%
  \BibitemOpen
  \bibfield  {author} {\bibinfo {author} {\bibfnamefont {D.}~\bibnamefont
  {Giannakis}}, \bibinfo {author} {\bibfnamefont {A.}~\bibnamefont {Ourmazd}},
  \bibinfo {author} {\bibfnamefont {P.}~\bibnamefont {Pfeffer}}, \bibinfo
  {author} {\bibfnamefont {J.}~\bibnamefont {Schumacher}},\ and\ \bibinfo
  {author} {\bibfnamefont {J.}~\bibnamefont {Slawinska}},\ }\bibfield  {title}
  {\bibinfo {title} {Embedding classical dynamics in a quantum computer},\
  }\href@noop {} {\bibfield  {journal} {\bibinfo  {journal} {Phys. Rev. A}\
  }\textbf {\bibinfo {volume} {105}},\ \bibinfo {pages} {052404} (\bibinfo
  {year} {2022})}\BibitemShut {NoStop}%
\bibitem [{\citenamefont {Jin}\ \emph {et~al.}(2022)\citenamefont {Jin},
  \citenamefont {Li},\ and\ \citenamefont {Liu}}]{Jin2022_Quantum}%
  \BibitemOpen
  \bibfield  {author} {\bibinfo {author} {\bibfnamefont {S.}~\bibnamefont
  {Jin}}, \bibinfo {author} {\bibfnamefont {X.}~\bibnamefont {Li}},\ and\
  \bibinfo {author} {\bibfnamefont {N.}~\bibnamefont {Liu}},\ }\bibfield
  {title} {\bibinfo {title} {Quantum simulation in the semi-classical regime},\
  }\href@noop {} {\bibfield  {journal} {\bibinfo  {journal} {{Quantum}}\
  }\textbf {\bibinfo {volume} {6}},\ \bibinfo {pages} {739} (\bibinfo {year}
  {2022})}\BibitemShut {NoStop}%
\bibitem [{\citenamefont {Pope}(2000)}]{Pope2000_Turbulent}%
  \BibitemOpen
  \bibfield  {author} {\bibinfo {author} {\bibfnamefont {S.~B.}\ \bibnamefont
  {Pope}},\ }\href@noop {} {\emph {\bibinfo {title} {Turbulent Flows}}}\
  (\bibinfo  {publisher} {Cambridge University Press},\ \bibinfo {year}
  {2000})\BibitemShut {NoStop}%
\bibitem [{\citenamefont {Moin}\ and\ \citenamefont
  {Mahesh}(1998)}]{Moin1998_Direct}%
  \BibitemOpen
  \bibfield  {author} {\bibinfo {author} {\bibfnamefont {P.}~\bibnamefont
  {Moin}}\ and\ \bibinfo {author} {\bibfnamefont {K.}~\bibnamefont {Mahesh}},\
  }\bibfield  {title} {\bibinfo {title} {Direct numerical simulation: a tool in
  turbulence research},\ }\href@noop {} {\bibfield  {journal} {\bibinfo
  {journal} {Annu. Rev. Fluid Mech.}\ }\textbf {\bibinfo {volume} {30}},\
  \bibinfo {pages} {539} (\bibinfo {year} {1998})}\BibitemShut {NoStop}%
\bibitem [{\citenamefont {Ishihara}\ \emph {et~al.}(2009)\citenamefont
  {Ishihara}, \citenamefont {Gotoh},\ and\ \citenamefont
  {Kaneda}}]{Ishihara2009_Study}%
  \BibitemOpen
  \bibfield  {author} {\bibinfo {author} {\bibfnamefont {T.}~\bibnamefont
  {Ishihara}}, \bibinfo {author} {\bibfnamefont {T.}~\bibnamefont {Gotoh}},\
  and\ \bibinfo {author} {\bibfnamefont {Y.}~\bibnamefont {Kaneda}},\
  }\bibfield  {title} {\bibinfo {title} {{Study of high-Reynolds number
  isotropic turbulence by direct numerical simulation}},\ }\href@noop {}
  {\bibfield  {journal} {\bibinfo  {journal} {Annu. Rev. Fluid Mech.}\ }\textbf
  {\bibinfo {volume} {41}},\ \bibinfo {pages} {165} (\bibinfo {year}
  {2009})}\BibitemShut {NoStop}%
\bibitem [{\citenamefont {Givi}\ \emph {et~al.}(2020)\citenamefont {Givi},
  \citenamefont {Daley}, \citenamefont {Mavriplis},\ and\ \citenamefont
  {Malik}}]{Givi2020_Quantum}%
  \BibitemOpen
  \bibfield  {author} {\bibinfo {author} {\bibfnamefont {P.}~\bibnamefont
  {Givi}}, \bibinfo {author} {\bibfnamefont {A.~J.}\ \bibnamefont {Daley}},
  \bibinfo {author} {\bibfnamefont {D.}~\bibnamefont {Mavriplis}},\ and\
  \bibinfo {author} {\bibfnamefont {M.}~\bibnamefont {Malik}},\ }\bibfield
  {title} {\bibinfo {title} {Quantum speedup for aeroscience and engineering},\
  }\href@noop {} {\bibfield  {journal} {\bibinfo  {journal} {AIAA J.}\ }\textbf
  {\bibinfo {volume} {58}},\ \bibinfo {pages} {8} (\bibinfo {year}
  {2020})}\BibitemShut {NoStop}%
\bibitem [{\citenamefont {Clader}\ \emph {et~al.}(2013)\citenamefont {Clader},
  \citenamefont {Jacobs},\ and\ \citenamefont
  {Sprouse}}]{Clader2013_Preconditioned}%
  \BibitemOpen
  \bibfield  {author} {\bibinfo {author} {\bibfnamefont {B.~D.}\ \bibnamefont
  {Clader}}, \bibinfo {author} {\bibfnamefont {B.~C.}\ \bibnamefont {Jacobs}},\
  and\ \bibinfo {author} {\bibfnamefont {C.~R.}\ \bibnamefont {Sprouse}},\
  }\bibfield  {title} {\bibinfo {title} {Preconditioned quantum linear system
  algorithm},\ }\href@noop {} {\bibfield  {journal} {\bibinfo  {journal} {Phys.
  Rev. Lett.}\ }\textbf {\bibinfo {volume} {110}},\ \bibinfo {pages} {250504}
  (\bibinfo {year} {2013})}\BibitemShut {NoStop}%
\bibitem [{\citenamefont {Cao}\ \emph {et~al.}(2013)\citenamefont {Cao},
  \citenamefont {Papageorgiou}, \citenamefont {Petras}, \citenamefont {Traub},\
  and\ \citenamefont {Kais}}]{Cao2013_Quantum}%
  \BibitemOpen
  \bibfield  {author} {\bibinfo {author} {\bibfnamefont {Y.}~\bibnamefont
  {Cao}}, \bibinfo {author} {\bibfnamefont {A.}~\bibnamefont {Papageorgiou}},
  \bibinfo {author} {\bibfnamefont {I.}~\bibnamefont {Petras}}, \bibinfo
  {author} {\bibfnamefont {J.}~\bibnamefont {Traub}},\ and\ \bibinfo {author}
  {\bibfnamefont {S.}~\bibnamefont {Kais}},\ }\bibfield  {title} {\bibinfo
  {title} {{Quantum algorithm and circuit design solving the Poisson
  equation}},\ }\href@noop {} {\bibfield  {journal} {\bibinfo  {journal} {New
  J. Phys.}\ }\textbf {\bibinfo {volume} {15}},\ \bibinfo {pages} {013021}
  (\bibinfo {year} {2013})}\BibitemShut {NoStop}%
\bibitem [{\citenamefont {Montanaro}\ and\ \citenamefont
  {Pallister}(2016)}]{Montanaro2016_Quantum}%
  \BibitemOpen
  \bibfield  {author} {\bibinfo {author} {\bibfnamefont {A.}~\bibnamefont
  {Montanaro}}\ and\ \bibinfo {author} {\bibfnamefont {S.}~\bibnamefont
  {Pallister}},\ }\bibfield  {title} {\bibinfo {title} {Quantum algorithms and
  the finite element method},\ }\href@noop {} {\bibfield  {journal} {\bibinfo
  {journal} {Phys. Rev. A}\ }\textbf {\bibinfo {volume} {93}},\ \bibinfo
  {pages} {032324} (\bibinfo {year} {2016})}\BibitemShut {NoStop}%
\bibitem [{\citenamefont {Costa}\ \emph {et~al.}(2019)\citenamefont {Costa},
  \citenamefont {Jordan},\ and\ \citenamefont {Ostrander}}]{Costa2019_Quantum}%
  \BibitemOpen
  \bibfield  {author} {\bibinfo {author} {\bibfnamefont {P.~C.~S.}\
  \bibnamefont {Costa}}, \bibinfo {author} {\bibfnamefont {S.}~\bibnamefont
  {Jordan}},\ and\ \bibinfo {author} {\bibfnamefont {A.}~\bibnamefont
  {Ostrander}},\ }\bibfield  {title} {\bibinfo {title} {Quantum algorithm for
  simulating the wave equation},\ }\href@noop {} {\bibfield  {journal}
  {\bibinfo  {journal} {Phys. Rev. A}\ }\textbf {\bibinfo {volume} {99}},\
  \bibinfo {pages} {012323} (\bibinfo {year} {2019})}\BibitemShut {NoStop}%
\bibitem [{\citenamefont {Lloyd}\ \emph {et~al.}(2020)\citenamefont {Lloyd},
  \citenamefont {Palma}, \citenamefont {Gokler}, \citenamefont {Kiani},
  \citenamefont {Liu}, \citenamefont {Marvian}, \citenamefont {Tennie},\ and\
  \citenamefont {Palmer}}]{Lloyd2020_Quantum}%
  \BibitemOpen
  \bibfield  {author} {\bibinfo {author} {\bibfnamefont {S.}~\bibnamefont
  {Lloyd}}, \bibinfo {author} {\bibfnamefont {G.~D.}\ \bibnamefont {Palma}},
  \bibinfo {author} {\bibfnamefont {C.}~\bibnamefont {Gokler}}, \bibinfo
  {author} {\bibfnamefont {B.}~\bibnamefont {Kiani}}, \bibinfo {author}
  {\bibfnamefont {Z.-W.}\ \bibnamefont {Liu}}, \bibinfo {author} {\bibfnamefont
  {M.}~\bibnamefont {Marvian}}, \bibinfo {author} {\bibfnamefont
  {F.}~\bibnamefont {Tennie}},\ and\ \bibinfo {author} {\bibfnamefont
  {T.}~\bibnamefont {Palmer}},\ }\href@noop {} {\bibinfo {title} {Quantum
  algorithm for nonlinear differential equations}} (\bibinfo {year} {2020}),\
  \Eprint {https://arxiv.org/abs/arXiv:2011.06571} {arXiv:2011.06571}
  \BibitemShut {NoStop}%
\bibitem [{\citenamefont {Lubasch}\ \emph {et~al.}(2020)\citenamefont
  {Lubasch}, \citenamefont {Joo}, \citenamefont {Moinier}, \citenamefont
  {Kiffner},\ and\ \citenamefont {Jaksch}}]{Lubasch2020_Variational}%
  \BibitemOpen
  \bibfield  {author} {\bibinfo {author} {\bibfnamefont {M.}~\bibnamefont
  {Lubasch}}, \bibinfo {author} {\bibfnamefont {J.}~\bibnamefont {Joo}},
  \bibinfo {author} {\bibfnamefont {P.}~\bibnamefont {Moinier}}, \bibinfo
  {author} {\bibfnamefont {M.}~\bibnamefont {Kiffner}},\ and\ \bibinfo {author}
  {\bibfnamefont {D.}~\bibnamefont {Jaksch}},\ }\bibfield  {title} {\bibinfo
  {title} {Variational quantum algorithms for nonlinear problems},\ }\href@noop
  {} {\bibfield  {journal} {\bibinfo  {journal} {Phys. Rev. A}\ }\textbf
  {\bibinfo {volume} {101}},\ \bibinfo {pages} {010301} (\bibinfo {year}
  {2020})}\BibitemShut {NoStop}%
\bibitem [{\citenamefont {Liu}\ \emph {et~al.}(2021{\natexlab{a}})\citenamefont
  {Liu}, \citenamefont {Kolden}, \citenamefont {Krovi}, \citenamefont
  {Loureiro}, \citenamefont {Trivisa},\ and\ \citenamefont
  {Childs}}]{Liu2021_Efficient}%
  \BibitemOpen
  \bibfield  {author} {\bibinfo {author} {\bibfnamefont {J.-P.}\ \bibnamefont
  {Liu}}, \bibinfo {author} {\bibfnamefont {H.~O.}\ \bibnamefont {Kolden}},
  \bibinfo {author} {\bibfnamefont {H.~K.}\ \bibnamefont {Krovi}}, \bibinfo
  {author} {\bibfnamefont {N.~F.}\ \bibnamefont {Loureiro}}, \bibinfo {author}
  {\bibfnamefont {K.}~\bibnamefont {Trivisa}},\ and\ \bibinfo {author}
  {\bibfnamefont {A.~M.}\ \bibnamefont {Childs}},\ }\bibfield  {title}
  {\bibinfo {title} {Efficient quantum algorithm for dissipative nonlinear
  differential equations},\ }\href@noop {} {\bibfield  {journal} {\bibinfo
  {journal} {Proc. Natl. Acad. Sci. U. S. A.}\ }\textbf {\bibinfo {volume}
  {118}},\ \bibinfo {pages} {e2026805118} (\bibinfo {year}
  {2021}{\natexlab{a}})}\BibitemShut {NoStop}%
\bibitem [{\citenamefont {Gaitan}(2020)}]{Gaitan2020_Finding}%
  \BibitemOpen
  \bibfield  {author} {\bibinfo {author} {\bibfnamefont {F.}~\bibnamefont
  {Gaitan}},\ }\bibfield  {title} {\bibinfo {title} {{Finding flows of a
  Navier--Stokes fluid through quantum computing}},\ }\href@noop {} {\bibfield
  {journal} {\bibinfo  {journal} {npj Quantum Inform.}\ }\textbf {\bibinfo
  {volume} {6}},\ \bibinfo {pages} {61} (\bibinfo {year} {2020})}\BibitemShut
  {NoStop}%
\bibitem [{\citenamefont {Ray}\ \emph {et~al.}(2019)\citenamefont {Ray},
  \citenamefont {Banerjee}, \citenamefont {Nadiga},\ and\ \citenamefont
  {Karra}}]{Ray2019_Towards}%
  \BibitemOpen
  \bibfield  {author} {\bibinfo {author} {\bibfnamefont {N.}~\bibnamefont
  {Ray}}, \bibinfo {author} {\bibfnamefont {T.}~\bibnamefont {Banerjee}},
  \bibinfo {author} {\bibfnamefont {B.}~\bibnamefont {Nadiga}},\ and\ \bibinfo
  {author} {\bibfnamefont {S.}~\bibnamefont {Karra}},\ }\href@noop {} {\bibinfo
  {title} {{Towards solving the Navier--Stokes equation on quantum computers}}}
  (\bibinfo {year} {2019}),\ \Eprint {https://arxiv.org/abs/arXiv:1904.09033}
  {arXiv:1904.09033} \BibitemShut {NoStop}%
\bibitem [{\citenamefont {Oz}\ \emph {et~al.}(2022)\citenamefont {Oz},
  \citenamefont {Vuppala}, \citenamefont {Kara},\ and\ \citenamefont
  {Gaitan}}]{Oz2022_Solving}%
  \BibitemOpen
  \bibfield  {author} {\bibinfo {author} {\bibfnamefont {F.}~\bibnamefont
  {Oz}}, \bibinfo {author} {\bibfnamefont {R.~K. S.~S.}\ \bibnamefont
  {Vuppala}}, \bibinfo {author} {\bibfnamefont {K.}~\bibnamefont {Kara}},\ and\
  \bibinfo {author} {\bibfnamefont {F.}~\bibnamefont {Gaitan}},\ }\bibfield
  {title} {\bibinfo {title} {{Solving Burgers' equation with quantum
  computing}},\ }\href@noop {} {\bibfield  {journal} {\bibinfo  {journal}
  {Quantum Inf. Process.}\ }\textbf {\bibinfo {volume} {21}},\ \bibinfo {pages}
  {30} (\bibinfo {year} {2022})}\BibitemShut {NoStop}%
\bibitem [{\citenamefont {Pfeffer}\ \emph {et~al.}(2022)\citenamefont
  {Pfeffer}, \citenamefont {Heyder},\ and\ \citenamefont
  {Schumacher}}]{Pfeffer2022_Hybrid}%
  \BibitemOpen
  \bibfield  {author} {\bibinfo {author} {\bibfnamefont {P.}~\bibnamefont
  {Pfeffer}}, \bibinfo {author} {\bibfnamefont {F.}~\bibnamefont {Heyder}},\
  and\ \bibinfo {author} {\bibfnamefont {J.}~\bibnamefont {Schumacher}},\
  }\bibfield  {title} {\bibinfo {title} {Hybrid quantum-classical reservoir
  computing of thermal convection flow},\ }\href@noop {} {\bibfield  {journal}
  {\bibinfo  {journal} {Phys. Rev. Res.}\ }\textbf {\bibinfo {volume} {4}},\
  \bibinfo {pages} {033176} (\bibinfo {year} {2022})}\BibitemShut {NoStop}%
\bibitem [{\citenamefont {Wen}\ \emph {et~al.}(2019)\citenamefont {Wen},
  \citenamefont {Kong}, \citenamefont {Wei}, \citenamefont {Wang},
  \citenamefont {Xin},\ and\ \citenamefont {Long}}]{Wen2019_Experimental}%
  \BibitemOpen
  \bibfield  {author} {\bibinfo {author} {\bibfnamefont {J.}~\bibnamefont
  {Wen}}, \bibinfo {author} {\bibfnamefont {X.}~\bibnamefont {Kong}}, \bibinfo
  {author} {\bibfnamefont {S.}~\bibnamefont {Wei}}, \bibinfo {author}
  {\bibfnamefont {B.}~\bibnamefont {Wang}}, \bibinfo {author} {\bibfnamefont
  {T.}~\bibnamefont {Xin}},\ and\ \bibinfo {author} {\bibfnamefont
  {G.}~\bibnamefont {Long}},\ }\bibfield  {title} {\bibinfo {title}
  {Experimental realization of quantum algorithms for a linear system inspired
  by adiabatic quantum computing},\ }\href@noop {} {\bibfield  {journal}
  {\bibinfo  {journal} {Phys. Rev. A}\ }\textbf {\bibinfo {volume} {99}},\
  \bibinfo {pages} {012320} (\bibinfo {year} {2019})}\BibitemShut {NoStop}%
\bibitem [{\citenamefont {Chen}\ \emph {et~al.}(2022)\citenamefont {Chen},
  \citenamefont {Xue}, \citenamefont {Chen}, \citenamefont {Lu}, \citenamefont
  {Wu}, \citenamefont {Ding}, \citenamefont {Huang},\ and\ \citenamefont
  {Guo}}]{Chen2022_Quantum}%
  \BibitemOpen
  \bibfield  {author} {\bibinfo {author} {\bibfnamefont {Z.-Y.}\ \bibnamefont
  {Chen}}, \bibinfo {author} {\bibfnamefont {C.}~\bibnamefont {Xue}}, \bibinfo
  {author} {\bibfnamefont {S.-M.}\ \bibnamefont {Chen}}, \bibinfo {author}
  {\bibfnamefont {B.-H.}\ \bibnamefont {Lu}}, \bibinfo {author} {\bibfnamefont
  {Y.-C.}\ \bibnamefont {Wu}}, \bibinfo {author} {\bibfnamefont {J.-C.}\
  \bibnamefont {Ding}}, \bibinfo {author} {\bibfnamefont {S.-H.}\ \bibnamefont
  {Huang}},\ and\ \bibinfo {author} {\bibfnamefont {G.-P.}\ \bibnamefont
  {Guo}},\ }\bibfield  {title} {\bibinfo {title} {Quantum approach to
  accelerate finite volume method on steady computational fluid dynamics
  problems},\ }\href@noop {} {\bibfield  {journal} {\bibinfo  {journal}
  {Quantum Inf. Process.}\ }\textbf {\bibinfo {volume} {21}},\ \bibinfo {pages}
  {137} (\bibinfo {year} {2022})}\BibitemShut {NoStop}%
\bibitem [{\citenamefont {Lapworth}(2022)}]{Lapworth2022_A}%
  \BibitemOpen
  \bibfield  {author} {\bibinfo {author} {\bibfnamefont {L.}~\bibnamefont
  {Lapworth}},\ }\href@noop {} {\bibinfo {title} {{A hybrid quantum-classical
  CFD methodology with benchmark HHL solutions}}} (\bibinfo {year} {2022}),\
  \Eprint {https://arxiv.org/abs/arXiv:2206.00419} {arXiv:2206.00419}
  \BibitemShut {NoStop}%
\bibitem [{\citenamefont {Demirdjian}\ \emph {et~al.}(2022)\citenamefont
  {Demirdjian}, \citenamefont {Gunlycke}, \citenamefont {Reynolds},
  \citenamefont {Doyle},\ and\ \citenamefont
  {Tafur}}]{Demirdjian2022_Variational}%
  \BibitemOpen
  \bibfield  {author} {\bibinfo {author} {\bibfnamefont {R.}~\bibnamefont
  {Demirdjian}}, \bibinfo {author} {\bibfnamefont {D.}~\bibnamefont
  {Gunlycke}}, \bibinfo {author} {\bibfnamefont {C.~A.}\ \bibnamefont
  {Reynolds}}, \bibinfo {author} {\bibfnamefont {J.~D.}\ \bibnamefont
  {Doyle}},\ and\ \bibinfo {author} {\bibfnamefont {S.}~\bibnamefont {Tafur}},\
  }\bibfield  {title} {\bibinfo {title} {Variational quantum solutions to the
  advection–diffusion equation for applications in fluid dynamics},\
  }\href@noop {} {\bibfield  {journal} {\bibinfo  {journal} {Quantum Inf.
  Process.}\ }\textbf {\bibinfo {volume} {21}},\ \bibinfo {pages} {322}
  (\bibinfo {year} {2022})}\BibitemShut {NoStop}%
\bibitem [{\citenamefont {Steijl}\ and\ \citenamefont
  {Barakos}(2018)}]{Steijl2018_Parallel}%
  \BibitemOpen
  \bibfield  {author} {\bibinfo {author} {\bibfnamefont {R.}~\bibnamefont
  {Steijl}}\ and\ \bibinfo {author} {\bibfnamefont {G.~N.}\ \bibnamefont
  {Barakos}},\ }\bibfield  {title} {\bibinfo {title} {Parallel evaluation of
  quantum algorithms for computational fluid dynamics},\ }\href@noop {}
  {\bibfield  {journal} {\bibinfo  {journal} {Comput. Fluids}\ }\textbf
  {\bibinfo {volume} {173}},\ \bibinfo {pages} {22} (\bibinfo {year}
  {2018})}\BibitemShut {NoStop}%
\bibitem [{\citenamefont {Aaronson}(2015)}]{Aaronson2015_Read}%
  \BibitemOpen
  \bibfield  {author} {\bibinfo {author} {\bibfnamefont {S.}~\bibnamefont
  {Aaronson}},\ }\bibfield  {title} {\bibinfo {title} {Read the fine print},\
  }\href@noop {} {\bibfield  {journal} {\bibinfo  {journal} {Nat. Phys.}\
  }\textbf {\bibinfo {volume} {11}},\ \bibinfo {pages} {291} (\bibinfo {year}
  {2015})}\BibitemShut {NoStop}%
\bibitem [{\citenamefont {Zylberman}\ \emph {et~al.}(2022)\citenamefont
  {Zylberman}, \citenamefont {Di~Molfetta}, \citenamefont {Brachet},
  \citenamefont {Loureiro},\ and\ \citenamefont
  {Debbasch}}]{Zylberman2022_Quantum}%
  \BibitemOpen
  \bibfield  {author} {\bibinfo {author} {\bibfnamefont {J.}~\bibnamefont
  {Zylberman}}, \bibinfo {author} {\bibfnamefont {G.}~\bibnamefont
  {Di~Molfetta}}, \bibinfo {author} {\bibfnamefont {M.}~\bibnamefont
  {Brachet}}, \bibinfo {author} {\bibfnamefont {N.~F.}\ \bibnamefont
  {Loureiro}},\ and\ \bibinfo {author} {\bibfnamefont {F.}~\bibnamefont
  {Debbasch}},\ }\bibfield  {title} {\bibinfo {title} {{Quantum simulations of
  hydrodynamics via the Madelung transformation}},\ }\href@noop {} {\bibfield
  {journal} {\bibinfo  {journal} {Phys. Rev. A}\ }\textbf {\bibinfo {volume}
  {106}},\ \bibinfo {pages} {032408} (\bibinfo {year} {2022})}\BibitemShut
  {NoStop}%
\bibitem [{\citenamefont {Joseph}(2020)}]{Joseph2020_Koopman}%
  \BibitemOpen
  \bibfield  {author} {\bibinfo {author} {\bibfnamefont {I.}~\bibnamefont
  {Joseph}},\ }\bibfield  {title} {\bibinfo {title} {{Koopman--von Neumann
  approach to quantum simulation of nonlinear classical dynamics}},\
  }\href@noop {} {\bibfield  {journal} {\bibinfo  {journal} {Phys. Rev. Res.}\
  }\textbf {\bibinfo {volume} {2}},\ \bibinfo {pages} {043102} (\bibinfo {year}
  {2020})}\BibitemShut {NoStop}%
\bibitem [{\citenamefont {Yepez}(2001)}]{Yepez2001_Quantum}%
  \BibitemOpen
  \bibfield  {author} {\bibinfo {author} {\bibfnamefont {J.}~\bibnamefont
  {Yepez}},\ }\bibfield  {title} {\bibinfo {title} {Quantum lattice-gas model
  for computational fluid dynamics},\ }\href@noop {} {\bibfield  {journal}
  {\bibinfo  {journal} {Phys. Rev. E}\ }\textbf {\bibinfo {volume} {63}},\
  \bibinfo {pages} {046702} (\bibinfo {year} {2001})}\BibitemShut {NoStop}%
\bibitem [{\citenamefont {Keating}\ \emph {et~al.}(2007)\citenamefont
  {Keating}, \citenamefont {Vahala}, \citenamefont {Yepez}, \citenamefont
  {Soe},\ and\ \citenamefont {Vahala}}]{Keating2007_Entropic}%
  \BibitemOpen
  \bibfield  {author} {\bibinfo {author} {\bibfnamefont {B.}~\bibnamefont
  {Keating}}, \bibinfo {author} {\bibfnamefont {G.}~\bibnamefont {Vahala}},
  \bibinfo {author} {\bibfnamefont {J.}~\bibnamefont {Yepez}}, \bibinfo
  {author} {\bibfnamefont {M.}~\bibnamefont {Soe}},\ and\ \bibinfo {author}
  {\bibfnamefont {L.}~\bibnamefont {Vahala}},\ }\bibfield  {title} {\bibinfo
  {title} {{Entropic lattice Boltzmann representations required to recover
  Navier--Stokes flows}},\ }\href@noop {} {\bibfield  {journal} {\bibinfo
  {journal} {Phys. Rev. E}\ }\textbf {\bibinfo {volume} {75}},\ \bibinfo
  {pages} {036712} (\bibinfo {year} {2007})}\BibitemShut {NoStop}%
\bibitem [{\citenamefont {Todorova}\ and\ \citenamefont
  {Steijl}(2020)}]{Todorova2020_Quantum}%
  \BibitemOpen
  \bibfield  {author} {\bibinfo {author} {\bibfnamefont {B.~N.}\ \bibnamefont
  {Todorova}}\ and\ \bibinfo {author} {\bibfnamefont {R.}~\bibnamefont
  {Steijl}},\ }\bibfield  {title} {\bibinfo {title} {{Quantum algorithm for the
  collisionless Boltzmann equation}},\ }\href@noop {} {\bibfield  {journal}
  {\bibinfo  {journal} {J. Comput. Phys.}\ }\textbf {\bibinfo {volume} {409}},\
  \bibinfo {pages} {109347} (\bibinfo {year} {2020})}\BibitemShut {NoStop}%
\bibitem [{\citenamefont {Gourianov}\ \emph {et~al.}(2022)\citenamefont
  {Gourianov}, \citenamefont {Lubasch}, \citenamefont {Dolgov}, \citenamefont
  {van~den Berg}, \citenamefont {Babaee}, \citenamefont {Givi}, \citenamefont
  {Kiffner},\ and\ \citenamefont {Jaksch}}]{Gourianov2022_A}%
  \BibitemOpen
  \bibfield  {author} {\bibinfo {author} {\bibfnamefont {N.}~\bibnamefont
  {Gourianov}}, \bibinfo {author} {\bibfnamefont {M.}~\bibnamefont {Lubasch}},
  \bibinfo {author} {\bibfnamefont {S.}~\bibnamefont {Dolgov}}, \bibinfo
  {author} {\bibfnamefont {Q.~Y.}\ \bibnamefont {van~den Berg}}, \bibinfo
  {author} {\bibfnamefont {H.}~\bibnamefont {Babaee}}, \bibinfo {author}
  {\bibfnamefont {P.}~\bibnamefont {Givi}}, \bibinfo {author} {\bibfnamefont
  {M.}~\bibnamefont {Kiffner}},\ and\ \bibinfo {author} {\bibfnamefont
  {D.}~\bibnamefont {Jaksch}},\ }\bibfield  {title} {\bibinfo {title} {A
  quantum-inspired approach to exploit turbulence structures},\ }\href@noop {}
  {\bibfield  {journal} {\bibinfo  {journal} {Nat. Comput. Sci.}\ }\textbf
  {\bibinfo {volume} {2}},\ \bibinfo {pages} {30} (\bibinfo {year}
  {2022})}\BibitemShut {NoStop}%
\bibitem [{\citenamefont {Fukagata}(2022)}]{Fukagata2022_Towards}%
  \BibitemOpen
  \bibfield  {author} {\bibinfo {author} {\bibfnamefont {K.}~\bibnamefont
  {Fukagata}},\ }\bibfield  {title} {\bibinfo {title} {Towards quantum
  computing of turbulence},\ }\href@noop {} {\bibfield  {journal} {\bibinfo
  {journal} {Nat. Comput. Sci.}\ }\textbf {\bibinfo {volume} {2}},\ \bibinfo
  {pages} {68} (\bibinfo {year} {2022})}\BibitemShut {NoStop}%
\bibitem [{Qis(2021)}]{Qiskit}%
  \BibitemOpen
  \href@noop {} {\bibinfo {title} {{Qiskit Version 0.24.1}}},\ \bibinfo
  {howpublished}
  {\url{https://qiskit.org/documentation/stable/0.24/release_notes.html}}
  (\bibinfo {year} {2021})\BibitemShut {NoStop}%
\bibitem [{\citenamefont {Griffiths}(2005)}]{Griffiths2005_Introduction}%
  \BibitemOpen
  \bibfield  {author} {\bibinfo {author} {\bibfnamefont {D.~J.}\ \bibnamefont
  {Griffiths}},\ }\href@noop {} {\emph {\bibinfo {title} {Introduction to
  Quantum Mechanics}}}\ (\bibinfo  {publisher} {Pearson Education
  International},\ \bibinfo {year} {2005})\BibitemShut {NoStop}%
\bibitem [{\citenamefont {Schr\"odinger}(1926)}]{Schrodinger1926_An}%
  \BibitemOpen
  \bibfield  {author} {\bibinfo {author} {\bibfnamefont {E.}~\bibnamefont
  {Schr\"odinger}},\ }\bibfield  {title} {\bibinfo {title} {An undulatory
  theory of the mechanics of atoms and molecules},\ }\href@noop {} {\bibfield
  {journal} {\bibinfo  {journal} {Phys. Rev.}\ }\textbf {\bibinfo {volume}
  {28}},\ \bibinfo {pages} {1049} (\bibinfo {year} {1926})}\BibitemShut
  {NoStop}%
\bibitem [{\citenamefont {Madelung}(1927)}]{Madelung1927_Quantentheorie}%
  \BibitemOpen
  \bibfield  {author} {\bibinfo {author} {\bibfnamefont {E.}~\bibnamefont
  {Madelung}},\ }\bibfield  {title} {\bibinfo {title} {Quantentheorie in
  hydrodynamischer form},\ }\href@noop {} {\bibfield  {journal} {\bibinfo
  {journal} {Zeitschrift für Physik}\ }\textbf {\bibinfo {volume} {40}},\
  \bibinfo {pages} {322–326} (\bibinfo {year} {1927})}\BibitemShut {NoStop}%
\bibitem [{\citenamefont {Sch\"onberg}(1954)}]{Schonberg1954_On}%
  \BibitemOpen
  \bibfield  {author} {\bibinfo {author} {\bibfnamefont {M.}~\bibnamefont
  {Sch\"onberg}},\ }\bibfield  {title} {\bibinfo {title} {On the hydrodynamical
  model of the quantum mechanics},\ }\href@noop {} {\bibfield  {journal}
  {\bibinfo  {journal} {Nuovo Cimento}\ }\textbf {\bibinfo {volume} {12}},\
  \bibinfo {pages} {103–133} (\bibinfo {year} {1954})}\BibitemShut {NoStop}%
\bibitem [{\citenamefont {Sorokin}(2001)}]{Sorokin2001_Madelung}%
  \BibitemOpen
  \bibfield  {author} {\bibinfo {author} {\bibfnamefont {A.~L.}\ \bibnamefont
  {Sorokin}},\ }\bibfield  {title} {\bibinfo {title} {Madelung transformation
  for vortex flows of a perfect liquid},\ }\href@noop {} {\bibfield  {journal}
  {\bibinfo  {journal} {Dokl. Phys.}\ }\textbf {\bibinfo {volume} {46}},\
  \bibinfo {pages} {576} (\bibinfo {year} {2001})}\BibitemShut {NoStop}%
\bibitem [{\citenamefont {Love}\ and\ \citenamefont
  {Boghosian}(2004)}]{Love2004_Quaternionic}%
  \BibitemOpen
  \bibfield  {author} {\bibinfo {author} {\bibfnamefont {P.~J.}\ \bibnamefont
  {Love}}\ and\ \bibinfo {author} {\bibfnamefont {B.~M.}\ \bibnamefont
  {Boghosian}},\ }\bibfield  {title} {\bibinfo {title} {{Quaternionic Madelung
  transformation and non-Abelian fluid dynamics}},\ }\href@noop {} {\bibfield
  {journal} {\bibinfo  {journal} {Physica A}\ }\textbf {\bibinfo {volume}
  {332}},\ \bibinfo {pages} {47} (\bibinfo {year} {2004})}\BibitemShut
  {NoStop}%
\bibitem [{\citenamefont {Mueller}\ and\ \citenamefont
  {Ho}(2002)}]{Mueller2002_Two}%
  \BibitemOpen
  \bibfield  {author} {\bibinfo {author} {\bibfnamefont {E.~J.}\ \bibnamefont
  {Mueller}}\ and\ \bibinfo {author} {\bibfnamefont {T.-L.}\ \bibnamefont
  {Ho}},\ }\bibfield  {title} {\bibinfo {title} {{Two-component Bose--Einstein
  condensates with a large number of vortices}},\ }\href@noop {} {\bibfield
  {journal} {\bibinfo  {journal} {Phys. Rev. Lett.}\ }\textbf {\bibinfo
  {volume} {88}},\ \bibinfo {pages} {180403} (\bibinfo {year}
  {2002})}\BibitemShut {NoStop}%
\bibitem [{\citenamefont {Kasamatsu}\ \emph {et~al.}(2003)\citenamefont
  {Kasamatsu}, \citenamefont {Tsubota},\ and\ \citenamefont
  {Ueda}}]{Kasamatsu2003_Vortex}%
  \BibitemOpen
  \bibfield  {author} {\bibinfo {author} {\bibfnamefont {K.}~\bibnamefont
  {Kasamatsu}}, \bibinfo {author} {\bibfnamefont {M.}~\bibnamefont {Tsubota}},\
  and\ \bibinfo {author} {\bibfnamefont {M.}~\bibnamefont {Ueda}},\ }\bibfield
  {title} {\bibinfo {title} {{Vortex phase diagram in rotating two-component
  Bose--Einstein condensates}},\ }\href@noop {} {\bibfield  {journal} {\bibinfo
   {journal} {Phys. Rev. Lett.}\ }\textbf {\bibinfo {volume} {91}},\ \bibinfo
  {pages} {150406} (\bibinfo {year} {2003})}\BibitemShut {NoStop}%
\bibitem [{\citenamefont {Werner}\ and\ \citenamefont
  {Castin}(2012)}]{Werner2012_General}%
  \BibitemOpen
  \bibfield  {author} {\bibinfo {author} {\bibfnamefont {F.}~\bibnamefont
  {Werner}}\ and\ \bibinfo {author} {\bibfnamefont {Y.}~\bibnamefont
  {Castin}},\ }\bibfield  {title} {\bibinfo {title} {General relations for
  quantum gases in two and three dimensions: two-component fermions},\
  }\href@noop {} {\bibfield  {journal} {\bibinfo  {journal} {Phys. Rev. A}\
  }\textbf {\bibinfo {volume} {86}},\ \bibinfo {pages} {013626} (\bibinfo
  {year} {2012})}\BibitemShut {NoStop}%
\bibitem [{\citenamefont {Bergmann}(1957)}]{Bergmann1957_Two}%
  \BibitemOpen
  \bibfield  {author} {\bibinfo {author} {\bibfnamefont {P.~G.}\ \bibnamefont
  {Bergmann}},\ }\bibfield  {title} {\bibinfo {title} {Two-component spinors in
  general relativity},\ }\href@noop {} {\bibfield  {journal} {\bibinfo
  {journal} {Phys. Rev.}\ }\textbf {\bibinfo {volume} {107}},\ \bibinfo {pages}
  {624} (\bibinfo {year} {1957})}\BibitemShut {NoStop}%
\bibitem [{\citenamefont {Sachs}(1982)}]{Sachs1982_Spinor}%
  \BibitemOpen
  \bibfield  {author} {\bibinfo {author} {\bibfnamefont {M.}~\bibnamefont
  {Sachs}},\ }\bibinfo {title} {Spinor-quaternion analysis in relativity
  theory},\ in\ \href@noop {} {\emph {\bibinfo {booktitle} {General Relativity
  and Matter: A Spinor Field Theory from Fermis to Light-Years}}}\ (\bibinfo
  {publisher} {Springer Netherlands},\ \bibinfo {address} {Dordrecht},\
  \bibinfo {year} {1982})\ pp.\ \bibinfo {pages} {40--69}\BibitemShut {NoStop}%
\bibitem [{\citenamefont {Adler}(1995)}]{Adler1995_Quaternionic}%
  \BibitemOpen
  \bibfield  {author} {\bibinfo {author} {\bibfnamefont {S.~L.}\ \bibnamefont
  {Adler}},\ }\href@noop {} {\emph {\bibinfo {title} {Quaternionic Quantum
  Mechanics and Quantum Fields}}},\ Vol.~\bibinfo {volume} {88}\ (\bibinfo
  {publisher} {Oxford University Press on Demand},\ \bibinfo {year}
  {1995})\BibitemShut {NoStop}%
\bibitem [{\citenamefont {Dreiner}\ \emph {et~al.}(2010)\citenamefont
  {Dreiner}, \citenamefont {Haber},\ and\ \citenamefont
  {Martin}}]{Dreiner2010_Two}%
  \BibitemOpen
  \bibfield  {author} {\bibinfo {author} {\bibfnamefont {H.~K.}\ \bibnamefont
  {Dreiner}}, \bibinfo {author} {\bibfnamefont {H.~E.}\ \bibnamefont {Haber}},\
  and\ \bibinfo {author} {\bibfnamefont {S.~P.}\ \bibnamefont {Martin}},\
  }\bibfield  {title} {\bibinfo {title} {{Two-component spinor techniques and
  Feynman rules for quantum field theory and supersymmetry}},\ }\href@noop {}
  {\bibfield  {journal} {\bibinfo  {journal} {Phys. Rep.}\ }\textbf {\bibinfo
  {volume} {494}},\ \bibinfo {pages} {1} (\bibinfo {year} {2010})}\BibitemShut
  {NoStop}%
\bibitem [{\citenamefont {Bjorken}\ and\ \citenamefont
  {Drell}(1964)}]{Bjorken1964_Relativistic}%
  \BibitemOpen
  \bibfield  {author} {\bibinfo {author} {\bibfnamefont {J.~D.}\ \bibnamefont
  {Bjorken}}\ and\ \bibinfo {author} {\bibfnamefont {S.~D.}\ \bibnamefont
  {Drell}},\ }\href@noop {} {\emph {\bibinfo {title} {Relativistic Quantum
  Mechanics}}}\ (\bibinfo  {publisher} {McGraw-Hill},\ \bibinfo {address} {New
  York},\ \bibinfo {year} {1964})\BibitemShut {NoStop}%
\bibitem [{\citenamefont {Davydov}(1965)}]{Davydov1965_Quantum}%
  \BibitemOpen
  \bibfield  {author} {\bibinfo {author} {\bibfnamefont {A.~S.}\ \bibnamefont
  {Davydov}},\ }\href@noop {} {\emph {\bibinfo {title} {Quantum Mechanics}}},\
  \bibinfo {edition} {2nd}\ ed.\ (\bibinfo  {publisher} {Pergamon},\ \bibinfo
  {address} {Oxford},\ \bibinfo {year} {1965})\BibitemShut {NoStop}%
\bibitem [{\citenamefont {Messiah}(1968)}]{Messiah1968_Quantum}%
  \BibitemOpen
  \bibfield  {author} {\bibinfo {author} {\bibfnamefont {A.}~\bibnamefont
  {Messiah}},\ }\href@noop {} {\emph {\bibinfo {title} {Quantum Mechanics}}},\
  Vol.~\bibinfo {volume} {II}\ (\bibinfo  {publisher} {Wiley},\ \bibinfo
  {address} {New York},\ \bibinfo {year} {1968})\BibitemShut {NoStop}%
\bibitem [{\citenamefont {Chern}\ \emph {et~al.}(2016)\citenamefont {Chern},
  \citenamefont {Knöppel}, \citenamefont {Pinkall}, \citenamefont
  {Schröder},\ and\ \citenamefont {Weißmann}}]{Chern2016_Schrodinger}%
  \BibitemOpen
  \bibfield  {author} {\bibinfo {author} {\bibfnamefont {A.}~\bibnamefont
  {Chern}}, \bibinfo {author} {\bibfnamefont {F.}~\bibnamefont {Knöppel}},
  \bibinfo {author} {\bibfnamefont {U.}~\bibnamefont {Pinkall}}, \bibinfo
  {author} {\bibfnamefont {P.}~\bibnamefont {Schröder}},\ and\ \bibinfo
  {author} {\bibfnamefont {S.}~\bibnamefont {Weißmann}},\ }\bibfield  {title}
  {\bibinfo {title} {{Schr\"odinger's smoke}},\ }\href@noop {} {\bibfield
  {journal} {\bibinfo  {journal} {ACM Trans. Graph.}\ }\textbf {\bibinfo
  {volume} {35}},\ \bibinfo {pages} {1} (\bibinfo {year} {2016})}\BibitemShut
  {NoStop}%
\bibitem [{\citenamefont {Chern}(2017)}]{Chern2017_Fluid}%
  \BibitemOpen
  \bibfield  {author} {\bibinfo {author} {\bibfnamefont {A.}~\bibnamefont
  {Chern}},\ }\emph {\bibinfo {title} {{Fluid dynamics with incompressible
  Schr\"odinger flow}}},\ \href@noop {} {\bibinfo {type} {Phd thesis}},\
  \bibinfo  {school} {California Institute of Technology, Pasadena, CA}
  (\bibinfo {year} {2017})\BibitemShut {NoStop}%
\bibitem [{\citenamefont {Gross}(1961)}]{Gross1961_Structure}%
  \BibitemOpen
  \bibfield  {author} {\bibinfo {author} {\bibfnamefont {E.~P.}\ \bibnamefont
  {Gross}},\ }\bibfield  {title} {\bibinfo {title} {{Structure of a quantized
  vortex in boson systems}},\ }\href@noop {} {\bibfield  {journal} {\bibinfo
  {journal} {Il Nuovo Cimento}\ }\textbf {\bibinfo {volume} {20}},\ \bibinfo
  {pages} {454} (\bibinfo {year} {1961})}\BibitemShut {NoStop}%
\bibitem [{\citenamefont {Pitaevskii}(1961)}]{Pitaevskii1961_Vortex}%
  \BibitemOpen
  \bibfield  {author} {\bibinfo {author} {\bibfnamefont {L.~P.}\ \bibnamefont
  {Pitaevskii}},\ }\bibfield  {title} {\bibinfo {title} {{Vortex lines in an
  imperfect Bose gas}},\ }\href@noop {} {\bibfield  {journal} {\bibinfo
  {journal} {Sov. Phys. JETP.}\ }\textbf {\bibinfo {volume} {13}},\ \bibinfo
  {pages} {451} (\bibinfo {year} {1961})}\BibitemShut {NoStop}%
\bibitem [{\citenamefont {Hopf}(1931)}]{Hopf1931_Uber}%
  \BibitemOpen
  \bibfield  {author} {\bibinfo {author} {\bibfnamefont {H.}~\bibnamefont
  {Hopf}},\ }\bibfield  {title} {\bibinfo {title} {{\"Uber die Abbildungen der
  Dreidimensionalen Sph\"are auf die Kugelfl\"ache}},\ }\href@noop {}
  {\bibfield  {journal} {\bibinfo  {journal} {Math. Ann.}\ }\textbf {\bibinfo
  {volume} {104}},\ \bibinfo {pages} {637–665} (\bibinfo {year}
  {1931})}\BibitemShut {NoStop}%
\bibitem [{\citenamefont {Moreau}(1961)}]{Moreau1961_Constantes}%
  \BibitemOpen
  \bibfield  {author} {\bibinfo {author} {\bibfnamefont {J.~J.}\ \bibnamefont
  {Moreau}},\ }\bibfield  {title} {\bibinfo {title} {Constantes d'un \^ilot
  tourbillonnaire en fluide parfait barotrope},\ }\href@noop {} {\bibfield
  {journal} {\bibinfo  {journal} {C. R. Acad. Sci. Paris}\ }\textbf {\bibinfo
  {volume} {252}},\ \bibinfo {pages} {2810} (\bibinfo {year}
  {1961})}\BibitemShut {NoStop}%
\bibitem [{\citenamefont {Moffatt}(1969)}]{Moffatt1969_The}%
  \BibitemOpen
  \bibfield  {author} {\bibinfo {author} {\bibfnamefont {H.~K.}\ \bibnamefont
  {Moffatt}},\ }\bibfield  {title} {\bibinfo {title} {The degree of knottedness
  of tangled vortex lines},\ }\href@noop {} {\bibfield  {journal} {\bibinfo
  {journal} {J. Fluid Mech.}\ }\textbf {\bibinfo {volume} {35}},\ \bibinfo
  {pages} {117} (\bibinfo {year} {1969})}\BibitemShut {NoStop}%
\bibitem [{\citenamefont {Meng}\ \emph {et~al.}(2023)\citenamefont {Meng},
  \citenamefont {Shen},\ and\ \citenamefont {Yang}}]{Meng2023_Evolution}%
  \BibitemOpen
  \bibfield  {author} {\bibinfo {author} {\bibfnamefont {Z.}~\bibnamefont
  {Meng}}, \bibinfo {author} {\bibfnamefont {W.}~\bibnamefont {Shen}},\ and\
  \bibinfo {author} {\bibfnamefont {Y.}~\bibnamefont {Yang}},\ }\bibfield
  {title} {\bibinfo {title} {Evolution of dissipative fluid flows with imposed
  helicity conservation},\ }\href@noop {} {\bibfield  {journal} {\bibinfo
  {journal} {J. Fluid Mech.}\ }\textbf {\bibinfo {volume} {954}},\ \bibinfo
  {pages} {A36} (\bibinfo {year} {2023})}\BibitemShut {NoStop}%
\bibitem [{\citenamefont {Yang}\ and\ \citenamefont
  {Pullin}(2010)}]{Yang2010_On}%
  \BibitemOpen
  \bibfield  {author} {\bibinfo {author} {\bibfnamefont {Y.}~\bibnamefont
  {Yang}}\ and\ \bibinfo {author} {\bibfnamefont {D.~I.}\ \bibnamefont
  {Pullin}},\ }\bibfield  {title} {\bibinfo {title} {{On Lagrangian and
  vortex-surface fields for flows with Taylor--Green and Kida--Pelz initial
  conditions}},\ }\href@noop {} {\bibfield  {journal} {\bibinfo  {journal} {J.
  Fluid Mech.}\ }\textbf {\bibinfo {volume} {661}},\ \bibinfo {pages} {446}
  (\bibinfo {year} {2010})}\BibitemShut {NoStop}%
\bibitem [{\citenamefont {Hao}\ \emph {et~al.}(2019)\citenamefont {Hao},
  \citenamefont {Xiong},\ and\ \citenamefont {Yang}}]{Hao2019_Tracking}%
  \BibitemOpen
  \bibfield  {author} {\bibinfo {author} {\bibfnamefont {J.}~\bibnamefont
  {Hao}}, \bibinfo {author} {\bibfnamefont {S.}~\bibnamefont {Xiong}},\ and\
  \bibinfo {author} {\bibfnamefont {Y.}~\bibnamefont {Yang}},\ }\bibfield
  {title} {\bibinfo {title} {Tracking vortex surfaces frozen in the virtual
  velocity in non-ideal flows},\ }\href@noop {} {\bibfield  {journal} {\bibinfo
   {journal} {J. Fluid Mech.}\ }\textbf {\bibinfo {volume} {863}},\ \bibinfo
  {pages} {513} (\bibinfo {year} {2019})}\BibitemShut {NoStop}%
\bibitem [{\citenamefont {Tao}\ \emph {et~al.}(2021)\citenamefont {Tao},
  \citenamefont {Ren}, \citenamefont {Tong},\ and\ \citenamefont
  {Xiong}}]{Tao2021_Construction}%
  \BibitemOpen
  \bibfield  {author} {\bibinfo {author} {\bibfnamefont {R.}~\bibnamefont
  {Tao}}, \bibinfo {author} {\bibfnamefont {H.}~\bibnamefont {Ren}}, \bibinfo
  {author} {\bibfnamefont {Y.}~\bibnamefont {Tong}},\ and\ \bibinfo {author}
  {\bibfnamefont {S.}~\bibnamefont {Xiong}},\ }\bibfield  {title} {\bibinfo
  {title} {{Construction and evolution of knotted vortex tubes in
  incompressible Schr\"odinger flow}},\ }\href@noop {} {\bibfield  {journal}
  {\bibinfo  {journal} {Phys. Fluids}\ }\textbf {\bibinfo {volume} {33}},\
  \bibinfo {pages} {077112} (\bibinfo {year} {2021})}\BibitemShut {NoStop}%
\bibitem [{\citenamefont {Xiong}\ and\ \citenamefont
  {Yang}(2019)}]{Xiong2019_Identifying}%
  \BibitemOpen
  \bibfield  {author} {\bibinfo {author} {\bibfnamefont {S.}~\bibnamefont
  {Xiong}}\ and\ \bibinfo {author} {\bibfnamefont {Y.}~\bibnamefont {Yang}},\
  }\bibfield  {title} {\bibinfo {title} {Identifying the tangle of vortex tubes
  in homogeneous isotropic turbulence},\ }\href@noop {} {\bibfield  {journal}
  {\bibinfo  {journal} {J. Fluid Mech.}\ }\textbf {\bibinfo {volume} {874}},\
  \bibinfo {pages} {952} (\bibinfo {year} {2019})}\BibitemShut {NoStop}%
\bibitem [{\citenamefont {Shen}\ \emph {et~al.}(2022)\citenamefont {Shen},
  \citenamefont {Yao}, \citenamefont {Hussain},\ and\ \citenamefont
  {Yang}}]{Shen2022_Topological}%
  \BibitemOpen
  \bibfield  {author} {\bibinfo {author} {\bibfnamefont {W.}~\bibnamefont
  {Shen}}, \bibinfo {author} {\bibfnamefont {J.}~\bibnamefont {Yao}}, \bibinfo
  {author} {\bibfnamefont {F.}~\bibnamefont {Hussain}},\ and\ \bibinfo {author}
  {\bibfnamefont {Y.}~\bibnamefont {Yang}},\ }\bibfield  {title} {\bibinfo
  {title} {Topological transition and helicity conversion of vortex knots and
  links},\ }\href@noop {} {\bibfield  {journal} {\bibinfo  {journal} {J. Fluid
  Mech.}\ }\textbf {\bibinfo {volume} {943}},\ \bibinfo {pages} {A41} (\bibinfo
  {year} {2022})}\BibitemShut {NoStop}%
\bibitem [{\citenamefont {Vahala}\ \emph {et~al.}()\citenamefont {Vahala},
  \citenamefont {Soe}, \citenamefont {Zhang}, \citenamefont {Yepez},
  \citenamefont {Vahala}, \citenamefont {Carter},\ and\ \citenamefont
  {Ziegeler}}]{Vahala2011_Unitary}%
  \BibitemOpen
  \bibfield  {author} {\bibinfo {author} {\bibfnamefont {G.}~\bibnamefont
  {Vahala}}, \bibinfo {author} {\bibfnamefont {M.}~\bibnamefont {Soe}},
  \bibinfo {author} {\bibfnamefont {B.}~\bibnamefont {Zhang}}, \bibinfo
  {author} {\bibfnamefont {J.}~\bibnamefont {Yepez}}, \bibinfo {author}
  {\bibfnamefont {L.}~\bibnamefont {Vahala}}, \bibinfo {author} {\bibfnamefont
  {J.}~\bibnamefont {Carter}},\ and\ \bibinfo {author} {\bibfnamefont
  {S.}~\bibnamefont {Ziegeler}},\ }\bibfield  {title} {\bibinfo {title}
  {Unitary qubit lattice simulations of multiscale phenomena in quantum
  turbulence},\ }in\ \href@noop {} {\emph {\bibinfo {booktitle} {SC '11:
  Proceedings of 2011 International Conference for High Performance Computing,
  Networking, Storage and Analysis}}},\ pp.\ \bibinfo {pages}
  {1--11}\BibitemShut {NoStop}%
\bibitem [{\citenamefont {Madeira}\ \emph {et~al.}(2020)\citenamefont
  {Madeira}, \citenamefont {Caracanhas}, \citenamefont {dos Santos},\ and\
  \citenamefont {Bagnato}}]{Madeira2020_Quantum}%
  \BibitemOpen
  \bibfield  {author} {\bibinfo {author} {\bibfnamefont {L.}~\bibnamefont
  {Madeira}}, \bibinfo {author} {\bibfnamefont {M.~A.}\ \bibnamefont
  {Caracanhas}}, \bibinfo {author} {\bibfnamefont {F.~E.~A.}\ \bibnamefont {dos
  Santos}},\ and\ \bibinfo {author} {\bibfnamefont {V.~S.}\ \bibnamefont
  {Bagnato}},\ }\bibfield  {title} {\bibinfo {title} {Quantum turbulence in
  quantum gases},\ }\href@noop {} {\bibfield  {journal} {\bibinfo  {journal}
  {Annu. Rev. Condens. Matter Phys.}\ }\textbf {\bibinfo {volume} {11}},\
  \bibinfo {pages} {37} (\bibinfo {year} {2020})}\BibitemShut {NoStop}%
\bibitem [{\citenamefont {M\"uller}\ \emph {et~al.}(2021)\citenamefont
  {M\"uller}, \citenamefont {Polanco},\ and\ \citenamefont
  {Krstulovic}}]{Muller2021_Intermittency}%
  \BibitemOpen
  \bibfield  {author} {\bibinfo {author} {\bibfnamefont {N.~P.}\ \bibnamefont
  {M\"uller}}, \bibinfo {author} {\bibfnamefont {J.~I.}\ \bibnamefont
  {Polanco}},\ and\ \bibinfo {author} {\bibfnamefont {G.}~\bibnamefont
  {Krstulovic}},\ }\bibfield  {title} {\bibinfo {title} {Intermittency of
  velocity circulation in quantum turbulence},\ }\href@noop {} {\bibfield
  {journal} {\bibinfo  {journal} {Phys. Rev. X}\ }\textbf {\bibinfo {volume}
  {11}},\ \bibinfo {pages} {011053} (\bibinfo {year} {2021})}\BibitemShut
  {NoStop}%
\bibitem [{\citenamefont {She}\ \emph {et~al.}(1990)\citenamefont {She},
  \citenamefont {Jackson},\ and\ \citenamefont
  {Orszag}}]{She1990_Intermittent}%
  \BibitemOpen
  \bibfield  {author} {\bibinfo {author} {\bibfnamefont {Z.-S.}\ \bibnamefont
  {She}}, \bibinfo {author} {\bibfnamefont {E.}~\bibnamefont {Jackson}},\ and\
  \bibinfo {author} {\bibfnamefont {S.~A.}\ \bibnamefont {Orszag}},\ }\bibfield
   {title} {\bibinfo {title} {Intermittent vortex structures in homogeneous
  isotropic turbulence},\ }\href@noop {} {\bibfield  {journal} {\bibinfo
  {journal} {Nature}\ }\textbf {\bibinfo {volume} {344}},\ \bibinfo {pages}
  {226} (\bibinfo {year} {1990})}\BibitemShut {NoStop}%
\bibitem [{\citenamefont {Cardesa}\ \emph {et~al.}(2017)\citenamefont
  {Cardesa}, \citenamefont {Vela-Martín},\ and\ \citenamefont
  {Jim\'enez}}]{Cardesa2017_The}%
  \BibitemOpen
  \bibfield  {author} {\bibinfo {author} {\bibfnamefont {J.~I.}\ \bibnamefont
  {Cardesa}}, \bibinfo {author} {\bibfnamefont {A.}~\bibnamefont
  {Vela-Martín}},\ and\ \bibinfo {author} {\bibfnamefont {J.}~\bibnamefont
  {Jim\'enez}},\ }\bibfield  {title} {\bibinfo {title} {The turbulent cascade
  in five dimensions},\ }\href@noop {} {\bibfield  {journal} {\bibinfo
  {journal} {Science}\ }\textbf {\bibinfo {volume} {357}},\ \bibinfo {pages}
  {782} (\bibinfo {year} {2017})}\BibitemShut {NoStop}%
\bibitem [{\citenamefont {Liu}\ \emph {et~al.}(2022)\citenamefont {Liu},
  \citenamefont {Chen}, \citenamefont {Shu}, \citenamefont {Chew},
  \citenamefont {Khoo}, \citenamefont {Zhao},\ and\ \citenamefont
  {Cui}}]{Liu2022_Application}%
  \BibitemOpen
  \bibfield  {author} {\bibinfo {author} {\bibfnamefont {Y.~Y.}\ \bibnamefont
  {Liu}}, \bibinfo {author} {\bibfnamefont {Z.}~\bibnamefont {Chen}}, \bibinfo
  {author} {\bibfnamefont {C.}~\bibnamefont {Shu}}, \bibinfo {author}
  {\bibfnamefont {S.~C.}\ \bibnamefont {Chew}}, \bibinfo {author}
  {\bibfnamefont {B.~C.}\ \bibnamefont {Khoo}}, \bibinfo {author}
  {\bibfnamefont {X.}~\bibnamefont {Zhao}},\ and\ \bibinfo {author}
  {\bibfnamefont {Y.~D.}\ \bibnamefont {Cui}},\ }\bibfield  {title} {\bibinfo
  {title} {Application of a variational hybrid quantum-classical algorithm to
  heat conduction equation and analysis of time complexity},\ }\href@noop {}
  {\bibfield  {journal} {\bibinfo  {journal} {Phys. Fluids}\ }\textbf {\bibinfo
  {volume} {34}},\ \bibinfo {pages} {117121} (\bibinfo {year}
  {2022})}\BibitemShut {NoStop}%
\bibitem [{\citenamefont {Patankar}\ and\ \citenamefont
  {Spalding}(1972)}]{Patankar1972_A}%
  \BibitemOpen
  \bibfield  {author} {\bibinfo {author} {\bibfnamefont {S.}~\bibnamefont
  {Patankar}}\ and\ \bibinfo {author} {\bibfnamefont {D.}~\bibnamefont
  {Spalding}},\ }\bibfield  {title} {\bibinfo {title} {A calculation procedure
  for heat, mass and momentum transfer in three-dimensional parabolic flows},\
  }\href@noop {} {\bibfield  {journal} {\bibinfo  {journal} {Int. J. Heat Mass
  Transf.}\ }\textbf {\bibinfo {volume} {15}},\ \bibinfo {pages} {1787}
  (\bibinfo {year} {1972})}\BibitemShut {NoStop}%
\bibitem [{\citenamefont {Kim}\ and\ \citenamefont
  {Moin}(1985)}]{Kim1985_Application}%
  \BibitemOpen
  \bibfield  {author} {\bibinfo {author} {\bibfnamefont {J.}~\bibnamefont
  {Kim}}\ and\ \bibinfo {author} {\bibfnamefont {P.}~\bibnamefont {Moin}},\
  }\bibfield  {title} {\bibinfo {title} {{Application of a fractional-step
  method to incompressible Navier--Stokes equations}},\ }\href@noop {}
  {\bibfield  {journal} {\bibinfo  {journal} {J. Comput. Phys.}\ }\textbf
  {\bibinfo {volume} {59}},\ \bibinfo {pages} {308} (\bibinfo {year}
  {1985})}\BibitemShut {NoStop}%
\bibitem [{\citenamefont {Issa}\ \emph {et~al.}(1986)\citenamefont {Issa},
  \citenamefont {Gosman},\ and\ \citenamefont {Watkins}}]{Issa1986_The}%
  \BibitemOpen
  \bibfield  {author} {\bibinfo {author} {\bibfnamefont {R.}~\bibnamefont
  {Issa}}, \bibinfo {author} {\bibfnamefont {A.}~\bibnamefont {Gosman}},\ and\
  \bibinfo {author} {\bibfnamefont {A.}~\bibnamefont {Watkins}},\ }\bibfield
  {title} {\bibinfo {title} {The computation of compressible and incompressible
  recirculating flows by a non-iterative implicit scheme},\ }\href@noop {}
  {\bibfield  {journal} {\bibinfo  {journal} {J. Comput. Phys.}\ }\textbf
  {\bibinfo {volume} {62}},\ \bibinfo {pages} {66} (\bibinfo {year}
  {1986})}\BibitemShut {NoStop}%
\bibitem [{\citenamefont {Benenti}\ and\ \citenamefont
  {Strini}(2008)}]{Benenti2008_Quantum}%
  \BibitemOpen
  \bibfield  {author} {\bibinfo {author} {\bibfnamefont {G.}~\bibnamefont
  {Benenti}}\ and\ \bibinfo {author} {\bibfnamefont {G.}~\bibnamefont
  {Strini}},\ }\bibfield  {title} {\bibinfo {title} {{Quantum simulation of the
  single-particle Schr\"odinger equation}},\ }\href@noop {} {\bibfield
  {journal} {\bibinfo  {journal} {Am. J. Phys.}\ }\textbf {\bibinfo {volume}
  {76}},\ \bibinfo {pages} {657} (\bibinfo {year} {2008})}\BibitemShut
  {NoStop}%
\bibitem [{\citenamefont {Ostrowski}(2017)}]{Ostrowski2017_Quantum}%
  \BibitemOpen
  \bibfield  {author} {\bibinfo {author} {\bibfnamefont {M.}~\bibnamefont
  {Ostrowski}},\ }\bibfield  {title} {\bibinfo {title} {{Quantum simulation of
  two interacting Schr\"odinger particles}},\ }\href@noop {} {\bibfield
  {journal} {\bibinfo  {journal} {Open Syst. Inf. Dyn.}\ }\textbf {\bibinfo
  {volume} {23}},\ \bibinfo {pages} {1650020} (\bibinfo {year}
  {2017})}\BibitemShut {NoStop}%
\bibitem [{\citenamefont {Bogdanov}\ \emph {et~al.}(2021)\citenamefont
  {Bogdanov}, \citenamefont {Bogdanova}, \citenamefont {Fastovets},\ and\
  \citenamefont {Lukichev}}]{Bogdanov2021_Solution}%
  \BibitemOpen
  \bibfield  {author} {\bibinfo {author} {\bibfnamefont {Y.~I.}\ \bibnamefont
  {Bogdanov}}, \bibinfo {author} {\bibfnamefont {N.~A.}\ \bibnamefont
  {Bogdanova}}, \bibinfo {author} {\bibfnamefont {D.~V.}\ \bibnamefont
  {Fastovets}},\ and\ \bibinfo {author} {\bibfnamefont {V.~F.}\ \bibnamefont
  {Lukichev}},\ }\bibfield  {title} {\bibinfo {title} {{Solution of the
  Schr\"odinger equation on a quantum computer by the Zalka--Wiesner method
  including quantum noise}},\ }\href@noop {} {\bibfield  {journal} {\bibinfo
  {journal} {Jetp Lett.}\ }\textbf {\bibinfo {volume} {114}},\ \bibinfo {pages}
  {354} (\bibinfo {year} {2021})}\BibitemShut {NoStop}%
\bibitem [{\citenamefont {Coppersmith}(1994)}]{Coppersmith1994_An}%
  \BibitemOpen
  \bibfield  {author} {\bibinfo {author} {\bibfnamefont {D.}~\bibnamefont
  {Coppersmith}},\ }\href@noop {} {\emph {\bibinfo {title} {{An approximate
  Fourier transform useful in quantum factoring}}}},\ \bibinfo {type} {Tech.
  Rep.}\ (\bibinfo  {institution} {IBM Research Report RC 19642},\ \bibinfo
  {year} {1994})\BibitemShut {NoStop}%
\bibitem [{\citenamefont {Jozsa}(1998)}]{Jozsa1998_Quantum}%
  \BibitemOpen
  \bibfield  {author} {\bibinfo {author} {\bibfnamefont {R.}~\bibnamefont
  {Jozsa}},\ }\bibfield  {title} {\bibinfo {title} {{Quantum algorithms and the
  Fourier transform}},\ }\href@noop {} {\bibfield  {journal} {\bibinfo
  {journal} {Proc. R. Soc. Lond. A.}\ }\textbf {\bibinfo {volume} {454}},\
  \bibinfo {pages} {323} (\bibinfo {year} {1998})}\BibitemShut {NoStop}%
\bibitem [{\citenamefont {Weinstein}\ \emph {et~al.}(2001)\citenamefont
  {Weinstein}, \citenamefont {Pravia}, \citenamefont {Fortunato}, \citenamefont
  {Lloyd},\ and\ \citenamefont {Cory}}]{Weinstein2001_Implementation}%
  \BibitemOpen
  \bibfield  {author} {\bibinfo {author} {\bibfnamefont {Y.~S.}\ \bibnamefont
  {Weinstein}}, \bibinfo {author} {\bibfnamefont {M.~A.}\ \bibnamefont
  {Pravia}}, \bibinfo {author} {\bibfnamefont {E.~M.}\ \bibnamefont
  {Fortunato}}, \bibinfo {author} {\bibfnamefont {S.}~\bibnamefont {Lloyd}},\
  and\ \bibinfo {author} {\bibfnamefont {D.~G.}\ \bibnamefont {Cory}},\
  }\bibfield  {title} {\bibinfo {title} {{Implementation of the quantum Fourier
  transform}},\ }\href@noop {} {\bibfield  {journal} {\bibinfo  {journal}
  {Phys. Rev. Lett.}\ }\textbf {\bibinfo {volume} {86}},\ \bibinfo {pages}
  {1889} (\bibinfo {year} {2001})}\BibitemShut {NoStop}%
\bibitem [{\citenamefont {Rodrigues}(2018)}]{Rodrigues2018_Validation}%
  \BibitemOpen
  \bibfield  {author} {\bibinfo {author} {\bibfnamefont {A.}~\bibnamefont
  {Rodrigues}},\ }\emph {\bibinfo {title} {Validation of quantum
  simulations}},\ \href@noop {} {\bibinfo {type} {Phd thesis}},\ \bibinfo
  {school} {Universidade do Minho} (\bibinfo {year} {2018})\BibitemShut
  {NoStop}%
\bibitem [{\citenamefont {Yang}\ \emph {et~al.}(2021)\citenamefont {Yang},
  \citenamefont {Xiong}, \citenamefont {Zhang}, \citenamefont {Feng},
  \citenamefont {Liu},\ and\ \citenamefont {Zhu}}]{Yang2021_Clebsch}%
  \BibitemOpen
  \bibfield  {author} {\bibinfo {author} {\bibfnamefont {S.}~\bibnamefont
  {Yang}}, \bibinfo {author} {\bibfnamefont {S.}~\bibnamefont {Xiong}},
  \bibinfo {author} {\bibfnamefont {Y.}~\bibnamefont {Zhang}}, \bibinfo
  {author} {\bibfnamefont {F.}~\bibnamefont {Feng}}, \bibinfo {author}
  {\bibfnamefont {J.}~\bibnamefont {Liu}},\ and\ \bibinfo {author}
  {\bibfnamefont {B.}~\bibnamefont {Zhu}},\ }\bibfield  {title} {\bibinfo
  {title} {Clebsch gauge fluid},\ }\href@noop {} {\bibfield  {journal}
  {\bibinfo  {journal} {ACM Trans. Graph.}\ }\textbf {\bibinfo {volume} {40}},\
  \bibinfo {pages} {1} (\bibinfo {year} {2021})}\BibitemShut {NoStop}%
\bibitem [{\citenamefont {Arrazola}\ \emph {et~al.}(2019)\citenamefont
  {Arrazola}, \citenamefont {Kalajdzievski}, \citenamefont {Weedbrook},\ and\
  \citenamefont {Lloyd}}]{Arrazola2019_Quantum}%
  \BibitemOpen
  \bibfield  {author} {\bibinfo {author} {\bibfnamefont {J.~M.}\ \bibnamefont
  {Arrazola}}, \bibinfo {author} {\bibfnamefont {T.}~\bibnamefont
  {Kalajdzievski}}, \bibinfo {author} {\bibfnamefont {C.}~\bibnamefont
  {Weedbrook}},\ and\ \bibinfo {author} {\bibfnamefont {S.}~\bibnamefont
  {Lloyd}},\ }\bibfield  {title} {\bibinfo {title} {Quantum algorithm for
  nonhomogeneous linear partial differential equations},\ }\href@noop {}
  {\bibfield  {journal} {\bibinfo  {journal} {Phys. Rev. A}\ }\textbf {\bibinfo
  {volume} {100}},\ \bibinfo {pages} {032306} (\bibinfo {year}
  {2019})}\BibitemShut {NoStop}%
\bibitem [{\citenamefont {Childs}\ and\ \citenamefont
  {Liu}(2020)}]{Childs2020_Quantum}%
  \BibitemOpen
  \bibfield  {author} {\bibinfo {author} {\bibfnamefont {A.~M.}\ \bibnamefont
  {Childs}}\ and\ \bibinfo {author} {\bibfnamefont {J.-P.}\ \bibnamefont
  {Liu}},\ }\bibfield  {title} {\bibinfo {title} {Quantum spectral methods for
  differential equations},\ }\href@noop {} {\bibfield  {journal} {\bibinfo
  {journal} {Commun. Math. Phys.}\ }\textbf {\bibinfo {volume} {375}},\
  \bibinfo {pages} {1427} (\bibinfo {year} {2020})}\BibitemShut {NoStop}%
\bibitem [{\citenamefont {Liu}\ \emph {et~al.}(2021{\natexlab{b}})\citenamefont
  {Liu}, \citenamefont {Wu}, \citenamefont {Wan}, \citenamefont {Pan},
  \citenamefont {Qin}, \citenamefont {Gao},\ and\ \citenamefont
  {Wen}}]{Liu2021_Variational}%
  \BibitemOpen
  \bibfield  {author} {\bibinfo {author} {\bibfnamefont {H.-L.}\ \bibnamefont
  {Liu}}, \bibinfo {author} {\bibfnamefont {Y.-S.}\ \bibnamefont {Wu}},
  \bibinfo {author} {\bibfnamefont {L.-C.}\ \bibnamefont {Wan}}, \bibinfo
  {author} {\bibfnamefont {S.-J.}\ \bibnamefont {Pan}}, \bibinfo {author}
  {\bibfnamefont {S.-J.}\ \bibnamefont {Qin}}, \bibinfo {author} {\bibfnamefont
  {F.}~\bibnamefont {Gao}},\ and\ \bibinfo {author} {\bibfnamefont {Q.-Y.}\
  \bibnamefont {Wen}},\ }\bibfield  {title} {\bibinfo {title} {{Variational
  quantum algorithm for the Poisson equation}},\ }\href@noop {} {\bibfield
  {journal} {\bibinfo  {journal} {Phys. Rev. A}\ }\textbf {\bibinfo {volume}
  {104}},\ \bibinfo {pages} {022418} (\bibinfo {year}
  {2021}{\natexlab{b}})}\BibitemShut {NoStop}%
\bibitem [{\citenamefont {Childs}\ \emph {et~al.}(2021)\citenamefont {Childs},
  \citenamefont {Liu},\ and\ \citenamefont {Ostrander}}]{Childs2021_High}%
  \BibitemOpen
  \bibfield  {author} {\bibinfo {author} {\bibfnamefont {A.~M.}\ \bibnamefont
  {Childs}}, \bibinfo {author} {\bibfnamefont {J.-P.}\ \bibnamefont {Liu}},\
  and\ \bibinfo {author} {\bibfnamefont {A.}~\bibnamefont {Ostrander}},\
  }\bibfield  {title} {\bibinfo {title} {High-precision quantum algorithms for
  partial differential equations},\ }\href@noop {} {\bibfield  {journal}
  {\bibinfo  {journal} {Quantum}\ }\textbf {\bibinfo {volume} {5}},\ \bibinfo
  {pages} {574} (\bibinfo {year} {2021})}\BibitemShut {NoStop}%
\bibitem [{\citenamefont {Wiebe}\ \emph {et~al.}(2010)\citenamefont {Wiebe},
  \citenamefont {Berry}, \citenamefont {H\o~yer},\ and\ \citenamefont
  {Sanders}}]{Wiebe2010_Higher}%
  \BibitemOpen
  \bibfield  {author} {\bibinfo {author} {\bibfnamefont {N.}~\bibnamefont
  {Wiebe}}, \bibinfo {author} {\bibfnamefont {D.}~\bibnamefont {Berry}},
  \bibinfo {author} {\bibfnamefont {P.}~\bibnamefont {H\o~yer}},\ and\ \bibinfo
  {author} {\bibfnamefont {B.~C.}\ \bibnamefont {Sanders}},\ }\bibfield
  {title} {\bibinfo {title} {Higher order decompositions of ordered operator
  exponentials},\ }\href@noop {} {\bibfield  {journal} {\bibinfo  {journal} {J.
  Phys. A-Math. Theor.}\ }\textbf {\bibinfo {volume} {43}},\ \bibinfo {pages}
  {065203} (\bibinfo {year} {2010})}\BibitemShut {NoStop}%
\bibitem [{\citenamefont {Fillion-Gourdeau}\ \emph {et~al.}(2017)\citenamefont
  {Fillion-Gourdeau}, \citenamefont {MacLean},\ and\ \citenamefont
  {Laflamme}}]{Fillion2017_Algorithm}%
  \BibitemOpen
  \bibfield  {author} {\bibinfo {author} {\bibfnamefont {F.}~\bibnamefont
  {Fillion-Gourdeau}}, \bibinfo {author} {\bibfnamefont {S.}~\bibnamefont
  {MacLean}},\ and\ \bibinfo {author} {\bibfnamefont {R.}~\bibnamefont
  {Laflamme}},\ }\bibfield  {title} {\bibinfo {title} {{Algorithm for the
  solution of the Dirac equation on digital quantum computers}},\ }\href@noop
  {} {\bibfield  {journal} {\bibinfo  {journal} {Phys. Rev. A}\ }\textbf
  {\bibinfo {volume} {95}},\ \bibinfo {pages} {042343} (\bibinfo {year}
  {2017})}\BibitemShut {NoStop}%
\bibitem [{\citenamefont {Koch}\ \emph {et~al.}(2019)\citenamefont {Koch},
  \citenamefont {Wessing},\ and\ \citenamefont
  {Alsing}}]{Koch2019_Introduction}%
  \BibitemOpen
  \bibfield  {author} {\bibinfo {author} {\bibfnamefont {D.}~\bibnamefont
  {Koch}}, \bibinfo {author} {\bibfnamefont {L.}~\bibnamefont {Wessing}},\ and\
  \bibinfo {author} {\bibfnamefont {P.~M.}\ \bibnamefont {Alsing}},\
  }\href@noop {} {\bibinfo {title} {{Introduction to coding quantum algorithms:
  A tutorial series using Qiskit}}} (\bibinfo {year} {2019}),\ \Eprint
  {https://arxiv.org/abs/arXiv:1903.04359} {arXiv:1903.04359} \BibitemShut
  {NoStop}%
\bibitem [{\citenamefont {Gardiner}\ \emph {et~al.}(1997)\citenamefont
  {Gardiner}, \citenamefont {Cirac},\ and\ \citenamefont
  {Zoller}}]{Gardiner1997_Quantum}%
  \BibitemOpen
  \bibfield  {author} {\bibinfo {author} {\bibfnamefont {S.~A.}\ \bibnamefont
  {Gardiner}}, \bibinfo {author} {\bibfnamefont {J.~I.}\ \bibnamefont
  {Cirac}},\ and\ \bibinfo {author} {\bibfnamefont {P.}~\bibnamefont
  {Zoller}},\ }\bibfield  {title} {\bibinfo {title} {{Quantum chaos in an ion
  trap: The delta-kicked harmonic oscillator}},\ }\href@noop {} {\bibfield
  {journal} {\bibinfo  {journal} {Phys. Rev. Lett.}\ }\textbf {\bibinfo
  {volume} {79}},\ \bibinfo {pages} {4790} (\bibinfo {year}
  {1997})}\BibitemShut {NoStop}%
\bibitem [{\citenamefont {Smithey}\ \emph {et~al.}(1993)\citenamefont
  {Smithey}, \citenamefont {Beck}, \citenamefont {Raymer},\ and\ \citenamefont
  {Faridani}}]{Smithey1993_Measurement}%
  \BibitemOpen
  \bibfield  {author} {\bibinfo {author} {\bibfnamefont {D.~T.}\ \bibnamefont
  {Smithey}}, \bibinfo {author} {\bibfnamefont {M.}~\bibnamefont {Beck}},
  \bibinfo {author} {\bibfnamefont {M.~G.}\ \bibnamefont {Raymer}},\ and\
  \bibinfo {author} {\bibfnamefont {A.}~\bibnamefont {Faridani}},\ }\bibfield
  {title} {\bibinfo {title} {Measurement of the wigner distribution and the
  density matrix of a light mode using optical homodyne tomography: Application
  to squeezed states and the vacuum},\ }\href@noop {} {\bibfield  {journal}
  {\bibinfo  {journal} {Phys. Rev. Lett.}\ }\textbf {\bibinfo {volume} {70}},\
  \bibinfo {pages} {1244} (\bibinfo {year} {1993})}\BibitemShut {NoStop}%
\bibitem [{\citenamefont {Breitenbach}\ \emph {et~al.}(1997)\citenamefont
  {Breitenbach}, \citenamefont {Schiller},\ and\ \citenamefont
  {Mlynek}}]{Breitenbach1997_Measurement}%
  \BibitemOpen
  \bibfield  {author} {\bibinfo {author} {\bibfnamefont {G.}~\bibnamefont
  {Breitenbach}}, \bibinfo {author} {\bibfnamefont {S.}~\bibnamefont
  {Schiller}},\ and\ \bibinfo {author} {\bibfnamefont {J.}~\bibnamefont
  {Mlynek}},\ }\bibfield  {title} {\bibinfo {title} {Measurement of the quantum
  states of squeezed light},\ }\href@noop {} {\bibfield  {journal} {\bibinfo
  {journal} {Nature}\ }\textbf {\bibinfo {volume} {387}},\ \bibinfo {pages}
  {471} (\bibinfo {year} {1997})}\BibitemShut {NoStop}%
\bibitem [{\citenamefont {James}\ \emph {et~al.}(2001)\citenamefont {James},
  \citenamefont {Kwiat}, \citenamefont {Munro},\ and\ \citenamefont
  {White}}]{James2001_Measurement}%
  \BibitemOpen
  \bibfield  {author} {\bibinfo {author} {\bibfnamefont {D.~F.~V.}\
  \bibnamefont {James}}, \bibinfo {author} {\bibfnamefont {P.~G.}\ \bibnamefont
  {Kwiat}}, \bibinfo {author} {\bibfnamefont {W.~J.}\ \bibnamefont {Munro}},\
  and\ \bibinfo {author} {\bibfnamefont {A.~G.}\ \bibnamefont {White}},\
  }\bibfield  {title} {\bibinfo {title} {Measurement of qubits},\ }\href@noop
  {} {\bibfield  {journal} {\bibinfo  {journal} {Phys. Rev. A}\ }\textbf
  {\bibinfo {volume} {64}},\ \bibinfo {pages} {052312} (\bibinfo {year}
  {2001})}\BibitemShut {NoStop}%
\bibitem [{\citenamefont {Vallone}\ and\ \citenamefont
  {Dequal}(2016)}]{Vallone2016_Strong}%
  \BibitemOpen
  \bibfield  {author} {\bibinfo {author} {\bibfnamefont {G.}~\bibnamefont
  {Vallone}}\ and\ \bibinfo {author} {\bibfnamefont {D.}~\bibnamefont
  {Dequal}},\ }\bibfield  {title} {\bibinfo {title} {Strong measurements give a
  better direct measurement of the quantum wave function},\ }\href@noop {}
  {\bibfield  {journal} {\bibinfo  {journal} {Phys. Rev. Lett.}\ }\textbf
  {\bibinfo {volume} {116}},\ \bibinfo {pages} {040502} (\bibinfo {year}
  {2016})}\BibitemShut {NoStop}%
\bibitem [{\citenamefont {Gilbert}(1955)}]{Gilbert1955_A}%
  \BibitemOpen
  \bibfield  {author} {\bibinfo {author} {\bibfnamefont {T.~L.}\ \bibnamefont
  {Gilbert}},\ }\bibfield  {title} {\bibinfo {title} {{A Lagrangian formulation
  of the gyromagnetic equation of the magnetization field}},\ }\href@noop {}
  {\bibfield  {journal} {\bibinfo  {journal} {Phys. Rev.}\ }\textbf {\bibinfo
  {volume} {100}},\ \bibinfo {pages} {1243} (\bibinfo {year}
  {1955})}\BibitemShut {NoStop}%
\bibitem [{\citenamefont {Landau}\ and\ \citenamefont
  {Lifshitz}(1935)}]{Landau1935_On}%
  \BibitemOpen
  \bibfield  {author} {\bibinfo {author} {\bibfnamefont {L.}~\bibnamefont
  {Landau}}\ and\ \bibinfo {author} {\bibfnamefont {E.}~\bibnamefont
  {Lifshitz}},\ }\bibfield  {title} {\bibinfo {title} {On the theory of the
  dispersion of magnetic permeability in ferromagnetic bodies},\ }\href@noop {}
  {\bibfield  {journal} {\bibinfo  {journal} {Phys. Zeitsch. Sow.}\ }\textbf
  {\bibinfo {volume} {8}},\ \bibinfo {pages} {153} (\bibinfo {year}
  {1935})}\BibitemShut {NoStop}%
\bibitem [{\citenamefont {Taylor}\ and\ \citenamefont
  {Green}(1937)}]{Taylor1937_Mechanism}%
  \BibitemOpen
  \bibfield  {author} {\bibinfo {author} {\bibfnamefont {G.~I.}\ \bibnamefont
  {Taylor}}\ and\ \bibinfo {author} {\bibfnamefont {A.~E.}\ \bibnamefont
  {Green}},\ }\bibfield  {title} {\bibinfo {title} {Mechanism of the production
  of small eddies from large ones},\ }\href@noop {} {\bibfield  {journal}
  {\bibinfo  {journal} {Proc. R. Soc. London Ser. A-Math. Phys. Eng. Sci.}\
  }\textbf {\bibinfo {volume} {158}},\ \bibinfo {pages} {499} (\bibinfo {year}
  {1937})}\BibitemShut {NoStop}%
\bibitem [{\citenamefont {Clebsch}(1859)}]{Clebsch1859_Ueber}%
  \BibitemOpen
  \bibfield  {author} {\bibinfo {author} {\bibfnamefont {A.}~\bibnamefont
  {Clebsch}},\ }\bibfield  {title} {\bibinfo {title} {{Ueber die integration
  der hydrodynamischen gleichungen}},\ }\href@noop {} {\bibfield  {journal}
  {\bibinfo  {journal} {J. Reine Angew. Math.}\ }\textbf {\bibinfo {volume}
  {56}},\ \bibinfo {pages} {1} (\bibinfo {year} {1859})}\BibitemShut {NoStop}%
\bibitem [{\citenamefont {Nore}\ \emph {et~al.}(1997)\citenamefont {Nore},
  \citenamefont {Abid},\ and\ \citenamefont {Brachet}}]{Nore1997_Decaying}%
  \BibitemOpen
  \bibfield  {author} {\bibinfo {author} {\bibfnamefont {C.}~\bibnamefont
  {Nore}}, \bibinfo {author} {\bibfnamefont {M.}~\bibnamefont {Abid}},\ and\
  \bibinfo {author} {\bibfnamefont {M.~E.}\ \bibnamefont {Brachet}},\
  }\bibfield  {title} {\bibinfo {title} {{Decaying Kolmogorov turbulence in a
  model of superflow}},\ }\href@noop {} {\bibfield  {journal} {\bibinfo
  {journal} {Phys. Fluids}\ }\textbf {\bibinfo {volume} {9}},\ \bibinfo {pages}
  {2644} (\bibinfo {year} {1997})}\BibitemShut {NoStop}%
\end{thebibliography}
\end{document}